\documentclass[letterpaper,12pt]{amsart}

\usepackage{latexsym, amsmath, amssymb,amsthm, natbib,verbatim, xy}
\xyoption{all}

\usepackage{tgpagella}
\usepackage{mathpazo}

\usepackage[left=1in,top=1in,right=1in,bottom=1in]{geometry}

\usepackage{setspace,color}
\usepackage{graphicx}
\usepackage{enumerate}
\usepackage[multiple]{footmisc}
\usepackage[leqno]{amsmath}
\usepackage{hyperref}
\usepackage{mathtools}

\usepackage{etoolbox}
\patchcmd{\section}{\scshape}{\bfseries}{}{}
\makeatletter
\renewcommand{\@secnumfont}{\bfseries}
\makeatother

\renewenvironment{quote}{%
   \list{}{%
     \leftmargin0.75cm   
     \rightmargin0.5cm
   }
   \item\relax
}
{\endlist}

\newenvironment{indquote}{
    \begin{quote}
    \fontfamily{lmtt}\selectfont\small
}
{
    \end{quote}
}

\usepackage{tikz}
\newcommand*\circled[1]{\tikz[baseline=(char.base)]{
    \node[shape=circle,draw,inner sep=1pt, scale=0.4] (char) {#1};}}

\newtheorem{theorem}{Theorem}
\newtheorem{corollary}{Corollary}
\newtheorem{lemma}{Lemma}

\newtheorem{claim}{Claim}

\theoremstyle{definition}

\newtheorem{example}{Example}

\newtheorem{definition}{Definition}

\def\cala{\mathcal{A}}   
 \def\calc{\mathcal{C}}  \def\calt{\mathcal{T}}
\def\calr{\mathcal{R}} \def\cali{\mathcal{I}}  \def\calv{\mathcal{V}} \def\calp{\mathcal{P}}
\def\calj{\mathcal{J}}

\newcommand{\abs}[1]{\left| #1 \right|}

\def\cenvcat{C_{\circled{M},j}^{2s,v}}

\renewenvironment{quote}{%
   \list{}{%
     \leftmargin0.75cm   
     \rightmargin0.5cm
   }
   \item\relax
}
{\endlist}

\newcommand*{\medcap}{\mathbin{\scalebox{1.2}{\ensuremath{\cap}}}}

\newcommand*{\medcup}{\mathbin{\scalebox{1.2}{\ensuremath{\cup}}}}

\begin{document}

\title{Constitutional Implementation of Affirmative Action Policies in India}

\author[S\"{o}nmez and Yenmez]{Tayfun S\"{o}nmez \and M. Bumin Yenmez}\thanks{This version subsumes and supersedes \cite{sonmez/yenmez:19b} and Section 6.1 of \cite{sonmez/yenmez:19c}.
We are grateful to the Editor for his exceptionally diligent review, very clear guidance, and most notably, for sending our paper to a lawyer and scholar of the Indian legal system 
(Referee 1) to assess the soundness of our interpretation of the relevant statutes. We are grateful to Referee 1 for their careful assessment of our legal analysis, and to Referees 2 and 3 for their comments, which improved the paper.  
S\"{o}nmez is affiliated with the Department of Economics, Boston College, 140 Commonwealth Ave, Chestnut Hill, MA, 02467. 
Yenmez is affiliated with the Department of Economics, Washington University in St. Louis, 1 Brookings Ave, St. Louis, MO 63130.  Emails: \texttt{sonmezt@bc.edu},
\texttt{bumin@wustl.edu}.}

\date{First version: March 2022. This version: February 2024.}

\begin{abstract}
India is home to a comprehensive affirmative action program that reserves a fraction of positions at governmental institutions for various disadvantaged groups. 
While there is a Supreme Court-endorsed mechanism to implement these reservation policies when all positions are identical, courts have refrained from 
endorsing explicit mechanisms when positions are heterogeneous. This lacuna has resulted in widespread adoption of unconstitutional mechanisms,
 countless lawsuits, and inconsistent court rulings. By formulating mandates from the landmark Supreme Court judgment \textit{Saurav Yadav} (2020) 
 as technical axioms, we show that the 2SMH-DA mechanism is uniquely suited to overcome these challenges.
\end{abstract}

\maketitle



\newpage

\section{Introduction}

The landmark Supreme Court judgment \textit{Indra Sawhney v. Union of India (1992)} 
formulates a number of affirmative action provisions built into the Constitution of India.\footnote{Commonly referred to as the \textit{Mandal Commission Case}, 
this ruling is widely regarded as one of the most significant 
judgments in the history of the Supreme Court of India. Accessible at \url{https://indiankanoon.org/doc/1363234/} (retrieved on 02/20/2022).} 
Allocation of seats in the country’s legislative bodies, public employment, and publicly funded educational
institutions are governed by the principles outlined in this ruling.

Under this judgment, equity is ingrained within a merit-based framework through two affirmative action policies: \textit{vertical reservations} (VR) and \textit{horizontal reservations} (HR). 
The VR policy, envisioned as the primary  affirmative action provision, is mandated to operate on an ``over-and-above" basis for each VR-protected group. 
In practical terms, this means that if a member of a VR-protected group qualifies for an open position based on merit, they must be awarded that position instead of using up a reserved one. 
Consequently, under the VR policy, reserved positions are exclusively granted to eligible individuals who do not qualify for open positions. 
The VR policy primarily targets historically oppressed classes, notably \textit{Scheduled Castes, Scheduled Tribes,} and \textit{Other Backward Classes}.

In contrast, the HR policy, granted to groups such as persons with disabilities or women, operates on a ``minimum guarantee" basis, 
serving as a secondary affirmative action provision. In practical terms, any position awarded to a member of an HR-protected group contributes towards HR protections. 
Typically, the HR policy is implemented separately within open positions and each category of VR-protected positions.

As they are stated in \textit{Indra Sawhney (1992)}, the formulation of VR and HR policies becomes airtight under the following three conditions:
\begin{enumerate}
\item \textit{Homogeneity}: All positions are identical.
\item \textit{Stand-alone implementation}: Only one of the reservation policies is implemented.
\item \textit{Non-overlapping protected groups}: No one belongs to multiple protected groups.
\end{enumerate}
However, despite its significance, this landmark judgment has not offered detailed guidance on scenarios where any combination of the three 
conditions fails—a situation that commonly arises in practical applications across the country. 
For over three decades, this ambiguity has led to the adoption of numerous flawed allocation mechanisms in India. 
Furthermore, it has sparked countless litigations and resulted in inconsistent rulings at all three levels of the country's judicial system: the Supreme Court, the High Courts, and District Courts.

In an effort to address the ambiguity arising when both VR and HR policies are implemented jointly, thus leading to overlapping protected groups, 
the \textit{SCI-AKG choice rule} was established and enforced nationwide in a subsequent Supreme Court judgment, 
\textit{Anil Kumar Gupta vs State Of Uttar Pradesh (1995)}.\footnote{The ruling can be accessed at \url{https://indiankanoon.org/doc/1055016/} (retrieved on 02/20/2024).} 
However, a critical flaw in this procedure exacerbated the crisis over the next 25 years, as documented by \cite{sonmez/yenmez:19a, sonmez/yenmez:22}. 
To address this shortcoming, a remedy was also proposed by the authors through a procedure called the \textit{two-step meritorious horizontal (2SMH) choice rule}. 
Parallel to the analysis and policy recommendations presented in \cite{sonmez/yenmez:19a, sonmez/yenmez:22}, the flawed SCI-AKG choice rule was rescinded, 
and the 2SMH choice rule—discovered independently by the judiciary—was endorsed in the Supreme Court judgment 
\textit{Saurav Yadav vs The State Of Uttar Pradesh (2020)}.\footnote{\textit{Saurav Yadav (2020)} assumes no possible overlap between HR-protected groups, 
leading to the 2SMH choice rule taking a simpler form known as the \textit{two-step minimum guarantee (2SMG) choice rule}. The ruling can be accessed at \url{https://indiankanoon.org/doc/27820739/} (retrieved on 02/20/2024).}

While \textit{Saurav Yadav (2020)} has resolved legal inconsistencies and provided a well-defined procedure for cases where all positions are identical, to the best of our knowledge, 
no mechanism has been judicially mandated or endorsed to date for allocating positions that are heterogeneous. This gap has compelled local agencies to devise their own mechanisms, 
none of which fully adhere to the principles outlined in \textit{Indra Sawhney (1992)}. Consequently, these mechanisms are frequently challenged in court, 
resulting in numerous inconsistent decisions across all three tiers of the Indian Judicial System. For a comprehensive legal analysis of three particularly problematic judgments, 
refer to  Sections  \ref{subsec-RameshRam}, \ref{subsec-Sharan} and \ref{subsec-Samiti} of the Online Appendix.

\subsection{Minimalist Market Design}

In this paper, we turn to \textit{minimalist market design} \citep{sonmez:23} to formulate and propose a mechanism for the general version of the problem, 
involving joint implementation of VR and HR policies, overlapping protected groups (to the extent possible in India),\footnote{In India, VR-protected groups do not overlap. 
However, this aspect of the Indian reservation system faced a challenge in 2019 when a VR-protected group was introduced based on economic deprivation. 
The Supreme Court addressed this issue, maintaining the non-overlapping nature of VR-protected groups with a 3-2 split verdict, as discussed in \cite{sonmez/unver:22}. 
However, HR-protected groups not only overlap with VR-protected groups but also often overlap with other HR-protected groups.} and heterogeneous positions.

Minimalist market design is an institution design framework that relies on the underlying objectives of an institution, namely, its ``mission'', along with the operational details of the existing institution, 
to formulate a reformed institution whenever the existing one falls short of meeting certain aspects of its mission. Under this approach, the root causes of the existing institution's failures are identified, 
and a new institution that aligns with the mission is designed by addressing only the flawed aspects of the existing system. This is the essence of the term ``minimalist'' in this institution design framework.

Naturally, this paradigm is most effective in environments where the mission of the institution is  clear and the root causes of institution failures can be identified. 
This renders our application an ideal candidate for adopting minimalist market design.

\subsection{The Mission: Implementing \textit{Saurav Yadav (2020)} Mandates}

Making minimalist market design especially promising to address the failures later presented in Section \ref{sec:migration-adjustment}, 
the objectives of the institution—specifically, the principles governing the joint operation of VR and HR policies—are already formulated in \textit{Saurav Yadav (2020)}. 
While the focus of this judgment pertains to scenarios with homogeneous positions, it effectively clarifies ambiguities introduced in \textit{Indra Sawhney (1992)} 
concerning the joint implementation of VR and HR policies across all problem variations.\footnote{See, for example,  paragraph 31 in \textit{Saurav Yadav (2020)} in 
Appendix \ref{sec:saurav}.} Our task now lies in formalizing the principles outlined in \textit{Saurav Yadav (2020)}—as presented in the judgment—into formal axioms. 
Expanding upon their formulation in \cite{sonmez/yenmez:22} for the case of homogeneous positions, in Section \ref{sec:desiderata}, 
we articulate these principles as formal axioms applicable to the broader context with heterogeneous positions.

\subsection{Root Causes of the Failures: Excessive Reliance on the Concept of Migration and Creation of Artificial Property Rights} \label{sec:migration-adjustment}

One of the fundamental principles guiding India's implementation of its reservation policy is the \textit{principle of inter se merit}, which dictates that allocation should be based on merit within the same category, 
with exceptions outlined by HR policy.\footnote{The effect of HR policy on this principle became clear in India with the \textit{Saurav Yadav (2020)} case.} 
This principle, pivotal in numerous Supreme Court rulings, often sparks legal disputes in India (see, for instance, 
Sections \ref{subsec-AnuragPatel}, \ref{subsec-RameshRam} and \ref{subsec-Sharan} in the Online Appendix). 
In this section, we identify the root causes of these challenges, focusing solely on VR policy as the core issue remains independent of HR policy.

The VR policy, designed as the primary affirmative action provision in the Constitution to rectify years of caste-based discrimination, aims to uplift historically disadvantaged groups. 
Since no individual belongs to more than one caste (i.e., the VR-protected groups cannot overlap), the stand-alone implementation of the VR policy is straightforward with homogeneous positions: 
Open-category positions are allocated based on merit first, and reserved positions are allocated next to the remaining members of each VR-protected category based on inter se merit. 
Under this benchmark procedure, called the \textit{over-and-above choice rule}, positions within each category are processed one at a time, starting with the open category.

In India, legal terminology fails to distinguish between categories of individuals and categories of positions. 
Although individuals not belonging to any VR-protected category are referred to as \textit{general-category} candidates, they are often also called \textit{open-category} candidates. 
Since open-category positions are not exclusively reserved for individuals in the general category, unlike positions in VR-protected categories that are exclusively reserved for their members, 
this terminology can lead to confusion. To help avoid any resulting confusion, the legal concept of \textit{migration} emerged, signifying that when a VR-protected individual secures an open-category position based on merit, 
they are considered to have ``migrated" to the open or general category. With this convention, open-category positions are awarded to members of the ``revised" open category.

Parallel to the case of homogeneous positions, traditions of allocating open-category positions before the VR-protected ones and utilizing the concept of migration persist in the allocation 
of heterogeneous positions. In a typical mechanism, open-category positions across all institutions are tentatively allocated via a \textit{simple serial dictatorship} (SSD). 
The highest-merit individual is assigned their top choice, the second-highest-merit individual is assigned their highest choice among the remaining open-category positions, and so on. 
Importantly, VR-protected individuals who tentatively receive an open position are granted a status called \textit{meritorious reserved candidate} (MRC), one that plays a key role in several important judgments.

Once open-category positions are provisionally managed, VR-protected positions are also tentatively allocated to eligible individuals in a similar way, again using SSD in a second phase. 
Since an MRC can receive two distinct tentative assignments in the two phases, based on inter se merit, they migrate to the institution-category pair associated with their more preferred position. 
Subsequently, positions vacated by MRCs must be reallocated while still adhering to principle of merit for open-category positions and the principle of inter se merit for VR-protected positions.

Enforcing directives based on the mechanics of the above-described (and misguided) class of SSD-based mechanisms rather than the underlying principles, 
several key Supreme Court judgments made fundamental errors at this juncture. 
For example, in \textit{Union of India vs Ramesh Ram (2010)},\footnote{The ruling can be accessed at \url{https://indiankanoon.org/doc/1368252/}, retrieved on 02/25/2024.}  
positions vacated by MRCs were exclusively awarded to general category candidates for government jobs, 
while in \textit{Tripurari Sharan vs Ranjit Kumar Yadav (2018)},\footnote{The ruling can be accessed at \url{https://indiankanoon.org/doc/102870864/}, retrieved on 02/25/2024.}  
they were exclusively awarded to VR-protected individuals from the category of vacating MRCs. 
Notably, the creation of such ``artificial" property rights for a specific category of individuals directly contradicts the \textit{principle of inter se merit}, 
as individuals deserving the vacated positions can be from any category, either unmatched or tentatively holding less-preferred positions.

Thus, the root causes of the legal challenges lies in the excessive reliance on the concept of migration, 
further exacerbated by the practice of awarding ``artificial" property rights to essentially arbitrary groups based on tentative assignments of MRCs, 
which itself is an ill-equipped tool for addressing applications with heterogeneous positions. 
This methodological failure lies at the core of the legal inconsistencies and litigations in India.\footnote{A simple search via Indian Kanoon, a free search engine for Indian Law, reveals that as of February 2024, 
there have been 869 cases related to the ``migration of meritorious reserved candidates" at the Supreme Court or the high courts.} 

As we address these challenges, it is important to observe that it is not possible to accommodate the principles outlined in \textit{Indra Sawhney (1992)} or 
their refinements in \textit{Saurav Yadav (2020)} solely by relying on the concept of migration, unless an arbitrary number of rounds of migrations are allowed through an 
iterative procedure such as the individual-proposing deferred acceptance (DA) algorithm \citep{gale/shapley:62}. 
The reason is analogous to the necessity of iterative procedures to find stable matchings in two-sided matching markets. 
This observation is at the heart of our proposed resolution. Before we formulate and propose a solution in Section \ref{intro-2SMH-DA} utilizing the DA algorithm, 
we first characterize the set of outcomes that satisfy the mandates of \textit{Saurav Yadav (2020)}.

\subsection{Characterizing \textit{Saurav Yadav (2020)}-Compliant Assignments}

For scenarios with homogeneous positions, the mandates outlined in \textit{Saurav Yadav (2020)} are uniquely satisfied by the 2SMH choice rule \citep{sonmez/yenmez:22}. 
This rule not only determines which candidates are awarded positions but also specifies the category from which they receive them.
This additional structure enables us to introduce a refinement of the traditional \textit{stability} axiom for  our setting.  

For each individual, an assignment specifies whether they receive a position and the following specifics of this position: 
Which institution offers the position and from which category. Given a strict preference ranking for each individual over all institutions and remaining unmatched, 
and a (multi-category) choice rule for each institution, an assignment is \textit{stable} if:
\begin{enumerate}
\item No individual prefers their institution assignment to remaining unmatched.
\item No institution would rather remove an individual from its admitted group or change the category of the position they are offered.
\item There exists no individual-institution pair where 
\begin{enumerate}
\item the individual prefers the institution to their assignment, and 
\item the institution would rather offer the individual a position 
when it is allowed to maintain some or all of the existing individuals from its admitted group.
\end{enumerate}
\end{enumerate}

There is only one subtle difference between our stability notion and the traditional stability notion in standard two-sided matching models, 
such as the \textit{matching with contracts} \citep{hatfield/milgrom:05}: In the second condition, in addition to maintaining its admitted group of individuals, 
an institution is also expected to retain their assigned categories. Since our notion of a choice rule is multidimensional, also specifying categories, 
it is only natural that an extension of the standard stability to our setting takes this additional specification into consideration. 
Thus, we believe our notion is a natural refinement of the traditional stability notion for our more structured environment.

This notion of stability is critical for our analysis: In Theorem \ref{thm:resrulechar}, we show that an assignment satisfies the mandates of \textit{Saurav Yadav (2020)} 
if and only if it is stable when each institution is endowed with the 2SMH choice rule. 
Thus, the path to adhering to the Supreme Court's mandates goes through a natural generalization of a well-established methodology in matching theory.

\subsection{A Resolution via 2SMH-DA Mechanism} \label{intro-2SMH-DA}

Unlike in the case of homogeneous positions, as shown by Theorem \ref{thm:resrulechar}, the Supreme Court's mandates do not prescribe a unique mechanism when positions are heterogeneous. 
Fortunately, earlier literature in two-sided matching and school choice guides us in a promising direction to formulate and propose a unique mechanism for this more general version of the problem.

Our proposed mechanism, \textit{2SMH-DA}, is based on extending the 2SMH choice rule to the general version of the problem with heterogeneous positions across multiple institutions 
through a joint implementation of the individual-proposing DA algorithm, where each institution is endowed with the 2SMH choice rule.

It is well-established that, in the standard school choice setting without categories \citep{abdulkadiroglu/sonmez:03}, there exists a stable outcome that \textit{Pareto dominates} 
any other stable outcome \citep{gale/shapley:62, balinski/sonmez:99}. Moreover, this outcome can be obtained with the individual-proposing DA algorithm \citep{balinski/sonmez:99}. 
Finally, not only does the resulting mechanism satisfy \textit{strategy-proofness} \citep{dubins/freedman:81, roth82}, it is the only one that also satisfies stability \citep{alcalde/barbera:94, balinski/sonmez:99}. 
Under some technical conditions on (single-category) choice rules, these results are very robust across a wide range of settings.

Our multi-category setting is not an exception. In Theorem \ref{thm:dominate}, we show that the 2SMH-DA mechanism Pareto dominates any other mechanism 
that satisfies the Supreme Court's mandates in \textit{Saurav Yadav (2020)}, and in Theorem \ref{thm:jointchar}, we show that it is the only \textit{strategy-proof} mechanism that satisfies these mandates.

Therefore, either one of the two fundamental principles in economic theory directly implies the 2SMH-DA mechanism when combined with \textit{Saurav Yadav (2020)} 
refinement of the principles in \textit{Indra Sawhney (1992)}. Hence, we argue that 2SMH-DA is the only natural mechanism to address numerous legal challenges 
faced by public institutions in India due to their flawed allocation mechanisms.

\subsection{Organization of the Rest of the Paper}

In Section \ref{sec:model}, we present the model, including the formulation of the directives of \textit{Saurav Yadav (2020)} regarding the joint implementation of VR and HR policies.
Section \ref{sec-MSCC} summarizes the key Supreme Court judgments on VR and HR policies, identifying several inconsistencies within and between these cases. 
This section also identifies the root causes of these failures.
In Section \ref{sec:stable}, we formulate a variant of the celebrated \textit{stability} axiom from the two-sided matching literature to our more structured setting. 
We show that, collectively, the mandates of the Supreme Court are equivalent to stability under the 2SMH choice rule by \cite{sonmez/yenmez:22}.
Section \ref{sec:2SMH+DA} presents our proposed mechanism \textit{2SMH-DA} and shows that, among all mechanisms that satisfy the Supreme Court's mandates, 
it (i) \textit{Pareto dominates} any other and (ii) is the only one that is \textit{strategy-proof}.
In Section \ref{sec:beyond-India}, we discuss additional real-life applications of our model, with a particular focus on school choice in Chile.
We conclude in Section \ref{sec:conclusion}. Legal analysis of several pivotal Supreme Court cases and all proofs of our formal results are provided in an Online Appendix.

\section{Model} \label{sec:model}

There exist a finite set $\cali$ of individuals and a finite set $\calj$ of institutions referred to as ``jobs'' throughout this section.\footnote{Our model
and notation build on \cite{sonmez/yenmez:22} where there is a single job with multiple identical positions.}
Each job $j\in \calj$ has $q_j$ identical positions. Each individual $i\in \cali$ has a strict preference ranking $\succ_i$
over all jobs and the outside option denoted by $\emptyset$, which could be being unemployed. We denote the set of all preference rankings for agent $i$ by $\mathcal{P}_i$.
A job $j\in \calj$ is \emph{acceptable} to individual $i\in \cali$ if it is stricty more preferred to the outside  option, that is, $j \mathrel{\succ_i} \emptyset$.
We denote the corresponding weak order by $\succeq_i$ and the indifference relation by $\sim_i$.
For any set of individuals $I\subseteq \cali$, we denote the profile of individual preferences by $\succ_I=(\succ_i)_{i\in I}$. In addition, we denote
the set of all preference profiles by $\mathcal{P}=(\mathcal{P}_i)_{i\in \cali}$.

Each individual $i \in \cali$ has a distinct merit score $\sigma_j(i) \in \mathbb{R}_{+}$ for any given job $j \in \calj$. 
While individuals with higher merit scores have higher claims for a job in the absence of affirmative action policies, various groups are protected by two types of affirmative action policies: 
(i) the \emph{vertical reservation (VR)} policy, which provides primary \emph{VR protections}; and (ii) the \emph{horizontal reservation (HR)} policy, which provides secondary \emph{HR protections}.

\subsection{VR Policy}\label{sec:ver}
There exist a  set $\calr$  of \textit{VR-protected} (or \textit{reserved}) categories and a  \emph{general category} $g \not\in \calr$. 
Each individual belongs to a single category in $\calr \medcup \{g\}$.
Individual memberships to VR-protected categories is given by a function $\rho: \cali \rightarrow \calr \medcup \{\emptyset\}$.
Here,  $\rho(i) = \emptyset$ indicates that individual $i$ is ineligible for VR protections, and thus she is a member of the
general category $g$.\footnote{To keep the notation at a minimum,
we assume that (i) the set of VR-protected categories $\calr$, (ii) the
general category $g$, and (iii) the category-membership function $\rho$  are all independent of a job.
This assumption is without any loss of generality and all these primitives can be made job-dependent
by a simple inclusion of a job index without interfering with any aspect of our analysis.}

At any given  job $j\in \calj$, there are $r^c_j\geq 0$ positions set aside exclusively for the
members of category $c\in \calr$.
We refer to these positions as \emph{category-c positions} or (\emph{VR-protected positions for category c}).
We assume that $\sum_{c\in \calr} r^c_j \leq q_j$. 
In contrast, members of the general category do not have any positions set aside for them under the VR policy.
Therefore, $ r^o_j = q_j-\sum_{c\in \calr} r^c_j$
positions are open for all individuals.  We refer to these positions as \emph{open-category positions}
(or \emph{category-o positions}).
Let $\calv = \calr \medcup \{o\}$ denote the set of \emph{vertical categories for positions}.

Given a VR-protected category $c \in \calr$, an individual $i \in \cali$ is considered \emph{eligible for positions in category $c$} if $\rho(i) = c$.
Each individual $i \in \cali$ is \emph{eligible for open-category positions}.
Given a category $v\in \calv$, let $\cali^{v} \subseteq \cali$ denote the set of individuals who are eligible for category-$v$ positions.

The defining characteristic of the VR protections is stated as follows in the landmark Supreme Court judgment
\textit{Indra Sawhney (1992)}: \smallskip

\begin{indquote}
It may well happen that some members belonging to, say Scheduled Castes get selected in the open competition
field on the basis of their own merit; they will not be counted against the quota
reserved for Scheduled Castes; they will be treated as open competition candidates.\smallskip
\end{indquote}

When there is a single job and the VR policy is the only affirmative action policy, the interpretation of this statement
becomes airtight: If a VR-protected individual deserves an open position on the basis of her  merit score only,
she should be awarded an open position and not use up a VR-protected position set aside for her category.
In this sense, VR protections are implemented on an ``over-and-above'' basis, a feature
which makes this policy the ``higher level'' affirmative action policy.
Unfortunately,  \textit{Indra Sawhney (1992)}  formulation of the VR policy given above loses its clarity
when it is implemented jointly with the HR policy which is formally introduced next in Section \ref{sec:hor}.

\subsection{HR Policy}\label{sec:hor}
In addition to the categories in $\calr$ that are associated with the primary VR protections,
there is a finite set  $\calt$  of traits associated with the secondary HR protections.
Each individual has a (possibly empty) subset of traits,  given by the function $\tau : \cali \rightarrow 2^{\calt}$.

HR protections are provided in the form of ``minimum guarantees''
within each vertical category $v \in \calv$.\footnote{Provision of HR protections within each vertical category
is not a federal mandate in India but rather a formal recommendation by
the Supreme Court judgment \textit{Anil Kumar Gupta (1995)\/}. The vast majority of the institutions in India follows this recommendation and implement
the HR policy in this form, which is also referred to as  \textit{interlocking reservations\/} or \textit{compartmentalized horizontal reservations\/}.}
For any job $j\in \calj$, VR-protected category $c\in \calr$, and trait $t\in \calt$,
subject to the availability of qualified individuals,
a minimum of $r^{c,t}_j$  category-$c$ positions are to be assigned to  individuals from category $c$ with trait $t$.
If there are not enough individuals from category $c$ with trait $t$ to fill these positions,
then the remaining empty seats are to be allocated to other individuals from category $c$.
We refer to these positions as \emph{category-c HR-protected positions for trait t}.
Similarly, for any trait $t\in \calt$ and subject to the availability of individuals with trait $t$,
a minimum of $r^{o,t}_j$ open-category positions are to be assigned to  individuals with trait $t$. If there are not enough individuals with trait $t$ to fill these positions,
then the remaining empty seats are to be allocated to other individuals.
We refer to these positions as \emph{open-category HR-protected positions for trait t}.

For each job $j\in \calj$ and vertical category $v \in \calv$, we assume that the total number of category-$v$ HR-protected positions
is no more than the number of positions in category $v$. That is,  for each
job $j\in \calj$ and category $v\in \calv$, we assume 
$\sum_{t\in \calt} r^{v,t}_j \leq r^v_j.$

As we already indicated, in contrast to VR protections, which are provided on an ``over-and-above" basis, 
HR protections are provided within each vertical category on a ``minimum guarantee" basis. This means that positions obtained without invoking any HR protection still accommodate the HR protections.

\subsection{Primary Assignment of Individuals to Jobs and Vertical Categories} \label{sec-primary}

In India, each position is classified by its job, vertical category (including the open category), and the associated trait (or its absence). 
Therefore, to describe an outcome, it may be compelling to assign individuals to a triple consisting of a job, a vertical category, and a trait or its absence. However, we will take a different approach for the reasons we elaborate below.

An outcome needs to indicate the job assignments of individuals because they have strict preferences over jobs. 
While the category assignment is not important for individual preferences, it is crucial for the implementation of the VR policy. 
This is because the laws clearly specify who should receive the open positions and who should receive the VR-protected positions. 
Therefore, at a minimum, an outcome needs to specify the job assignment and the category assignment of each individual who receives a position.

The specification of a trait assignment (or its absence), on the other hand, offers some flexibility in terms of modeling an outcome. 
While an outcome can explicitly specify the trait (or its absence) for a position received, this modeling choice results in immaterial multiplicities under our axioms, which formulate affirmative action legislation in India. 
For example, if there is a minimum guarantee of two positions for women in the open category of a given job, and five women receive open-category positions, under Indian laws, 
there is no meaningful way to specify which two of these five women receive the HR-protected positions. Therefore, to avoid any arbitrary conditions that fail to capture Indian legislation, 
in our model, an outcome simply assigns individuals to job-vertical category pairs or leaves them unassigned.

An implicit trait assignment will still be important to verify that the HR protections are honored to the extent possible, 
and it will be captured in our model through a secondary assignment introduced in Section \ref{sec-secondary}.\footnote{This modeling choice allows us to relegate 
any immaterial multiplicities to a secondary assignment within the primary assignment.}

\begin{definition}
An \emph{assignment} is a function $\alpha:\cali \to (\calj\times \calv) \medcup \{\emptyset\}$  such that,  for each job-category pair $(j,v)\in \calj\times\calv$,
\begin{center}
$\alpha^{-1}(j,v) \subseteq \cali^v $ \; and \; $|\alpha^{-1}(j,v)|\leq r^v_j$.
\end{center}
We denote the set of all assignments by $\mathcal{A}$.
\end{definition}

An assignment specifies the job offering the position received by each individual, if any, and the vertical category through which it is obtained. 

Given an assignment $\alpha\in \cala$, let
\[\alpha^{-1}(j)=\bigcup_{v\in \calv} \alpha^{-1}(j,v)\] denote the set of individuals who receive a position at job $j$.

Since individual preferences are defined over $\calj \medcup \{\emptyset\}$, rather than over $\big(\calj\times\calv\big)\medcup \{\emptyset\}$, 
we trivially extend the domain $\calp$ of the preferences to $\big(\calj\times\calv\big)\medcup\calj\medcup \{\emptyset\}$. 

For any preference profile $\succ_{\cali}\in\calp$, individual $i\in\cali$, job $j\in \calj$, and category $v \in \calv$, we have 
\[ j \mathrel{\succeq_i} (j,v) \mbox{ and } (j,v) \mathrel{\succeq_i} j,
\] or equivalently 
\[ (j,v) \sim_i j.
\] 
Therefore, an individual has preferences over jobs only and is indifferent between categories of the same job.

\begin{definition}
A \emph{mechanism} $\varphi: \mathcal{P} \to \mathcal{A} $ is a function that selects an assignment $\varphi(\succ_{\cali})\in\cala$
for each preference profile $\succ_{\cali}\in \mathcal{P}$.
\end{definition}

Given a mechanism $\varphi$ and a profile  $\succ_{\cali}$ of preferences, the assignment for individual $i\in \cali$ is denoted by $\varphi(\mathcal{\succ_{\cali}})(i)$. 
Likewise, the set of individuals assigned to job-category pair $(j,v)\in \calj \times \calv$ is denoted by $\varphi^{-1}(\succ_{\cali})(j,v)$, 
and the set of individuals assigned to job $j$ is denoted by $\varphi^{-1}(\succ_{\cali})(j)$.

\subsection{Secondary Assignment of Individuals to Traits within their Primary Assignments} \label{sec-secondary}

In order to manage the secondary HR policy, in this section, we discuss a technical tool called \textit{trait-matching} \citep{sonmez/yenmez:22}. 
Given an assignment $\alpha\in\cala$, a trait-matching can be thought of as a secondary assignment of HR-protected individuals to traits 
within any given pair $(j,v)\in \calj\times\calv$, and its size provides us with a natural metric to assess the extent to which the HR protections are honored at pair $(j,v)$.

Fix an assignment $\alpha \in \cala$, a job $j\in\calj$, and a category $v\in\calv$.
Let $I\subseteq \cali$ be the set of individuals  who receive category-$v$ positions at job $j$ under assignment $\alpha$.
Hence, $I = \alpha^{-1}(j,v)$.

First, consider a simpler version of the problem where each individual has at most one trait. For any trait $t\in\calt$, 
the set of individuals in $I$ who have trait $t$ is given by the set $\{i\in I : t\in\tau(i)\}$. 
Therefore, within category $v$ of job $j$, trait-$t$ HR protections are fully honored if $\left|\{i\in I : t\in\tau(i)\}\right| \geq r^{v,t}_j$, whereas $(r^{v,t}_j - \left|\{i\in I : t\in\tau(i)\}\right|)$ of them are left dishonored otherwise.

For the latter case, an individual $i\in\cali^v\setminus I$ can object to the allocation of  category-$v$ positions at job $j$ under assignment $\alpha$,
provided that she has trait $t$ and desires to receive a position at job $j$.
Also observe that  maximum number of HR-protected positions that can be honored within category $v$ at job $j$ by the set of individuals $I$
is given by
\[ n_j^v(I) = \sum_{t\in\calt} \min\Big\{\big|\{i\in I : t\in\tau(i)\}\big|, r^{v,t}_j\Big\}.
\]
Hence, any individual  $i\in\cali^v\setminus I$ can object to the allocation of  category-$v$ positions at job $j$ under assignment $\alpha$,
provided that  $n_j^v\big(I\medcup\{i\}\big) > n_j^v(I)$ and they desire to receive a position at job $j$.
This observation plays a key role in several of our formal axioms, later introduced in Section \ref{sec:desiderata}.

The same idea can also be extended to the more general version of the problem when individuals can have multiple traits. 
However, the secondary assignment of individuals to traits requires additional care in this case.

For example, suppose there is a single HR-protected position for women and another single HR-protected position for persons with disabilities. 
In such a scenario, a disabled woman can receive positive discrimination for either of the two HR-protected positions. 
However, if the only other individual possessing either of the two traits is a disabled man, it would be implausible to award the HR-protected position for 
persons with disabilities to the disabled woman and consequently deny an HR-protected position to the disabled man. 
Both HR-protected positions can be honored by assigning the HR-protected position for women to the disabled woman and the HR-protected position for persons with disabilities to the disabled man. 
We will now build on this simple observation to extend the above-given function $n_j^v$ to the general version of the problem.

Fix a job $j\in \calj$, a category $v \in \calv$, and a set of individuals $I \subseteq \cali^v$.
Let $H_j^{v,t}$ denote the set of HR-protected  positions for  trait-$t$  within category $v$ at  job $j$, and
$H^v_j = \bigcup_{t \in \calt} H_j^{v,t}$ denote the set of all HR-protected  positions within category $v$ at  job $j$. 
Construct the following bipartite
\emph{HR graph}: Individuals in $I$ are on one side of the graph and positions in $H^v_j$ are
on the other side. For any trait $t\in\calt$,
an individual $i \in I$ and a position $p \in H_j^{v,t}$ are \emph{connected} in this graph if  and only if individual $i$ has trait $t$.

\begin{definition}
Given a job $j\in \calj$, a category $v \in \calv$, and a set of individuals
$I \subseteq \cali^v$, a \emph{trait-matching} of individuals in $I$ with HR-protected  positions in $H^v_j$  is a function
$\mu: I \rightarrow H^v_j \medcup \{\emptyset\}$ such that,
\begin{enumerate}
\item for any $i \in I$ and $t\in \calt$,
\[ \mu(i)\in H_j^{v,t} \; \implies \; t \in \tau(i),  \]
\item for any $i,j \in I$,
\[ \mu(i) = \mu(j) \not= \emptyset \; \implies \; i=j.\]
\end{enumerate}
\end{definition}

\begin{definition}
Given a job $j\in \calj$, a category $v \in \calv$, and a set of individuals $I \subseteq \cali^v$,
a trait-matching $\mu$ of individuals in $I$  with the HR-protected  positions in $H^v_j$
\emph{has maximum cardinality in the HR graph} if
there exists no other trait-matching that assigns a strictly higher number of HR-protected  positions to individuals.
\end{definition}

Let $n^v_j(I)$ denote the maximum number of job-$j$ category-$v$
HR-protected positions in $H^v_j$ that can be assigned to individuals in $I$.\footnote{This number can be found in polynomial time by the famous
\textit{Hungarian maximum matching algorithm\/}, which is originally published in \cite{kuhn55} and based on the earlier work of the Hungarian
mathematicians  D\'{e}nes K\"{o}nig and Jen\"{o} Egerv\'{a}ry.}
For any job $j\in\calj$ and category $v\in\calv$,
this number identifies how many of its HR-protected positions are \textit{honored} when category-$v$ positions
of job $j$  are  awarded to  individuals  in $I$.
As such, it serves as a key summary statistic on compliance with the HR policy,\footnote{This observation is the main
reason why it it not necessary to explicitly include a trait matching in our formulation of an assignment.}
reflected in three of our formal axioms introduced next in Section  \ref{sec:desiderata}.

For illustration, consider the case where each individual has at most one trait. In addition to singleton nodes associated with individuals who have no trait, 
the rest of the HR graph for this case consists of $\left|\calt\right|$ disjoint components, one for each trait. Moreover, each component is a complete bipartite graph. 
That is, each node on one side of any component is connected to each node on the other side. Thus, given a job $j\in \calj$ and a category $v\in \calv$, for any given component associated with trait $t\in \calt$, 
the size of the maximum cardinality trait-matching is equal to the size of the smaller of the two sides of the component; i.e., $\min\Big\{\big|{i\in I : t\in\tau(i)}\big|, r^{v,t}_j\Big\}$. 
Therefore, just as we found before, the maximum number of job-$j$ category-$v$ HR-protected positions that can be honored can be derived as
\[ n_j^v(I) = \sum_{t\in\calt} \min\Big\{\big|{i\in I : t\in\tau(i)}\big|, r^{v,t}_j\Big\}.
\]

\subsection{\textit{Saurav Yadav (2020)} Axioms} \label{sec:desiderata}

In this section, we introduce our primary axioms on assignments and mechanisms.
Our main objective in formulating these axioms is giving a mathematically precise meaning to the mandates of the Supreme Court of India on VR and HR policies.
We will organize our axioms into two groups.

\begin{enumerate}
\item In the first group, we have two axioms that are so benign that, although they were implicitly intended, they are not explicitly discussed 
in the court rulings.\footnote{We are unaware of any practical mechanism in India where either of these two axioms fails or any court ruling that pertains to these axioms.}

\item  In the second group, we have three core axioms that formulate the Supreme Court's explicit mandates on the concurrent implementation of VR and HR policies. 
As far as we can tell, the inability to design mechanisms that satisfy these axioms, and the legislative confusion on their formulation are the primary reasons for the challenges in India. 
The formulation of these three axioms in \textit{Saurav Yadav (2020)} finally clears the second of these reasons; i.e., the legal confusion on their formulation.
\end{enumerate}

With a slight abuse of terminology, we refer to the five axioms in these  two groups as  \textit{Saurav Yadav (2020) axioms\/}.

Our first axiom states that no position should be awarded to an individual who has no desire to receive this position.

\begin{definition}
An assignment $\alpha\in \mathcal{A}$ satisfies \emph{individual rationality\/} if, for every $i\in \cali$,
\[ \alpha(i) \succeq_i \emptyset.
\]
A mechanism $\varphi$ satisfies  \emph{individual rationality} if its outcome $\varphi(\succ_{\cali})$ satisfies individual rationality for
each $\succ_{\cali}\in \mathcal{P}$.
\end{definition}

Our second axiom states that, a position can be left idle only if no individual who desires to receive it is eligible for the position.

\begin{definition}
An assignment $\alpha\in \mathcal{A}$ satisfies \emph{non-wastefulness\/} if, for every $j\in \calj$, $v\in \calv$, and $i\in \cali$,
\[j \succ_i \alpha(i) \; \mbox{ and } \; \big|\alpha^{-1}(j,v)\big|<r_j^v   \; \implies \; i\notin \cali^v.\]
A mechanism $\varphi$ satisfies  \emph{non-wastefulness} if its outcome $\varphi(\succ_{\cali})$ satisfies non-wastefulness for
each $\succ_{\cali}\in \mathcal{P}$.
\end{definition}

Our third axiom formulates the positive discrimination given to HR-protected individuals. 
It states that an individual who remains unassigned cannot be denied a position at any job-category pair $(j,v)\in\calj\times\calv$, 
if her recruitment increases the number of HR-protected positions honored at pair $(j,v)$.

\begin{definition}
An assignment $\alpha\in \mathcal{A}$ satisfies \emph{maximal accommodation of HR protections\/} if, for
every $j\in \calj$, $v\in \calv$, and $i\in \cali^v$,
\[j \succ_i \alpha(i) \; \implies \;  n_j^v\Big(\alpha^{-1}(j,v) \medcup \{i\}\Big) \not > n_j^v\Big(\alpha^{-1}(j,v)\Big).\]
A mechanism $\varphi$ satisfies  \emph{maximal accommodation of HR protections} if its outcome $\varphi(\succ_{\cali})$ satisfies maximal accommodation of HR protections
for  each $\succ_{\cali}\in \mathcal{P}$.
\end{definition}

Our fourth axiom formulates the following equity principle given in the paragraph 31 of the Supreme Court judgment \textit{Saurav Yadav (2020)\/}:

\begin{indquote}
[\dots] Subject to any permissible reservations i.e. either Social (Vertical) or Special (Horizontal), opportunities to public employment and selection of candidates must purely be based on merit.

Any selection which results in candidates getting selected against Open/General category with less merit than the other available candidates will certainly be opposed to principles of equality. 
There can be special dispensation when it comes to candidates being considered against seats or quota meant for reserved categories and in theory it is possible that a more meritorious candidate coming from Open/General category may not get selected. But the converse can never be true and will be opposed to the very basic principles which have all the while been accepted by this Court. Any view or process of interpretation which will lead to incongruity as highlighted earlier, must be rejected.
\end{indquote}

\begin{definition} \label{def:nje}
An assignment $\alpha\in \mathcal{A}$ satisfies \emph{no justified envy\/} if, for every
$i\in \cali$, $j\in \calj$, $v\in\calv$, and $i'\in\cali^v$,
\[ \left. \begin{array}{c}
  \alpha(i)=(j,v) \; \mbox{ and}\\
  j \succ_{i'} \alpha(i') \end{array}  \right\}
\implies \;
\sigma_j(i)>\sigma_j(i') \;  \mbox{ or } \; n_j^v\Big(\alpha^{-1}(j,v)\Big) > n_j^v\Big(\big(\alpha^{-1}(j,v)\setminus \{i\}\big)\medcup \{i'\}\Big).
\]
A mechanism $\varphi$ satisfies  \emph{no justified envy\/} if its outcome $\varphi(\succ_{\cali})$ satisfies no justified envy for
each $\succ_{\cali}\in \mathcal{P}$.
\end{definition}
That is,  if an individual $i$ receives a category-$v$ position at job $j$ while another individual $i'$ receives a less-desired assignment,
it is either because individual $i$ has a higher merit score under $\sigma_j$ than individual $i'$
or because replacing individual $i$ with individual $i'$ decreases the number of HR-protected positions that
are honored.

When applied to open-category positions where every individual is eligible, the axiom of \textit{no justified envy} transforms into a condition known as the \textit{principle of merit} in India. 
Similarly, when applied to positions in a VR-protected category $c \in \calr$ where only category-$c$ individuals are eligible, the axiom of \textit{no justified envy} 
transforms into a condition known as the \textit{principle of inter se merit}.
These principles were originally established in the landmark Supreme Court judgment \textit{Indra Sawhney} (1992). 
However, the precise role of HR protections for open positions under the \textit{principle of merit} only became clear with the recent judgment 
\textit{Saurav Yadav} (2020).\footnote{In particular, the role of individuals who are both HR and VR protected for the open positions
were unclear prior to \textit{Saurav Yadav (2020)\/}. See, for example, the following quote from the  01/25/2021 \textit{The Leaflet\/} article 
``Supreme Court strikes down policy of excluding the reserved community from competing for general and open category.''
\begin{quote}
``Until now, the specific question of whether female candidates belonging to any of the vertically reserved categories can be selected 
on `merit' against the vacancies horizontally reserved for general/open category was a res integra before the Supreme Court.''
\end{quote}
The story is available in \url{https://tinyurl.com/z6y7wwfn}, 
last accessed on 02/09/2022.}
The failure of \textit{no justified envy} is one of the primary reasons for litigations in India regarding the implementation of VR and HR policies.

Our fifth axiom formulates the defining characteristic of VR protections as the ``higher-level'' reservation policy. It states that VR-protected positions shall not be awarded to individuals 
who ``deserve an open-category position based on merit.'' Instead, they should be left for VR-protected individuals who are truly in need of positive discrimination. 
Critically, with the clarification in \textit{Saurav Yadav (2020)}, an individual may deserve an open-category position not only due to their merit score 
but also because they may increase the number of honored open-category HR protected positions.\footnote{For the exact statement of this clarification, 
refer to Paragraph 36 from \textit{Saurav Yadav (2020)}, provided in Section \ref{sec:saurav} of the Online Appendix.}

\begin{definition} \label{def:VR}
An assignment $\alpha\in \mathcal{A}$ satisfies \emph{compliance with VR protections\/} if, for every $j\in \calj$, $c\in\calr$,  and $i\in\cali^c$,
the following three conditions hold whenever $\alpha(i) = (j,c)$:
\begin{enumerate}
  \item $\abs{\alpha^{-1}(j,o)}=r_j^o$,
  \item for every $i'\in \cali$ with $\alpha(i')=(j,o)$,
\[\sigma_j(i') > \sigma_j(i) \; \mbox{ or } \; n_j^o\left(\alpha^{-1}(j,o)\right) > n_j^o\left((\alpha^{-1}(j,o)\setminus \{i'\})\medcup \{i\}\right), \mbox{ and}\]
  \item $n_j^o\Big(\alpha^{-1}(j,o)\medcup \{i\}\Big) \ngtr n_j^o\Big(\alpha^{-1}(j,o)\Big)$.
\end{enumerate}
A mechanism $\varphi$ satisfies  \emph{compliance with VR protections\/} if its outcome $\varphi(\succ_{\cali})$ satisfies compliance with VR protections for
each $\succ_{\cali}\in \mathcal{P}$.
\end{definition}

For an individual $i$ to receive a VR-protected position at job $j$, a prerequisite is that they do not merit an open-category position at job $j$. That means:
\begin{enumerate}
\item There is no idle open-category position left at job $j$.
\item Each individual $i'$ who received an open-category position at job $j$ either has a higher merit score than $i$ or admitting $i'$ rather than $i$ 
increases the number of honored open-category HR protected positions at job $j$.
\item Replacing individual $i$ with any recipient of an open-category position at job $j$ does not increase the number of honored open-category HR protected positions at job $j$.
\end{enumerate}
If any of the three conditions fail, with the clarification in \textit{Saurav Yadav (2020)} that formulates how ``open-category merit'' accounts for HR protections, 
individual $i$ would deserve an open-category position at job $j$ based on merit.

\section{Addressing Inconsistencies within and between Supreme Court Judgments through Minimalist Market Design} \label{sec-MSCC}

In this section and Sections \ref{subsec-AnuragPatel} through \ref{subsec-Samiti} of the Online Appendix, we discuss several key Supreme Court judgments on VR and HR policies. 
We present a series of inconsistencies that have emerged within and between these judgments and identify a methodological flaw largely responsible for these inconsistencies. These inconsistencies
often result in litigations, interruption of the recruitment processes, and
reversals of recruitment decisions in India. For example, a March 2017
\textit{The Times of India} story reports the likely consequences of a ruling by the  High Court of Gujarat as follows:\footnote{The \textit{The Times of India} story  is available at
\url{https://timesofindia.indiatimes.com/city/ahmedabad/general-seat-vacated-by-quota-candidate-remains-general-hc/articleshow/57658109.cms}. Retrieved on 02/19/2024.}

\begin{quote}
``The advertisement was issued in 2010 and recruitment took place in 2016 amid too many litigations over the issue of reservation [$\ldots$]
With the recent observation by the HC, the merit list will now be changed for the third time. Those already selected and at present under training might lose their jobs, and half a dozen new candidates might find their names on the new list. However, all appointments have been made by the HC conditionally and subject to final outcome of these multiple litigations.''
\end{quote}
We then formulate these issues as a case study in minimalist market design \citep{sonmez:23}, 
a framework developed for joint efforts in research and policy, and address them in Sections \ref{sec:stable} and \ref{sec:2SMH+DA}.

\subsection{Key Judgments and Issues on Implementation of VR and HR Policies} \label{subsec:keyjudgments}

\subsubsection{Indra Sawhney vs Union of India (1992)}\label{subsubsec:Sawhney} This judgment is considered the primary ruling on VR and HR policies. 
It establishes VR policies as the specific framework for implementing the higher-level affirmative action provisions outlined in Articles 15(4) and 16(4) of the Constitution of India, 
while designating HR policies for implementing the lower-level affirmative action provisions specified in Articles 15(3) and 16(1). 
Importantly, whether positions are homogeneous or heterogeneous, this key judgment does not specify how the two policies are affected when implemented together, 
nor does it provide a procedure for their joint implementation. Therefore, although both policies have clear formulations as described in this judgment when implemented individually, 
their joint implementation allows for multiple interpretations, even for the simpler version of the problem with homogeneous positions.

\subsubsection{Anil Kumar Gupta vs State Of U. P. (1995) \& Saurav Yadav vs State Of U. P. (2020)}\label{subsubsec:GuptaYadav} Until it was rescinded by \textit{Saurav Yadav (2020)}, 
a subsequent judgment of the Supreme Court, \textit{Anil Kumar Gupta (1995)} served as a primary ruling on the joint implementation of VR and HR policies in cases where positions are homogeneous. 
Although it introduced and mandated a procedure (SCI-AKG choice rule) to implement these two policies jointly, a flaw in this procedure resulted in countless litigations, 
inconsistent judgments, and interruptions in recruitment processes for 25 years. The main issue was the failure of the \textit{no justified envy} axiom (cf. Definition \ref{def:nje} in Section \ref{sec:desiderata}), 
although this principle was never explicitly defined or mandated by court rulings until \textit{Saurav Yadav (2020)}.

In a working paper, \cite{sonmez/yenmez:19a} documented the failures of the SCI-AKG choice rule and tied its shortcomings to its failure to satisfy the \textit{no justified envy} axiom. 
Through minimalist market design, they also proposed the 2SMH choice rule as a remedy. Before their analysis and policy proposal were published in \cite{sonmez/yenmez:22}, 
the SCI-AKG choice rule was rescinded, \textit{no justified envy} was formulated and mandated, and the 2SMH choice 
rule—discovered independently by members of the judiciary—was endorsed by the landmark Supreme Court judgment \textit{Saurav Yadav (2020)}. 
\cite{sonmez:23} interprets the parallel with \cite{sonmez/yenmez:19a, sonmez/yenmez:22} as external validity for minimalist market design. 
The current paper aims to fulfill a similar role as \cite{sonmez/yenmez:19a, sonmez/yenmez:22}, but for the more general and complex version of the problem with heterogeneous positions, where the issues are deeper.

\subsubsection{Allocation of Heterogenous Positions \& the Status of Meritorious Reserved Candidate} \label{subsec-MRC}

In contrast to the version of the problem with homogeneous positions, to the best of our knowledge, no court has ever established or mandated an 
explicit procedure for the more complex version with heterogeneous positions. Consequently, government agencies have devised their own mechanisms. 
Although the specific mechanisms vary across applications, the initial step often involves tentatively allocating the $\sum_{j\in J} r_j^o$ open-category positions to candidates 
using a mechanism commonly known as the \textit{simple serial dictatorship} (SSD) in the literature: The highest merit-ranking candidate is tentatively assigned a position in their top-choice institution, 
followed by the second-highest merit-ranking candidate receiving their top-choice position at an institution with available open positions, and so forth. In legal terminology, 
each reserved category candidate who tentatively receives an open position at this stage is referred to as a \textit{meritorious reserved candidate} (MRC).

Consider an individual $i$ who is an MRC from a VR-protected category, such as SC. Note that while candidate $i$ tentatively receives an open-category position based on their own merit, 
thus without utilizing the benefits of VR protections, they might prefer to receive an SC-category position at a more preferred institution. 
At this point, the following important questions arise, the answers to which guide the mechanics of the rest of the mechanism:

\begin{enumerate}
\item Should an MRC who is tentatively assigned an open-category position be allowed to \textit{migrate\/} to a higher-choice institution and receive a position set aside for their VR-protected category?
\item If the answer to the first question is affirmative, then what should happen to the open-category position vacated by the MRC?
\end{enumerate}

These two questions and their answers lie at the heart of countless lawsuits in India. While the Supreme Court has not mandated an explicit mechanism when positions are heterogeneous, 
depending on the specific setting, it has mandated how these two questions on MRCs are answered when open positions are tentatively allocated with SSD as the first step of a mechanism.

\subsubsection{Anurag Patel vs U.P. Public Service Commission (2004)}  \label{sec:AnuragPatel-maintext}
Considered a primary ruling on allocation of heterogenous positions, this judgment answered the first question on MRCs in the affirmative. 
An MRC who is tentatively assigned an open-category position is allowed to migrate to a higher-choice institution and receive a position set aside for their VR-protected category. 
Moreover, it reaffirmed that any mechanism used for the allocation of government jobs or seats at public educational institutions must respect \textit{inter se merit}:\footnote{\textit{Inter se merit} is implied by 
\textit{no justified envy\/} (cf. Definition \ref{def:nje} in Section \ref{sec:desiderata}).}
Barring exceptions due to HR policy, no VR-protected individual can receive a less-preferred position than an individual of lower merit from the same category. 
See Section \ref{subsec-AnuragPatel} in the Online Appendix for more on \textit{Anurag Patel (2004)}.

\subsubsection{Union of India vs Ramesh Ram (2010)} Considered the principal opinion for allocations of government jobs for heterogeneous positions, 
this judgment answered the second question on MRCs as follows: Open positions vacated when MRCs migrate to their more-preferred VR-protected positions are to be offered to general category candidates. 
See Section \ref{subsec-RameshRam}  in the Online Appendix for a thorough analysis of \textit{Ramesh Ram (2010)}.

\subsubsection{Tripurari Sharan vs Ranjit Kumar Yadav (2018)} Developing a fundamentally different jurisprudence than \textit{Ramesh Ram Ors (2010)}, 
this judgment answered the second question on MRCs as follows for allocation of medical college seats: 
Open positions vacated when MRCs migrate to their more-preferred VR-protected positions are to be offered to candidates from the vacating MRC's VR-protected category. 
See Section \ref{subsec-Sharan} in the Online Appendix for a thorough analysis of \textit{Tripurari Sharan (2018)}.\medskip

In Sections  \ref{subsec-RameshRam} and \ref{subsec-Sharan} in the Online Appendix, 
we present thorough analyses of \textit{Ramesh Ram (2010)} and  \textit{Tripurari Sharan (2018)}, illustrating  
that the methodology of using migrations and adjustments through MRCs is fundamentally flawed, 
and it constitutes the root cause of legal conflict and confusion not only in these cases but also in countless others.\footnote{Each Supreme Court case in 
Sections \ref{subsec-AnuragPatel} through \ref{subsec-Samiti} in the Online Appendix involves the handling of MRCs under litigated mechanisms in India. 
The descriptions of the mechanisms we present in these sections are based on their descriptions in these court cases. Not all aspects of the actual mechanisms are relevant for these cases, 
and they only provide details that relate to the case. In particular, all the cases focus on VR protections to SC/ST/OBC, and none of them give details on the handling of HR protections as they are not focal to these cases. 
This means that the mechanisms we present may correspond to a simplified case, abstracting away from HR protections. In actual implementation, 
HR protections are likely accommodated through adjustments at various steps of the procedures, as is traditional in India. Since we present failures of these mechanisms even in the absence of HR protections, 
the details provided in the cases are sufficient for our purposes. However, there are other cases where the litigation involves both HR protections and heterogeneous positions across multiple institutions. 
See, for example, the Patna High Court case \textit{The Controller Of Exam., Bihar vs Nidhi Sinha \& Anr}, available at \url{https://indiankanoon.org/doc/180601564/} (retrieved on 02/18/2024).}

\subsection{Root Cause of Judicial Inconsistencies for Allocation of Heterogenous Positions} \label{sec-MRC-not}

In Sections \ref{subsec-RameshRam} and \ref{subsec-Sharan} in the Online Appendix, we argue that not only do the allocation mechanisms employed by various Indian institutions have important shortcomings, 
but also the Supreme Court judgments on these mechanisms have a number of inconsistencies. In this section, we argue that the root cause of 
these difficulties lies in the excessive reliance on the concept of migration to solve more complex versions of the problem, further exacerbated by the introduction of the status of an 
MRC as an especially ill-equipped tool to facilitate a solution for applications with heterogeneous positions.

Since open positions are allocated prior to VR-protected positions when positions are homogeneous, decision-makers may be tempted to adopt a similar approach and allocate them 
via SSD in heterogeneous scenarios. However, unlike in the homogeneous case, an open-category position secured by a VR-protected individual may not be preferred 
over a reserved position they can obtain within their VR-protected category. Consequently, this widespread practice has led to the establishment of the 
MRC status in affirmative action jurisprudence and the formulation of the two questions posed in Section \ref{subsec-MRC} to manage their assignments. 
It is critical to note that these questions focus not on the fundamentals of the problem but rather on the mechanics of a specific class of mechanisms.

The root cause of the challenges faced by MRC-based mechanisms can be traced to the following observation: 
once an MRC vacates an open position to secure a more-preferred position reserved for their VR-protected category, the next deserving candidate can be:
\begin{enumerate}
\item a member of the general category who either holds a less-preferred open position from earlier phases or remains unassigned,
\item another MRC who holds a less-preferred position from earlier phases, or
\item another member of a reserved category who remains unassigned from earlier phases.
\end{enumerate}

Thus, the widespread practice of tentatively allocating open positions in the first phase results in the creation of an artificial interim allocation, 
one that is often given undue weight despite being a technical construct. This, in turn, leads to the exclusive awarding of the ``property rights'' of a vacated open position to 
members of a specific category, creating an open invitation for litigation. This misguided and artificial construction of property rights is the primary source of dispute in the vast majority of legal conflicts involving MRCs.

Indeed, the judges of the Central Administrative Tribunal, Chennai Bench (CAT-CB), correctly identified this root cause of challenges in 
a lower court decision preceding the Supreme Court judgment in \textit{Ramesh Ram (2010)}. The judges of the CAT-CB included the following statement in their ruling:

\begin{indquote}
In doing so, the respondents also would notice that the
steps taken by them in accordance with the Rules 16 (3)(-)(5) are redundant once they issue the result of recruitment
in one phase, instead of two as they have become primary cause for the litigation and avoidable confusion in the minds of the candidates seeking recruitment.
\end{indquote}

Therefore, the judges directed the Union of India to announce their outcome in one phase in a manner that respects \emph{inter se merit}, 
without relying on the artificial concept of migration.\footnote{See Online Appendix \ref{sec:ramesh} for a comprehensive quote from this case.} 
However, despite being accurate, this ruling was ignored by the Union of India, and the case moved all the way to the Supreme Court. 
One possible explanation for the Union of India's refusal to follow the decision of CAT-CB may be their technical inability to construct a mechanism that complies with the court's order.

With the subsequent clarification on the joint implementation of VR and HR policies provided by \textit{Saurav Yadav (2020)}, in Section \ref{sec:2SMH+DA}, 
we will formulate a particularly compelling mechanism that complies with the tribunal's orders.

\subsection{Formulation as an Application in Minimalist Market Design}

Having identified the root cause of the failures in Supreme Court judgments on the allocation of heterogeneous positions in Section \ref{sec-MRC-not},
we turn to minimalist market design \citep{sonmez:23} as a framework to address them.

Think of an institution like a machine with many parts or a body with many organs, each meant to perform specific jobs. 
Sometimes, these parts break or fail to work well together, causing the machine or body to fail in its tasks. 
To address these issues, a design economist can learn from how experts in other fields, such as medicine, handle similar problems. 
For instance, when a doctor encounters a problem in the body, they identify the cause, such as removing diseased tissue or organs, fixing different body systems, or performing organ transplants. 
A design economist operates analogously within minimalist market design, which involves identifying the root causes of failure and making minimal changes to rectify them. 
In simpler terms, the paradigm works like a ``minimally invasive'' procedure performed by a physician.

Under this paradigm, three main tasks emerge:
\begin{enumerate}
\item Identify the institution's mission: What are the primary objectives of policymakers, system operators, and other stakeholders?
\item Determine whether the existing institution satisfies these primary objectives: If it doesn't, there's potential for policy impact with a compelling alternative design. 
To realize this potential, the root causes of the failures should be identified.
\item Address the failures of the deficient institution by interfering only with its flawed features.
\end{enumerate}

As discussed in Section \ref{subsubsec:Sawhney}, the primary objectives of VR policy are rigorously and fully described in 
\textit{Indra Sawhney (1992)} for the basic version of the problem in the absence of HR protections. 
Moreover, the primary root cause of the failures in subsequent legislation and real-life institutions for the case of heterogeneous positions is identified in Section \ref{sec-MRC-not}. 
One possibility would be to completely ignore HR policy and deploy minimalist market design to address the challenges identified in MRC-based mechanisms 
(e.g., those in Sections \ref{subsec-RameshRam}, \ref{subsec-Sharan} and \ref{subsec-Samiti} of the Online Appendix), 
using the root cause of the failures of these mechanisms identified in Section \ref{sec-MRC-not} and the objectives mandated in \textit{Indra Sawhney (1992)}.

While this may be plausible for purely academic purposes, it generally will not produce mechanisms that could be implemented in India because virtually all current applications involve HR protections. 
This is because the \textit{Union Of India vs National Federation Of the Blind (2013)} case grants HR protections to persons with disabilities throughout India.\footnote{The ruling  is available at
\url{https://indiankanoon.org/doc/178530295/}, retrieved on 02/26/2024.} 
Additionally, other groups, notably women, are granted HR protections in many states through high court decisions. 
Examples include Bihar and  Madhya Pradesh with 35\% of positions each, Andhra Pradesh, Gujarat and Telangana with 33\% each, 
and Madhya Pradesh, Uttarakhand, Chhattisgarh, Rajasthan, and Sikkim with 30\% each.

Fortunately, as we already discussed in Section \ref{sec:desiderata}, the primary objectives of VR and HR policies and how they must be integrated are rigorously 
described in \textit{Saurav Yadav (2020)}. Thus, in Sections \ref{sec:stable} and \ref{sec:2SMH+DA}, 
we will deploy minimalist market design for the most general version of the problem with VR and HR policies and heterogeneous positions.

\section{Affirmative Action Policies in India as a Stable Assignment Problem}  \label{sec:stable}

As we discussed in Section \ref{sec-MRC-not}, 
the root cause of inconsistencies within and between Supreme Court judgments on the implementation of reservation policies for heterogeneous positions lies in the 
artificial creation of ``property rights'' through the flawed status of the MRC with repeated applications of the SSD.
Fortunately, \textit{Saurav Yadav (2020)} mandates imply a unique way to implement VR and HR policies when positions are homogeneous, 
through a procedure called 2SMH \citep{sonmez/yenmez:22}.

Following minimalist market design, in this section, we explore the implications of Saurav Yadav (2020) axioms when positions are heterogeneous, 
establishing a close link between them and the celebrated \textit{stability} axiom for two-sided matching markets \citep{gale/shapley:62, hatfield/milgrom:05}.

\subsection{Single-Job Solution Concepts} \label{subsec-choicerule}
To establish this link, we first formulate three related solution concepts for a single job, as introduced in \cite{sonmez/yenmez:22}.

Assuming \textit{individual rationality\/} and that a mechanism only relies on preferences over acceptable positions, 
the set of applicants contains all the necessary information given in a preference profile when all positions are identical.
That is because, individuals who prefer remaining unmatched over receiving a position can be ignored, and the remaining individuals all have the same preference relation.
Therefore, the domain of single-job solution concepts can be given as $2^\cali$ rather than $\mathcal{P}$.

\begin{definition} \label{def:single-category}
Given a job $j\in \calj$ and a category $v \in \calv$, a \emph{single-category choice rule}
is a function $C^v_j: 2^{\cali} \rightarrow 2^{\cali^v}$ such that,  for any $I\subseteq \cali$,
\[ C^v_j(I)\subseteq I \medcap \cali^v  \quad \mbox{ and  } \quad  |C^v_j(I)|\leq r^v_j.\]
\end{definition}
For any set of applicants, a single-category choice rule  specifies which applicants are chosen 
by a given job-category pair.\footnote{The inverse mapping of  a single-category choice rule  is a single-job and single-category mechanism.}

\begin{definition}  \label{def:multi-category}
Given a job $j\in \calj$, a \emph{multi-category choice rule} is a multidimensional function
$\vec{C}_j = (C^v_j)_{v \in \calv} :  2^{\cali} \rightarrow \prod_{v \in \calv}2^{\cali^v}$ such that,  for any $I\subseteq \cali$,
\begin{enumerate}
\item for any category $v \in \calv$,
\[ C^v_j(I)\subseteq I \medcap \cali^v  \quad \mbox{ and  } \quad  |C^v_j(I)| \leq r^v_j,
\]
\item for any two distinct categories $v,v'\in \calv$,
\[ C_j^v(I)\medcap C^{v'}_j(I)=\emptyset. \]
\end{enumerate}
\end{definition}
For any set of applicants, multi-category choice rule specifies which applicants are chosen for each of its categories by a given job.

\begin{definition}  \label{def:aggregate}
For any multi-category choice rule $\vec{C}_j=(C^v_j)_{v \in \calv}$,  the resulting \emph{aggregate choice rule}
$\widehat{C}_j: 2^{\cali} \rightarrow 2^{\cali}$ is such that, for any  $I\subseteq \cali$,
\[ \widehat{C}_j(I)=\bigcup_{v\in \calv} C^v_j(I).\]
\end{definition}
For any set of individuals, the aggregate choice rule yields the set of chosen individuals across all categories.

\subsection{2-Step Meritorious Horizontal Choice Rule}

The following single-category choice rule is introduced in \cite{sonmez/yenmez:22}.  

Consider a job $j\in \calj$ and a category $v\in \calv$. Let $I\subseteq \cali^v$
be a set of individuals who are eligible for category $v$.

	\begin{quote}
        \noindent{}{\bf Meritorious Horizontal Choice Rule} {\boldmath$C^{v}_{\circled{M},j}$}\smallskip

		\noindent{}{\bf Step 1.1}
		Assuming such an individual exists, let $i_1$ be the
		the highest merit-score individual (with respect to $\sigma_j$) in $I$ who has a trait for an  HR-protected  position.
		Choose individual $i_1$ for an HR-protected  position.
		 Let $I_1 = \{i_1\}$, and proceed with Step 1.2.
            If no such individual exists, proceed to Step 2. \smallskip
	
		\noindent{}{\bf Step 1.k} {\boldmath($k\in \{2, \ldots, \sum_{t\in \calt} r^{v,t}_j\}$)}
		Assuming such an individual exists, let $i_k$ be the
		the highest merit-score individual in $I \setminus I_{k-1}$
		with
		\[ n^v_j( I_{k-1} \medcup\{i_k\}) = n^v_j( I_{k-1})+1.\]
		Choose  individual $i_k$ for an HR-protected position.
		Let $I_k=I_{k-1}\medcup \{i_k\}$, and proceed with Step 1.(k+1).
		If no such individual exists, proceed to Step 2.\smallskip

        \noindent{}{\bf Step 2.}
           For unfilled positions, choose unassigned individuals with highest merit scores until either all positions
           are filled or  all individuals are selected. 
	\end{quote}

The following multi-category choice rule uses the meritorious horizontal single-category
choice rule multiple times;  first, to allocate  open-category positions, and next
for each VR-protected category to allocate VR-protected positions.
\smallskip

	\begin{quote}
        \noindent{}{\bf 2-Step Meritorious Horizontal (2SMH) Choice Rule} {\boldmath$\vec{C}^{2s}_{\circled{M},j} = (C^{2s,v}_{\circled{M},j})_{v \in \calv}$}\smallskip

\noindent For each set of individuals  $I \subseteq \cali$,
\begin{eqnarray*}
&& C^{2s,o}_{\circled{M},j}(I) = C^{o}_{\circled{M},j}(I), \mbox{ and}\\
&& C^{2s,c}_{\circled{M},j}(I) = C^{c}_{\circled{M},j}\Big(\big(I\setminus C^{o}_{\circled{M},j}(I)\big)\medcap{\cali}^c\Big) \quad \mbox{ for any } c \in \calr.
\end{eqnarray*}
    \end{quote}

\subsection{Stability of an Assignment}

The following axiom is a generalization of the celebrated \textit{stability\/} axiom \citep{gale/shapley:62, hatfield/milgrom:05} for two-sided matching markets. 

\begin{definition}
An assignment $\alpha$ is \emph{stable} with respect to a profile of multi-category
choice rules $(\vec{C}_j)_{j\in \calj}$ if the following three conditions hold:
    \begin{enumerate}
        \item \textit{Individual rationality\/}: For each $i\in \cali$, \; $\alpha(i) \succeq_i \emptyset$,
        \item \textit{Job-and-category rationality\/}: For each $j \in \calj$, \; 
        $\vec{C}_j(\alpha^{-1}(j))=(\alpha^{-1}(j,v))_{v\in \calv}$. 
        \item \textit{No blocking pairs\/}: There exist no $i\in \cali$ and $j\in \calj$ such that
        $j \succ_i \alpha(i)$  and $i \in \widehat{C}_j(\alpha^{-1}(j) \medcup \{i\})$.
    \end{enumerate}
\end{definition}

Note that, while in the same spirit, our concept of stability diverges from its conventional definition in the two-sided matching literature 
due to the additional specification of position categories in assignments. In traditional two-sided matching literature (e.g., \citealp{hatfield/milgrom:05}), 
\textit{stability\/} is understood within a single-category context. However, in our model, stability takes into account the significance of position categories, 
thereby assigning different roles to positions of varying categories.

Essentially, when a job ``blocks'' an outcome (potentially along with other agents), all its positions assume symmetric roles in the standard definition. 
In contrast, in our model, the roles of positions from different categories differ, reflecting affirmative action legislation in India.

In particular, the ``job-and-category rationality'' condition renders our stability notion stronger than its standard counterpart. 
This condition requires that a job not only maintains all its assigned individuals, as per standard stability, but also ensures the preservation of the categories of these individuals within the assignment. 
Otherwise, when there is a single category, our notion of stability is equivalent to its conventional definition.

Our first result establishes an important equivalence.  

\begin{theorem} \label{thm:resrulechar}
An assignment $\alpha$ is stable with respect to $(\vec{C}_{\circled{M},j}^{2s})_{j\in \calj}$ if and only if it satisfies
(i) individual rationality, (ii) non-wastefulness, (iii) maximal accommodation of HR protections,
(iv) no justified envy,  and
(v) compliance with VR protections. 
\end{theorem}

Thus, given the 2SMH choice rule for each institution,  collectively, the Saurav Yadav (2020) axioms  
are equivalent to a natural refinement of the celebrated \textit{stability} axiom for two-sided matching problems.

\section{Proposed Mechanism for India: 2SMH-DA}  \label{sec:2SMH+DA}

In this section, we  introduce a mechanism for the general  version of the problem with heterogeneous positions and both reservation policies, 
and show that it is uniquely compelling among those satisfying the Supreme Court's mandates. 

\subsection{Multiplicity of Stable Assignments with Heterogenous Positions}

When all positions are identical at a single job,
there is a unique  assignment--one that can be obtained with the inverse mapping of the 2SMH choice rule--that satisfies the Saurav Yadav (2020) axioms \citep{sonmez/yenmez:22}.
As we show in the next example, this is not the case when positions are heterogenous.

\begin{example} \label{ex:inefficiency}
There are two jobs, $x$ and $y$, each with one position open in the general category. Additionally, there are two traits: $t_1$ and $t_2$. 
The position at job $x$ is HR-protected for trait $t_1$, while the position at job $y$ is HR-protected for trait $t_2$. 
Two individuals, $a$ and $b$, possess traits $\tau(a) = {t_1}$ and $\tau(b) = {t_2}$, respectively. The preferences of these individuals and their merit rankings at the jobs are provided as follows:
\[
	\begin{array}{cc}
	\succ_{a} & \succ_{b}\\ \hline
	y       & x     \\
	x        & y     \\
	\emptyset & \emptyset
	\end{array}
\qquad \qquad \qquad
	\begin{array}{cc}
	\sigma_{x} \; & \sigma_{y}\\ \hline
	a       &  a    \\
	b        &  b    \\
	\mbox{} & \mbox{}
	\end{array}
	\]
Observe that, both of the following two assignments satisfy all five axioms:
\[ \alpha = \left(\begin{array}{cc}
	a & b\\
	(y,o)  &  (x,o)
	\end{array}\right)
\qquad \mbox {and} \qquad	
 \beta = \left(\begin{array}{cc}
	a & b\\
	(x,o)  &  (y,o)
	\end{array}\right).
\]
While both individuals receive their first choices under assignment $\alpha$,
they receive their second choices under assignment $\beta$.
\qed
\end{example}

Interestingly, Example \ref{ex:inefficiency} reveals that, in the absence of additional considerations, the HR policy may be detrimental to the very groups it is supposed to help. 
Specifically, if assignment $\beta$ is chosen by a central planner (potentially motivated by maximizing the number of honored HR-protected positions), 
the positive discrimination given to individual $a$ for the position at job $x$ due to their trait $t_1$ not only ends up harming individual $b$ but also individual $a$ themselves. 
Indeed, as individual $a$ has the highest merit score for both jobs, they do not require positive discrimination. 
Therefore, an excessive effort to honor HR-protected positions without considering individual preferences can be harmful to the very groups the HR policy is meant to protect. 
Fortunately, there is another assignment that not only satisfies Saurav Yadav (2020) axioms but also serves the best interests of all individuals, despite not honoring any of the HR-protected positions.

While our next axiom does not correspond to any
desiderata formulated by the Supreme Court, it is among the most fundamental principles in economic theory.

\begin{definition}
An assignment $\alpha\in \mathcal{A}$ \textit{Pareto dominates\/} assignment  $\beta\in \mathcal{A}$ if, for all  $i\in\cali$,
\[ \alpha(i)  \succeq_i \beta(i),\] 
which holds strictly for at least one individual.  

A mechanism $\varphi$ \textit{Pareto dominates\/} a mechanism $\phi$ if,
(i) the assignment $\varphi(\succ_{\cali})$ either Pareto dominates or is equal to the assignment  $\phi(\succ_{\cali})$ for each $\succ_{\cali}\in \mathcal{P}$, and
(ii) the assignment $\varphi(\succ_{\cali})$ Pareto dominates  the assignment  $\phi(\succ_{\cali})$ for some $\succ_{\cali}\in \mathcal{P}$.
\end{definition}

\begin{definition}
An assignment $\alpha\in \mathcal{A}$ is \textit{Pareto efficient\/} if, there is no other assignment $\beta \in \mathcal{A}$ such that
\begin{enumerate}
\item $\beta(i)  \succeq_i \alpha(i)$ for all $i\in\cali$, and
\item $\beta(i)  \succ_i \alpha(i)$ for some $i\in\cali$.
\end{enumerate}
A mechanism $\varphi$ is \emph{Pareto efficient\/} if its outcome $\varphi(\succ_{\cali})$ is Pareto efficient for
each $\succ_{\cali}\in \mathcal{P}$.
\end{definition}

It is well-established that, even in the absence of VR and HR protections, 
the axioms of \textit{individual rationality\/}, \textit{non-wastefulness\/},
and \textit{no-justified envy} are incompatible with \textit{Pareto efficiency\/} \citep{alcalde/barbera:94,balinski/sonmez:99}. 
However, this incompatibility disappears when the merit ranking of individuals is identical across all institutions  \citep{balinski/sonmez:99}.
Example \ref{ex:inefficient} in Section \ref{appendix:Chileanexamples} of the Online Appendix establishes 
an analogous but stronger incompatibility between the Saurav Yadav (2020) axioms and Pareto efficiency in our more general 
model – one that does not disappear even when all institutions have the same merit ranking of individuals.

While \textit{Pareto efficiency\/} is incompatible with \textit{individual rationality\/}, \textit{non-wastefulness\/},
and \textit{no-justified envy} in the absence of VR and HR policies, of all outcomes that satisfy these axioms, 
there exists one that \textit{Pareto dominates\/} any other  \citep{gale/shapley:62, balinski/sonmez:99}. 
We next show that this result extends to our model, thus motivating our proposed mechanism.  

\subsection{2SMH-DA Mechanism} \label{sec:2SMH-DA}

We are ready to present our proposed mechanism, which extends the (inverse mapping of the) 2SMH choice rule--defined for identical positions--to
the general version of the problem with heterogenous positions using the celebrated individual-proposing DA algorithm \citep{gale/shapley:62}.

	\begin{quote}
        \noindent{}{\bf 2SMH-DA Mechanism {\boldmath$\varphi^{2s}_{\circled{M}}$}}

        \noindent For each preference profile $\succ_{\cali}\;\in\calp$, the outcome $\varphi^{2s}_{\circled{M}}(\succ_{\cali})$ of the 2SMH-DA mechanism is obtained
        as follows:

		\noindent{}{\bf Step 1.}
		Assuming such a job exists,
            each individual $i$ applies to her most preferred acceptable job under $\succ_i$.
            Let $I_j^1$ be the set of individuals who apply to job $j$. Each job $j$
            tentatively assigns individuals in $\widehat C_{\circled{M},j}^{2s}(I_j^1)$ to its
            categories based on $\vec{C}_{\circled{M},j}^{2s}(I_j^1)$,
            and (permanently) rejects any remaining applicants. If there is no rejection by any job, then the procedure is terminated and
            the tentative assignments are finalized. Otherwise, proceed to Step 2.\smallskip
	
		\noindent{}{\bf Step k.}
		Assuming such a job exists,
			each individual $i$ who is rejected in Step $(k-1)$ applies to her next preferred
			acceptable job under $\succ_i$. For any job $j$,
			let $I_j^k$ be the set of new applicants in Step $k$ along with individuals who are tentatively
            assigned to categories of job $j$ in Step $(k-1)$.
            Each job $j$ tentatively assigns individuals in $\widehat C_{\circled{M},j}^{2s}(I_j^k)$
            to its categories based on $\vec{C}_{\circled{M},j}^{2s}(I_j^k)$, and (permanently) rejects
            any remaining applicants.
            If there is no rejection by any job, then the procedure is terminated  and
            the tentative assignments are finalized. Otherwise, proceed to Step (k+1).
            	\end{quote}
Since there is a finite number of jobs and individuals, this mechanism terminates at a finite round for
every preference profile.

We next present two additional results. The general message of these results is that while the Saurav Yadav (2020) 
axioms do not single out a mechanism when positions are heterogeneous, they come very close. 
Either one of two fundamental axioms in economic theory leads to a unique prescription.

\begin{theorem}\label{thm:dominate}
2SMH-DA  Pareto dominates any other mechanism that satisfies
(i) individual rationality,
(ii) non-wastefulness,
(iii) maximal accommodation of HR protections,
(iv) no justified envy, and
(v) compliance with VR protections.
\end{theorem}

The primary objective of the Indian reservation system is to enhance the social and educational status of underprivileged communities. 
Given Theorem \ref{thm:dominate}, we believe that 2SMH-DA is the only plausible mechanism to pursue that objective.

While the Pareto principle is perhaps the most fundamental axiom in economic theory, it is not the only condition that bridges the gap between the Saurav Yadav (2020) axioms and our proposed mechanism, 
2SMH-DA. The following highly sought-after incentive compatibility condition, defined exclusively for mechanisms (and not for assignments), also serves a similar role.

\begin{definition} \label{def:sp}
A mechanism $\varphi: \mathcal{P} \to \mathcal{A} $ is \emph{strategy-proof\/}
if, for each $\succ_{\cali}\in \mathcal{P}$, individual $i\in\cali$, and $\succ'_i\in \mathcal{P}_i$,
\[\varphi(\succ_{\cali})(i) \; \mathrel{\succeq_i} \; \varphi(\succ'_i,\succ_{\cali\setminus \{i\}})(i).\]
\end{definition}
Truthful preference revelation is always weakly more preferred than reporting any other preference ranking
for every individual under a strategy-proof mechanism.

\begin{theorem}\label{thm:jointchar}
A mechanism satisfies
(i) individual rationality,
(ii) non-wastefulness,
(iii) maximal accommodation of HR protections,
(iv) no justified envy,
(v) compliance with VR protections, and
(vi) strategy-proofness
if and only if it is 2SMH-DA.
\end{theorem}

\subsection{Related Literature in Matching Theory and Market Design}

To the best of our knowledge, our paper is the first to propose a mechanism that complies with the 
Supreme Court's mandates for the joint implementation of VR and HR policies in India when positions are heterogeneous. 
Our analysis builds upon \cite{sonmez/yenmez:22}, which addressed a simpler version of the problem with identical positions. While our work shares motivation with \cite{EhlersMorrill2020}, 
which also considers legal desiderata, our formulation directly relies on explicit mandates outlined in \textit{Saurav Yadav (2020)}.

Theorem \ref{thm:resrulechar}, which establishes an equivalence between assignments satisfying the \textit{Saurav Yadav (2020)} axioms and 
those satisfying a refinement of the traditional ``two-sided matching'' axiom of stability, has no antecedent in the literature.

The existence of an ``individual-optimal" stable outcome is a fairly robust result under a wide range of assumptions in two-sided matching and school choice \citep{gale/shapley:62, balinski/sonmez:99}. 
Theorem \ref{thm:dominate} generalizes this result to our model under our more refined definition of stability.

Theorem \ref{thm:jointchar} generalizes characterizations in \cite{sonmez/yenmez:22} derived for homogeneous positions only, as well as the basic characterizations in \cite{alcalde/barbera:94} and \cite{balinski/sonmez:99}, 
which were derived in the absence of any form of reservation policy. At a broader level, Theorem \ref{thm:jointchar} is also in the same spirit as the ``similar'' results in \cite{Hirata/Kasuya:17} and \cite{Hatfield/Kominers/Westkamp:21}, 
which characterize stable and strategy-proof mechanisms for the matching with contracts model.

Thus, one might be compelled to invoke these results by formulating our model as an application of the matching with contracts model, 
where each contract specifies an individual, a job, a vertical category, and a trait (or its absence). However, we cannot do this for two reasons. 
First, whereas the single-category choice rules in \cite{Hirata/Kasuya:17} and \cite{Hatfield/Kominers/Westkamp:21} are exogenous, our multi-category choice rule 2SMH is endogenous to the Supreme Court's mandates. 
Additionally, since many of the theoretical constructs and results in the matching with contracts literature, including \cite{Hirata/Kasuya:17} and \cite{Hatfield/Kominers/Westkamp:21}, 
assume strict preferences for individuals over the set of contracts, we cannot directly derive our results through this approach.

Interestingly, building on \cite{kurata17}, the team that designed the K-12 school choice mechanism for Chile adopted this approach \citep{correa19}. 
As we discuss in Section \ref{sec:chile}, their practical application is a special case of our model with HR policy only. 
In order to invoke the ``cumulative offer mechanism'' from the matching with contracts literature with ``minimum guarantee'' reserves, they imposed arbitrary preferences over different traits for students. 
In Examples \ref{ex1} and \ref{ex2}, presented in Section \ref{appendix:Chileanexamples} of the Online Appendix, 
we show that this approach results in not only losing \textit{no justified envy} but also \textit{maximal accommodation of HR protections}.

Subsequent to our paper, \cite{dogan/imamura/yenmez:22} discuss a generalization of our single-category analysis to broader distributional constraints beyond HR policy, 
where distributional objectives 
are given by rank functions of some matroids. In our single-category analysis, the function that gives the number of HR-protected positions 
that can be filled corresponds to the rank function of a transversal matroid. 
In particular, \cite{dogan/imamura/yenmez:22} apply their general theory to the school choice in Chile like we do in Section  \ref{sec:chile}, 
but they also accommodate policies that guarantee enrollment for returning students.

Prior to our paper, others have also proposed mechanisms based on the DA algorithm for practical applications in India. 
For instance, \cite{thakur18} adopts a similar approach for allocating government positions by the Union Public Service Commission, 
while both \cite{baswana18} and \cite{aygunturhan2021} utilize similar methods for seat allocation at engineering colleges.
However, unlike our paper, none of these prior works properly account for the HR policy. In \cite{thakur18} and \cite{aygunturhan2021}, the HR policy is entirely disregarded. 
In \cite{baswana18}, the authors detail their design and implementation of a seat allocation process for several technical universities in India. 
Despite incorporating both VR-protected and HR-protected groups into their design, they failed to distinguish between the two policies within their mechanism.
Consequently, their design is in direct violation of  \textit{Indra Sawhney (1992)\/}.\footnote{The mechanism devised by \cite{baswana18, baswana2019b} 
treats persons with disabilities (PwD) as a VR-protected group, despite them being explicitly granted HR protections by the Supreme Court's judgment in the 
\textit{National Federation Of The Blind (2013)} case. 
For instance, a candidate from the general category with a disability is first considered for an open-category position, and only afterward for a PwD position in the open category. 
This approach implies that a position awarded to a disabled individual does not count toward the HR-protected positions for PwD, contradicting the fundamental 
mandate outlined in the \textit{Indra Sawhney (1992)} case regarding the implementation of HR protections.}

Abstracting away from the specific laws in India, there's extensive research on priority-based resource allocation mechanisms that rely on the DA algorithm. Several papers in this field, such as those by
 \cite{hafalir/yenmez/yildirim:13},  \cite{ehlers/hafalir/yenmez/yildirim:14}, \cite{echenique/yenmez:15},  
 \cite{dur/kominers/pathak/sonmez:18},  \cite{dur/pathak/sonmez:20},  and \cite{abdulkadiroglu/grigoryan:21}, are particularly relevant to our work as they incorporate various reservation policies.

Finally, with our focus on a concrete real-life resource allocation problem, we contribute to the literature in market design.  
Studies in this vein include those on entry-level labor markets 
\citep{roth/peranson:99}, school choice \citep{abdulkadiroglu/sonmez:03, erdil/ergin:08, kesten:10, chenkesten17, hafalir/kojima/yenmez22, reny22}, 
spectrum auctions \citep{milgrom:00}, kidney exchange \citep{roth/sonmez/unver:04, roth:sonmez:unver:2005}, liver exchange \citep{ergin/sonmez/unver:20, Yilmaz:2023}, 
internet auctions \citep{edelman/ostrovsky/schwarz:07,varian:07}, 
cadet-branch matching \citep{sonmez/switzer:13,sonmez:13, greenberg/pathak/sonmez:23}, 
assignment of airport arrival slots \citep{schummer/vohra:13,schummer/abizada:17}, rent control \citep{andersson/svensson:14},
refugee resettlement \citep{jones/teytelboym:17,Andersson2019,anderson/ehlers/20, hadad/teytelboym:22, delacretaz/kominers/teytelboym:23}, 
and pandemic medical  resource allocation  \citep{pathak/sonmez/unver/yenmez:20}.

\section{Applications Beyond India} \label{sec:beyond-India}

While our analysis is motivated by India's legal and implementation challenges for its reservation system, 
our analytical results have policy relevance for applications in other countries as well. We are unaware of any institution in other countries which implements the VR and HR policies concurrently. 
However, either one of these policies is implemented on a stand-alone basis in several applications worldwide. 
For example, for the version of the problem with heterogeneous positions, the VR policy is implemented for allocation of seats at Chicago's elite high schools \citep{dur/pathak/sonmez:20}, 
and the HR policy with overlapping protected groups is implemented in all cities of Chile for allocation of K-12 public school seats \citep{correa19}. 
For the version of the problem with identical positions and overlapping protected groups, 
the Jordanian House of Representatives use a reservation system with 15 of the 130 seats reserved for women and 12 reserved for 
minorities,\footnote{Refer to \url{https://data.ipu.org/content/jordan?chamber_id=13434}, retrieved on 02/25/2024)} and the National Assembly of Pakistan use a reservation system with 
60 of the 342 seats reserved for women and 10 reserved for minorities\footnote{Refer to \url{https://na.gov.pk/en/content.php?id=2}, retrieved on 02/25/2024.}

We next present what our analysis reveals about allocation of K-12 public school seats in Chile, and how their student assignment mechanism can be improved.

\subsection{School Choice in Chile}\label{sec:chile}

Since 2020, Chile has implemented a centralized school choice system following the enactment of the \textit{School Inclusion Law} in 2015 \citep{correa19}. 
This system, covering Pre-K to grade 12, is a collaboration between the Chilean Ministry of Education and a team of economists and operations researchers. 
Similar to its predecessors, the Chilean school choice system is based on the individual-proposing DA algorithm. 
The following three features in its design make it a special case of our model, with only open category and overlapping HR protections:
\begin{enumerate}
\item Affirmative action is granted
for financially disadvantaged students and children with special needs, implemented through reserved seats at each school.
In addition, some schools are allowed to reserve seats for high-achieving students.

\item Reserved seats are implemented in the form of a ``minimum guarantee". Therefore, in our terminology, there are three ``traits'': 
\textit{Financially disadvantaged, Special needs}, and \textit{High-achieving}. A student can have any subset of these traits, or none at all. 
Students with none of the three traits are referred to as \textit{Regular}.

\item Students with multiple traits are eligible for reserved seats corresponding to each trait, but, as in India, they only count towards one reserved seat upon acceptance. 
\end{enumerate}

For this special case, our axiom of \textit{compliance with VR protections} becomes vacuous, and the 2SMH choice rule reduces to the \textit{meritorious horizontal} choice rule. 
Otherwise, our entire formal analysis holds. Let us refer to the resulting special case of the 2SMH-DA mechanism as the \textit{MH-DA} mechanism. 
Given Theorems \ref{thm:dominate} and \ref{thm:jointchar}, naturally, our proposed mechanism for Chilean school choice is the MH-DA mechanism. 
However, this is not the mechanism used in Chile. Adopting a methodology developed by \cite{kurata17}, the Chilean school choice mechanism is equivalent to using the 
DA algorithm together with a choice rule different from the meritorious horizontal choice rule.

Following \cite{kurata17}, the team which designed  the Chilean school choice mechanism
formulated their problem as an application of the \textit{matching with contracts\/} model of \cite{hatfield/milgrom:05}, where the
contractual term between a school and a student specifies which of the four types of seats
(i.e., open seats, reserved seats for financially disadvantaged students, reserved seats
for special needs students, and reserved seats for high-achieving students) the student receives.
However, the theory of matching with contracts is developed under the assumption that
students have strict preferences over all their contracts, which in this context corresponds to them having strict preferences on
the specific type of seats they receive at each school.
Since students declare preferences over only schools in Chile, a tie-breaking rule is used to construct
student preferences over specific type of seats at each school.
In \cite{correa19}, the designers emphasize that the choice of a tie-breaking rule is not straightforward,
and it has distributional consequences.
In order to implement the reserves in the form of a minimum guarantee, a design requirement by the policy makers,  they break ties in a way each student is assumed to prefer
reserved seats for any of their traits to open seats.
When each student has at most one trait, this construction assures that the reserves are implemented
as a minimum guarantee (\citealp{hafalir/yenmez/yildirim:13, dur/pathak/sonmez:20}).

Given Theorem \ref{thm:jointchar}, the Chilean school choice mechanism must violate some of our axioms. In Section \ref{appendix:Chileanexamples} of the Online Appendix, 
we present two simple examples demonstrating that the Chilean school choice mechanism fails to satisfy \textit{no justified envy} and \textit{maximal accommodation of HR protections}.  
Example \ref{ex1} reveals that the Chilean school choice mechanism violates \textit{no justified envy}, while Example \ref{ex2} shows that it fails to satisfy \textit{maximal accommodation of HR protections}.

\section{Conclusion} \label{sec:conclusion}

Public institutions in India have long struggled with implementing their constitutionally protected VR and HR policies, especially when either (i) the two policies are implemented together or (ii) positions are heterogeneous. 
While virtually all current field applications in India have the first feature, many others also have the second.

VR policy is designed to award reserved units to members of the protected group who could not receive an unreserved unit based on merit. 
Implementing this policy is straightforward when positions are homogeneous: first, allocate the unreserved units based on merit, 
and then allocate the VR-protected units to remaining members of the protected group again based on merit. Any position awarded to members of an HR-protected group, 
in contrast, counts towards accommodating the HR policy. Thus, it merely guarantees a minimum number of units for members of the target group.

Although the distinction between the two policies is complex, they are clearly defined in the landmark judgment \textit{Indra Sawhney (1992)}. 
This ruling mandated VR policy for historically oppressed groups like Scheduled Castes and Scheduled Tribes. 
Reservations for other groups, such as persons with disabilities or women, had to be implemented in the form of HR policy.

However, how should these two policies be implemented together, or when positions are heterogeneous? 
Since these are highly technical considerations, Indra Sawhney (1992) did not provide guidance on these operational aspects. 
Instead, these were left to government officials. However, these considerations are not any easier for government officials either. 
Mechanisms to implement these policies were regularly challenged in court. As a result, an explicit procedure—the SCI-AKG choice rule—was formulated in a subsequent Supreme Court judgment 
\textit{Anil Kumar Gupta (1995)} for the case with homogeneous positions, and mandated in the country.

In 2019, we observed that the mandated procedure had a critical flaw. Individuals who were eligible for both VR and HR protections, such as a women from Scheduled Castes,  
were losing their HR protections upon claiming their VR protections. This resulted in widespread failure of \textit{no justified envy}. 
For example, women from Scheduled Castes were losing positions to lower-merit women from more privileged upper castes. 
At this point, \textit{no justified envy} was not mandated in the country. Upon making these observations on the root causes of the challenges, in \cite{sonmez/yenmez:19a}, 
we proposed the 2SMH choice rule as a ``minimalist'' reform of the flawed procedure where HR protections are not lost upon claiming VR protections.

Shortly after discovering the issues in \textit{Anil Kumar Gupta (1995)}, in \cite{sonmez/yenmez:19b}, the precursor of this paper, we observed that the failures are much bigger when the positions are heterogeneous. 
Even though the Supreme Court never mandated a specific mechanism for this case, it gave a number of judgments based on rudimentary mechanisms designed by decision-makers 
using flawed concepts such as migration or meritorious reserved candidates. Unsurprisingly, this process led to inconsistencies within and between Supreme Court judgments. 
As long as mechanisms are designed using these flawed concepts, the challenges in India will continue. 
Fortunately, an easy fix exists for institutions through the 2SMA-DA mechanism, which simply implements the individual-proposing DA algorithm with the 2SMH choice rule for each institution.

While \cite{sonmez/yenmez:19a} was under editorial consideration, something remarkable happened. In \textit{Saurav Yadav (2020)}, 
the Supreme Court rescinded the flawed SCI-AKG choice rule and endorsed the 2SMH choice rule, independently discovered by the judiciary. Additionally, 
not only was \textit{no justified envy} mandated with this judgment, but it also formulated what it means to merit a position in the presence of HR protections. 
Since the judgment parallels \cite{sonmez/yenmez:19a}, later published in \cite{sonmez/yenmez:22}, we interpret this development as external validity for minimalist market design: 
Its prescription, the 2SMH choice rule, is the procedure that was originally intended but could not be formulated by the decision-makers for 25 years.

Even though the issues for the case of heterogeneous positions are much deeper, thanks to the celebrated DA algorithm, their solution is not much harder. 
With the clarified mandates of the Supreme Court in \textit{Saurav Yadav (2020)}, our results in Theorems \ref{thm:dominate} and \ref{thm:jointchar} 
suggest that the 2SMH-DA mechanism is uniquely suited to implement VR and HR policies when positions are heterogeneous. 
Just as in the case of homogeneous positions, we believe the prescription of minimalist market design is also the ``intended'' mechanism for this more complex version of the problem.

\bibliographystyle{aer}
\bibliography{India-Constitutional-v15-ArXiv}

\newpage

\appendix

\begin{center}
\textbf{\large{Online Appendix}}
\end{center}

\section{Comprehensive Legal Analysis and Supplemental Insights into Pivotal Affirmative Action Rulings in India} \label{sec:additional-judgments}

\subsection{Summary of \textit{Anurag Patel vs U.P. Public Service Commission (2004)}} \label{subsec-AnuragPatel}

The Uttar Pradesh Public Service Commission (UPPSC) conducted an
examination in 1990, 
and used the following mechanism to allocate 358 positions at various jobs:
\begin{quote}
\noindent \textit{Step 1\/}. Allocate the $\sum_{j\in J} r_j^o$ units of open-category positions using
the SSD induced by the given merit ranking: The highest merit ranking candidate  receives his top choice, the second
highest merit ranking candidate receives his top choice among the remaining open positions, and so on.

All assignments in this step are final.\smallskip

\noindent \textit{Step 2\/}. For each VR-protected category $c\in \{SC, ST, OBC\}$,  consider only category-$c$ candidates who have not received an assignment in Step 1,
and allocate the $\sum_{j\in J} r_j^c$ units of category-$c$ positions to these candidates using the SSD induced by the given merit ranking.

All assignments in this step are final.\smallskip
\end{quote}

At least one of the shortcomings of this mechanism is immediately apparent:
MRC candidates who receive their assignments in Step 1
are not given an opportunity to migrate and be considered for
any of the VR-protected positions for their categories, and as such
they often receive positions at less-preferred jobs compared to lower merit ranking candidates from their own categories.
Therefore, the UPPSC mechanism fails to respect \textit{inter se merit\/},
an important principle that plays a key role in all Supreme Court cases we discuss in
Sections \ref{subsec-AnuragPatel}, \ref{subsec-RameshRam} and \ref{subsec-Sharan}.
This shortcoming of the UPPSC mechanism resulted in a lawsuit at the High Court of Allahabad, and consequently
the UPPSC was ordered to come up with a reallocation that respects \emph{inter se} merit.
This reallocation, in turn, resulted in an appeal at the Supreme Court by a candidate
who was adversely affected by the high court's decision.
The appeal was dismissed by the Supreme Court, and the high court's decision was sustained,
reaffirming that the mechanism has to respect \emph{inter se} merit.
The following quote is from this important judgment:
\begin{indquote}
In the instant case, as noticed earlier, out of 8 petitioners in writ petition No. 22753/93, two of them who had secured ranks 13 and 14 in the merit list,
were appointed as Sales Tax Officer-ll whereas the persons who secured rank Nos. 38, 72 and 97, ranks lower to them,
got appointment as Deputy Collectors and the Division Bench of the High Court held that it is a clear injustice to the persons who are more meritorious and directed that a list of all selected backward class candidates shall be prepared separately including those candidates selected in the general category and their appointments to the posts shall be made strictly in accordance with merit as per the select list and preference of a person higher in the select list will be seen first and appointment given accordingly, while preference of a person lower in the list will be seen only later.
\end{indquote}
\textit{Anurag Patel (2004)\/}
is best known for reaffirming that any mechanism used for allocation of government jobs or seats at
public educational institutions has to respect \emph{inter se} merit.\footnote{\textit{Anurag Patel (2004)\/} also supports our position that,
the principles on implementation of VR and HR policies clarified in \textit{Saurav Yadav (2020)\/} is not limited to applications with identical positions,
but they apply more broadly for applications with heterogenous positions as well.}
Therefore, an MRC is entitled by law to migrate to
a higher choice job claiming a position vertically reserved for his reserved category,
answering the first question in Section \ref{subsec-MRC} in the positive.

\subsection{Legal Analysis of \textit{Union of India vs Ramesh Ram (2010)}} \label{subsec-RameshRam}

Selection to three \textit{All India Services} (Indian Administrative Service, Indian Foreign Service, and Indian Police Service), 
as well as to eighteen other services across various government departments, is overseen by the Union Public Service Commission (UPSC) through periodic Civil Service Examinations (CSE). 
Based on the merit ranking generated by the CSE and the candidates' submitted preferences, positions are allocated with the following mechanism.

\begin{quote}

\noindent{}{\bf UPSC Mechanism}\smallskip

\noindent{}{\bf Step 1.}  Tentatively allocate the $\sum_{j\in J} r_j^o$ units of open-category positions using the SSD induced by the given merit ranking. 
Promote the VR-protected candidates who secured tentative positions at this step to the status of an MRC.

Finalize all tentative assignments except those received by the MRCs.\smallskip

\noindent{}{\bf Step 2.} 
For each VR-protected category $c \in \{SC, ST, OBC\}$, consider all category-$c$ candidates
(including MRCs who each received a tentative assignment in Step 1), and tentatively
allocate the $\sum_{j\in J} r_j^c$ units of category-$c$ positions to these candidates using the SSD induced by the given merit ranking.

Finalize all tentative assignments except those received by the MRCs. \smallskip

\noindent{}{\bf Step 3.} 
Let $m^c$ denote the number of MRCs from category  $c \in \{SC, ST, OBC\}$.
Restricting attention to candidates who have not received an assignment (tentative or final) in Step 1 or Step 2, prepare the following four waitlists:
\begin{enumerate}
\item General category waitlist: $(m^{SC}+m^{ST}+m^{OBC})$ highest merit ranking general category candidates.
\item Category-SC waitlist: $m^{SC}$ highest merit ranking candidates from SC.
\item Category-ST waitlist: $m^{ST}$ highest merit ranking candidates from ST.
\item Category-OBC waitlist: $m^{OBC}$ highest merit ranking candidates from OBC.\smallskip
\end{enumerate}

\noindent{}{\bf Step 4.} 
Finalize the assignment of each MRC with the more-preferred of the (at most) two tentative assignments received in Steps 1 and 2. 
In cases where the two tentative assignments correspond to the same job, finalize the open-category position received in Step 1. \smallskip

\noindent{}{\bf Step 5.}  For each MRC, (at most) one position may be vacated at Step 4 and become available for reassignment. They are allocated to waitlisted candidates as follows:\smallskip
\begin{itemize}
\item[(i)] For each MRC whose assignment is finalized as the VR-protected position they received in Step 2, the open-category position they received in Step 1 becomes vacant. 
Allocate these vacated open-category positions to candidates in the general category waitlist using the SSD induced by the merit ranking.

\item[(ii)] For each MRC from category $c$ whose assignment is finalized as the open-category position they received in Step 1, 
a category-$c$ position may be vacated if the MRC tentatively received one in Step 2. Allocate these vacated category-$c$ positions to candidates in the category-$c$ waitlist using the SSD induced by the merit ranking.\smallskip
\end{itemize}
\end{quote}

UPSC declares the results in two stages: Steps 1-3 in the first stage, and Steps 4 and 5 in the second stage. 
Under their mechanism, the MRC-related questions posed in Section \ref{subsec-MRC} are handled as follows:

\begin{enumerate}
\item Consistent with \textit{Anurag Patel (2004)}, an MRC is allowed to migrate to a preferred job, claiming a VR-protected position for their category.
\item An open-category position tentatively assigned to an MRC in Step 1 is awarded in Step 5 to a waitlisted candidate from the general category if  the MRC receives a more-preferred position in Step 2 that is VR protected.
\end{enumerate}

The legality of the UPSC mechanism was scrutinized at each of the three levels of the Indian Judicial System. Initially, a number of OBC candidates, 
each of whom failed to receive an assignment despite being waitlisted, filed several applications at various branches of the Central Administrative Tribunal, challenging the UPSC mechanism. 
They argued that MRCs should not be allowed to migrate to a higher choice job, claiming positions that are VR protected for SC/ST/OBC candidates.

Of course, the petitioners' position is against the principle of \emph{inter se merit} and in direct conflict with \textit{Anurag Patel (2004)} 
(cf. Section \ref{sec:AnuragPatel-maintext} and Section \ref{subsec-AnuragPatel} in the Online Appendix). Despite the unsustainable position taken by the petitioners, 
their case was not dismissed. The Tribunal instead ruled that, while the MRCs can be allowed to migrate to a higher choice job claiming positions that are vertically reserved for their categories, 
this shall not be done at the expense of consuming VR-protected positions for categories SC, ST, and OBC. In other words, while the petitioners challenged 
Step 1 of the UPSC mechanism, the Tribunal required the UPSC to change Steps 2, 3, and 5 of its mechanism.

This ruling was challenged by the Union of India at the Madras High Court.
Not only did the Union of India lose their appeal in a judgment upholding the Tribunal's decision, the High Court ruled
the following aspect of the UPSC mechanism  to be unconstitutional:

\begin{indquote}
Rule 16.(2): While making service allocation, the candidates belonging to the Scheduled Castes, the Scheduled Tribes or Other Backward Classes
recommended against unreserved vacancies may be adjusted against reserved vacancies by the Govt. if by this process they get a service of
higher choice in the order of their preference.
\end{indquote}

This corresponds to ruling Steps 2, 3, and 5 of the UPSC mechanism as unconstitutional. 
Consequently, the High Court directed the Government of India and UPSC to repeat the allocation process in the absence of their Rule 16(2).

The judgment of the Madras High Court, in turn, was challenged by the Union of India at the Supreme Court in \textit{Ramesh Ram (2010)}. 
In a decree that became a main reference for the allocation of government positions, the appeal was allowed, the judgment of the Madras High Court was set aside, and the UPSC mechanism was ruled constitutional.
The following statement is from the conclusion of this historical decree:
\begin{indquote}
We sum up our answers-:

i) MRC candidates who avail the benefit of Rule 16 (2) and adjusted in the reserved category should be counted as part of the reserved pool for the purpose of
computing the aggregate reservation quotas. The seats vacated by MRC candidates in the General Pool will be offered to general category candidates.

ii) By operation of Rule 16 (2), the reserved status of an MRC is protected so that his/her better performance does not deny him of the chance to be allotted to a more-preferred service.

iii) The amended Rule 16 (2) only seeks to recognize the inter se merit between two classes of
candidates i.e. a) meritorious reserved category candidates b) relatively lower ranked reserved category candidates,
for the purpose of allocation to the various Civil Services with due regard for the preferences indicated by them.

iv) The reserved category candidates ``belonging to OBC, SC/ ST categories'' who are selected on merit and placed in the
list of General/Unreserved category candidates can choose to migrate to the respective reserved category at the time of
allocation of services. Such migration as envisaged by Rule 16 (2) is not inconsistent with
Rule 16 (1) or Articles 14, 16 (4) and 335 of the Constitution.
\end{indquote}

Therefore, in the context of allocation of government jobs, the Supreme Court judgment \textit{Ramesh Ram (2010)}
provides the following answers to the questions posed in Section \ref{subsec-MRC}:
\begin{enumerate}
\item An MRC is entitled to migrate to a higher choice job claiming a VR-protected position for his category.
\item The open-category positions vacated by MRCs are to be offered to the general category candidates.
\end{enumerate}

The judges of the Supreme Court justified this crucial decision based on the principle of \emph{inter se merit}, reaffirming the judgment in \textit{Anurag Patel (2004)}. 
However, their judgment overlooked a critical aspect, rendering the UPSC mechanism unconstitutional. Despite overruling the Madras High Court's judgment and justifying 
their decision based on the principle of \emph{inter se merit}, the Supreme Court judges failed to observe that the UPSC mechanism itself does not satisfy this fundamental principle. 

\begin{example} \label{ex-intersemerit}

There are three jobs $x, y, z$ and one VR-protected category $c$. Each job has one open-category position. In addition, job $x$ has a VR-protected position for category $c$. 
There are five candidates $a_1, a_2, a_3, b_1, b_2$. Candidates $b_1$ and $b_2$ are members of category $c$ and hence are eligible for the single VR-protected position. 
Candidates $a_1, a_2, a_3$ are members of the general category and therefore ineligible for the VR-protected position. 
All candidates have the same preferences: $x$ is their first choice, $y$ is their second choice, $z$ is their third choice, and remaining unmatched is their last choice.

All jobs have the same merit ranking of the candidates based on the  merit function $\sigma$ as follows:
\[ \sigma(a_1) > \sigma(b_1) > \sigma(a_2)  > \sigma(b_2) > \sigma(a_3).\]

We next find the outcome of the UPSC mechanism:

\textbf{Step 1.} The highest merit ranking candidate $a_1$ tentatively receives an open position at job $x$, the second highest merit ranking
candidate $b_1$ receives an open position at job $y$, and  the third highest merit ranking
candidate $a_2$ receives an open position at job $z$.

Candidate $b_1$ is given the status of an MRC. Assignments for candidates $a_1$ and $a_2$ are finalized as open positions at jobs $x$ and $z$, respectively. \smallskip

\textbf{Step 2.} Candidates $b_1$ and $b_2$ are the only ones  eligible for the category-$c$ position at job $x$.
 Having higher merit ranking than candidate $b_2$, candidate $b_1$ tentatively receives this position. \smallskip

\textbf{Step 3.} A waitlist each is prepared for the general category and category-$c$.
Since there is only one MRC candidate, there is a single candidate in each waitlist.
Candidates $a_3$ and $b_2$ are waitlisted at the general category waitlist and
category-$c$ waitlist respectively. \smallskip

\textbf{Step 4.} Having the status of an MRC, candidate $b_1$'s assignment is finalized as the more-preferred position he tentatively received from Steps 1 and 2, namely the category-$c$ position at his first-choice job $x$.\smallskip

\textbf{Step 5.} The position vacated by candidate $b_1$ is an open-category position at job $y$, which is then assigned to candidate $a_3$ as the only individual in the general category waitlist.\smallskip

\noindent Therefore, the final assignment is given as follows:

\[	\left(\begin{array}{ccccc}
	a_{1} & a_{2} & a_{3} & b_{1} & b_{2}\\
	(x,o)        & (z,o)       & (y,o)        & (x,c)        & \emptyset
	\end{array}\right). \smallskip
	\]
	
Observe that this assignment does not respect \emph{inter se merit}.
Candidate $a_2$ receives a less-preferred assignment than candidate $a_3$,
despite being a member of the same category (i.e., the general category) and having a higher merit score.
\qed
\end{example}

Indeed, a close inspection of Example \ref{ex-intersemerit} reveals a number of additional issues with the judgment in \textit{Ramesh Ram (2010)}.
The Supreme Court ruled that:\smallskip

\begin{indquote}
The seats vacated by MRC candidates in the General Pool will be offered to general category candidates.\smallskip
\end{indquote}

This decision would be plausible only if candidates from the general category are more meritorious than those in the VR-protected categories. 
As it is seen in Example \ref{ex-intersemerit}, this may not always be the case. In our view, offering the vacated position to the lowest merit ranking candidate $a_3$ is not justified 
when the higher merit ranking candidate $b_2$ remains unassigned simply because they are a member of a VR-protected category. 
A system that is intended as positive discrimination for candidate $b_2$ results in their discrimination. 
Equivalently, the minimum score needed for a position is higher in this example for the category-$c$ candidates than for the general category candidates.\footnote{October 2019 \textit{ThePrint\/} story ``Why civil services
exams in some states have had higher cut-offs for SC/ST \& OBC applicants'' gives a real-life example of this failure. 
See  \url{https://theprint.in/opinion/why-civil-services-exams-in-some-states-have-had-higher-cut-offs-for-sc-st-obc-applicants/},
(retrieved on 02/18/2024).}

These types of scenarios result in some other related anomalies as well.
In the absence of affirmative action, the outcome of the UPSC mechanism would have been
	\[
	\left(\begin{array}{ccccc}
	a_{1} & a_{2} & a_{3} & b_{1} & b_{2}\\
	x        & y       &  \emptyset        & x        & z
	\end{array}\right),
	\]
and the sole VR-protected candidate $b_2$ would have been better off.
Or, alternatively, had candidate $b_2$ not claimed his VR protections,
he would have received a position at job $z$.

\subsection{Legal Analysis of \textit{Tripurari Sharan vs Ranjit Kumar Yadav (2018)}} \label{subsec-Sharan}

The judgment in  \textit{Ramesh Ram (2010)},  discussed in Section \ref{subsec-RameshRam}, is now considered a main reference for allocation of government jobs when positions are heterogeneous.
Based on this reference judgment,  open-category seats vacated by MRC candidates are to be offered
to general-category candidates
for allocation of government jobs. We emphasize \emph{government jobs}, because the Supreme Court has taken a contrary position for the allocation
of seats at medical colleges. While the main reference for this application is considered to be \textit{Shri Ritesh R. Sah vs Dr. Y.L. Yamul \& Ors (1996)},
we instead discuss the more recent Supreme Court case
\textit{Tripurari Sharan  (2018)}, for it is more illuminating for our purposes.

Citing the  judgment in \textit{Ramesh Ram (2010)}, the petitioners appealed in  \textit{Tripurari Sharan (2018)} an earlier decision by the Patna High Court, which ruled:
\begin{indquote}
In case of admission to medical institutions, an MRC can have in, for the purpose of allotment of institutions, of his choice,
the option of taking admission in a college, where a seat in his category is reserved. Though admitted against a reserved seat,
for the purpose of computation of percentage of reservation, he will be deemed to have admitted as an open category candidate,
rather he remains an MRC. He cannot be treated to have occupied a seat reserved for the category of reservation he belongs to.
Resultantly, this movement will not lead to ouster of the reserved candidate at the bottom on the list of that reserved category.
While his/her selection as reserved category candidate shall remain intact, he/she will have to adjusted against remaining seats,
because of movement of an MRC against reserved seats, only for the purpose of allotment of seats.
\end{indquote}
Aware of the contradictory judgment in \textit{Ramesh Ram (2010)},
the judges of the Patna High Court justified their decision as follows:
\begin{indquote}
\noindent (i) There is an obvious distinction between qualifying through a common entrance test for securing admission
to medical courses in various institutions vis-a-vis a common competitive examination held for filling up vacancies in various services.

\noindent (ii) This distinction arises because all candidates receive, in a case of common entrance test held for securing admission in
medical institutions, the same benefits of securing admission in one of the medical institutions, in a particular course,
whereas in the case common selection process adopted for filling up vacancies in various services, there are variations,
which accrue to the successful candidates, because the services may differ in terms of status and conditions of service
including pay scale, promotional avenues, etc. Consequence of migration of an MRC to the concerned reserved category
shall be, therefore, different in case of the admission to various medical institutions vis-a-vis selection to various posts.
\end{indquote}
According to the judges, while the benefits from securing different jobs may vary, the benefits from securing admission to
different medical institutions are uniform. We do not agree with this assessment; however, even if that is the case,
then why bother migrating an MRC to a higher choice medical institution?

The appeal was declined by the Supreme Court in  \textit{Tripurari Sharan (2018)}, reaffirming the Patna High Court's decision.
Furthermore, the Supreme Court judgment also specified the exact manner in which the open-category seats vacated by MRC candidates
are to be filled in allocation to medical institutions:
\begin{indquote}
\hspace{2mm} i) An MRC can opt for a seat earmarked for the reserved category, so as to not disadvantage him against less
meritorious reserved category candidates. Such MRC shall be treated as part of the general category only.

ii) Due to the MRC's choice, one reserved category seat is occupied, and one seat among the choices available to
general category candidates remains unoccupied. Consequently, one lesser-ranked reserved category candidate who had
choices among the reserved category is affected as he does not get any choice anymore.

To remedy the situation i.e. to provide the affected candidate a remedy, the 50th seat [intended as the last reserved position]
which would have been allotted
to X-MRC, had he not opted for a seat meant for the reserved category to which he belongs, shall now be filled up by that
candidate in the reserved category list who stands to lose out by the choice of the MRC.
\end{indquote}
So an MRC is allowed to migrate to a VR-protected seat at a higher choice college in order to respect \emph{inter se} merit,
and the open-category seat vacated by the MRC  is to be awarded to the VR-protected candidate
who is displaced due to this migration.
There are numerous issues with this judgment, including its contradiction with \textit{Ramesh Ram (2010)}.
But perhaps the most striking one is, the following inconsistency in the final judgment quoted above:
While the judges  justify part (i) above on the basis of \emph{inter se} merit, they fail to observe that
their mandate in part (ii) results in a potential compromise of \emph{inter se} merit!
As such, this judgment contradicts with \textit{Anurag Patel (2004)} as well.
This is the main point made in our next example.

\begin{example}
There are two colleges $x$ and $y$.
College $x$ has two open-category seats and  two VR-protected seats for category $c$.
College $y$  has one open-category seat only.
There are five candidates $a_1, a_2, b_1, b_2, b_3$.
Candidates  $b_1, b_2, b_3$ are members of category-$c$ and hence they are
eligible for the VR-protected position.
Candidates $a_1, a_2$ are members
of the general category, and therefore ineligible for the VR-protected position.

Preferences of the candidates are are given as follows.
\[
	\begin{array}{ccccc}
	\succ_{a_{1}} & \succ_{a_{2}} & \succ_{b_{1}} & \succ_{b_{2}} & \succ_{b_{3}}\\ \hline
	x        & x       &  x        & y        & y\\
	y        & y       &  y        & x        & x
	\end{array}
	\]
Both schools have the same merit ranking of the candidates,  given by the  merit function $\sigma$ as follows:
\[ \sigma(a_1) > \sigma(a_2) > \sigma(b_1)  > \sigma(b_2) > \sigma(b_3).\]

While the mechanisms of various medical colleges may differ, they all produce the same  assignment
in this example, provided that they comply with the judgment in \textit{Tripurari Sharan (2018)}.
The three open-category seats are allocated  to the highest merit score candidates,
where the general category candidates $a_1, a_2$ each receive an open-category seat at college $x$, and the
category-$c$ candidate $b_1$ tentatively receives an open-category seat at college $y$.
Receiving a seat on his own merit,
category-$c$ candidate $b_1$ is promoted to the status of an MRC.
The two category-$c$ seats at college $x$ are tentatively allocated to the two remaining candidates $b_2$ and $b_3$ from category $c$.
At this stage, the court decision in  \textit{Tripurari Sharan (2018)} kicks in.
Candidate $b_1$ who is promoted to the status of an MRC
prefers a seat at college $x$ to his tentative assignment at college $y$.
Therefore, he is assigned one of these seats at the expense of the lowest merit ranking category-$c$ candidate $b_3$.
Again, by  \textit{Tripurari Sharan (2018)},  category-$c$ candidate $b_3$ receives the open-category seat at college $y$
that is vacated by $b_1$, ironically profiting from this adjustment.
The assignment dictated by the Supreme Court's decision is:
\[
	\left(\begin{array}{ccccc}
	a_{1} & a_{2} & b_{1} & b_{2} & b_{3}\\
	(x,o)        & (x,o)       &  (x,c)        & (x,c)      & (y,o)
	\end{array}\right).
\]
This outcome fails \emph{inter se\/} merit, because category-$c$ candidate $b_2$ receives a less-preferred assignment than
the assignment of the lower merit ranking category-$c$ candidate $b_3$.
\qed
\end{example}

\subsection{Legal Analysis of \textit{Samta Aandolan Samiti vs Union of India (2013)}} \label{subsec-Samiti}

As we have presented in Sections \ref{subsec-AnuragPatel},  \ref{subsec-RameshRam} and \ref{subsec-Sharan}, allocation of positions at government jobs
and publicly funded educational institutions in India typically relies on the use of the SSD mechanism in two stages,
first for the open-category positions, and then in parallel for each category of reserved positions.  The outcome obtained in this way is almost always tentative, and
the mechanics at the final phases of the individual mechanisms differ,
depending on the  MRC-related adjustments carried out.
One very convenient feature of a SSD is that, not only it can be implemented as a direct preference revelation mechanism where the candidates submit their preferences,
but it can also be used as a sequential mechanism where the candidates pick their choices one at a time following their merit rankings.
Indeed, this feature of the SSD is utilized in some of the applications in India.
The lawsuit brought to the Supreme Court in \textit{Samta Aandolan Samiti (2013)}\footnote{The case is available at \url{https://indiankanoon.org/doc/60144106/} (last accessed on 04/01/2019).}  is about one of these applications.

As in  \textit{Tripurari Sharan (2018)\/}, discussed in Section \ref{subsec-Sharan}, the petition in  \textit{Samta Aandolan Samiti (2013)}
also concerns the allocation of seats at medical colleges, and as such the precedent for this case is also
\textit{Shri Ritesh R. Sah (1996)}.
The following sequential mechanism is used to jointly allocate seats at seven campuses of
\textit{The All India Institute of Medical Sciences\/} (AIIMS):
\begin{quote}
\noindent \textit{Step 1\/}. Following their merit ranking, the open-category positions are allocated to
candidates one at a time,
where each candidate picks an available open-category position. Candidates from the categories SC/ST/OBC who receive
positions at this step earn the status of MRC, and their assignments are tentative.
Assignments  to the general-category candidates, on the other hand, are final. \smallskip

\noindent \textit{Step 2\/}. For each category $X \in \{SC, ST, OBC\}$,
consider all category-$X$ candidates, including the MRC candidates, who have
been tentatively holding one open-category seat each. Category-$X$ candidates
sequentially pick one reserved seat at a time following
their merit rankings, until all category-$X$ seats are exhausted.\footnote{Observe that MRC candidates from category-$X$
make their picks before the remaining members of category-$X$ due to their higher merit rankings.}
In addition to choosing among colleges with available reserve seats, an MRC candidate
is also allowed to keep the open position he is tentatively assigned. If an MRC candidate
keeps his tentative assignment from Step 1, this becomes his final  assignment.
If an MRC candidate opts for a position at another college, the open position vacated  by the MRC candidate
(i.e., the open-category position which was his tentative assignment)
is transferred to pool of reserved seats for category-$X$.
\end{quote}
It is easy to see that,
unlike the mechanisms presented  in Sections \ref{subsec-RameshRam}, \ref{subsec-AnuragPatel} and \ref{subsec-Sharan},
the \textit{AIIMS mechanism\/} respects \emph{inter se} merit.
This is because a candidate in any given category has an opportunity to pick a seat before all lower merit ranking candidates of
his own category. However, this mechanism suffers from another (highly visible) shortcoming:
it is vulnerable to \textit{collusion\/} between the members of any one of the categories SC, ST, and OBC.
Moreover, this vulnerability is not very subtle.
Any MRC has an opportunity in Step 2 to increase the number of reserved positions earmarked for his category by one unit,
by simply claiming in Step 1 a seat he does not intend to keep.
Not surprisingly, this vulnerability of the AIIMS mechanism was exploited not only by its participants, but also
by its administrators,
which was one of the reasons this mechanism was challenged in   \textit{Samta Aandolan Samiti (2013)}.
The following quote from the court proceedings illustrates the extent of this collusion:
\begin{indquote}
The petitioners aver that the respondents had conducted the counseling in strict adherence of the procedure quoted hereinabove.
However, the respondents forced reserve candidates to obtain the unreserved
(UR) seats by note (4.2.a) in counseling call letter.
In this way the respondents deliberately tried to convert UR seats to reserve category seat because of note 4.2. Otherwise the candidates
would have been provided freedom to opt seats under UR seats or category seats of their choice in different AIIMS.
\end{indquote}
In this way, members of OBC secured 45\% of the seats even though they were reserved 27\% of the seats.
Ironically, the Supreme Court did not find any merit in the petition, dismissing the case.

\subsection{Additional Key Excerpts from \textit{Union Of India vs Ramesh Ram (2010)}}\label{sec:ramesh}

There are court rulings in India where the judges have observed the failure of the principle of \textit{inter se merit\/},
one of the failures of the MRC-based mechanisms presented in Section \ref{subsec-RameshRam} of the Online Appendix,
and demanded institutions to design mechanisms which avoid this failure.
The following quote is given in \textit{Ramesh Ram (2010)}:
\begin{indquote}
Central Administrative Tribunal, Chennai Bench in O.A. No. 690 of 2006 and 775 of 2006 had given the following directions:

``(i) The impugned Rule 16 (2) is declared as valid so long as it is confined to allocation of services and confirms
to the ratio of Paras 4 to 6 of Anurag Patel order of the Hon'ble Apex Court.

(ii) The Supplementary List issued by the second respondent to the first respondent dated 3.4.2007 is set aside.
This would entail issue of a fresh supplementary result from the reserved list of 64 in such a way that adequate number of
OBCs are announced in lieu of the OBCs who have come on merit and brought under General Category.
The respondents are directed to rework the result in such a way the select list for all the 457 candidates are announced in
one lot providing for 242-general, 117 OBC, 57 SC and 41 ST and also ensure that the candidates in OBC, SC \& ST who
come on merit and without availing any reservation are treated as general candidates and ensure that on equal number of
such reserved candidates who are of merit under General Category, are recruited for OBC, SC \& ST respectively and
complete the select list for 457. Having done this exercise, the respondents should apply Rule 16 (2) to ensure that allocation
of the service is in accordance with rank-cum- preference with priority given to meritorious reserved candidates for service
allocation by virtue of Rule 16 (2) which is as per para 5 of Anurag Patel order. The entire exercise, as directed above, should be completed as per the order.

(iii) Applying the ratio of Anurag Patel decision of Hon'ble Apex Court (Paras 6 \& 7), if there is need for re-allocation of services,
 the respondents will take appropriate measures to that extent and complete this process also within two months from the date of receipt of a copy of this order."

The CAT had also issued the following direction as to how the results of the UPSC examinations (2005) should have been announced:

``If the UPSC had followed the decision of the Hon'ble Apex Court cited supra and released the select list in one go for all the
457 vacancies then it would have ensured that the select list contained not only 117 OBCs but also an additional number of OBC
candidates by this number, in additional to 117 under 27\% reservation, while simultaneously be number of general candidates
recruited will be less to the extent of OBCs recruited on merit and included in the general list in the result of Civil Services Examination, 2005.
Once this order is met, the successful candidates list will include 242 candidates in the General Category which is inclusive of all those Reserved Category candidates coming on merit plus 117
OBC, 57 SC and 41 ST exclusively from these respective reserved categories by applying relaxed norms for them..
If such a list is subjected to Rule 16(2) of Civil Services Examination, 2005 in present form for making service allocation
only and then services are allotted based on Rule 16(2) in this context, then the announcement of recruitment result
and allocation services will be both in accordance with law as per various judgments the Hon'ble Apex Court and in accordance
with the extent orders issued by the Respondent No.1 and also in keeping with spirit of Rule 16 (2) so that, the meritorious reserved candidates get higher preference service
as compared to their lower ranked counter parts in OBC, ST,SC. In doing so, the respondents also would notice that the
steps taken by them in accordance with the Rules 16 (3)(-)(5) are redundant once they issue the result of recruitment
in one phase, instead of two as they have become primary cause for the litigation and avoidable confusion in the minds of the candidates seeking recruitment.''
\end{indquote}

\subsection{Key Excerpts from \textit{Saurav Yadav vs The State Of Uttar Pradesh (2020)}}\label{sec:saurav}

The following  paragraphs in \textit{Saurav Yadav (2020)\/} clarifies that,  any VR-protected individual who
deserves an open-category position on the basis of merit should be assigned an open-category position (and not a VR-protected position),
including VR-protected individuals who deserve an HR-protected position at open-category.
The clarification is important for it removes an ambiguity in the original formulation of VR protections in the landmark Supreme Court judgment \textit{Indra Sawhney (1992)\/}.

\begin{indquote}
24. Thus, according to the second view, different principles must be adopted at two stages; in that:-.

(I) At the initial stage when the ``Open or General Category'' seats are to be filled, the claim of all reserved category candidates based on merit must be considered and if any candidates from such reserved categories, on their own merit, are entitled to be selected against Open or General Category seats, such placement of the reserved category candidate is not to affect in any manner the quota reserved for such categories in vertical reservation.

(II) However, when it comes to adjustment at the stage of horizontal reservation, even if, such reserved category candidates are entitled, on merit, to be considered and accommodated against Open or General Seats, at that stage the candidates from any reserved category can be adjusted only and only if there is scope for their adjustment in their own vertical column of reservation.

Such exercise would be premised on following postulates: -

(A) After the initial allocation of Open General Category seats is completed, the claim or right of reserved category candidates to be admitted in Open General Category seats on the basis of their own merit stands exhausted and they can only be considered against their respective column of vertical reservation.

(B) If there be any resultant adjustment on account of horizontal reservation in Open General Category, only those candidates who are not in any of the categories for whom vertical reservations is provided, alone are to be considered.

(C) In other words, at the stage of horizontal reservation, Open General Category is to be construed as category meant for candidates other than those coming from any of the categories for whom vertical reservation is provided.

25. The second view may lead to a situation where, while making adjustment for horizontal reservation in Open or General Category seats, less meritorious candidates may be adjusted, as has happened in the present matter. Admittedly, the last selected candidates in Open General female category while making adjustment of horizontal reservation had secured lesser marks than the Applicants. The claim of the Applicants was disregarded on the ground that they could claim only and only if there was a vacancy or chance for them to be accommodated in their respective column of vertical reservation.\medskip

[\dots]\medskip

31. The second view is thus neither based on any authoritative pronouncement by this Court nor does it lead to a situation where the merit is given precedence. Subject to any permissible reservations i.e. either Social (Vertical) or Special (Horizontal), opportunities to public employment and selection of candidates must purely be based on merit.

Any selection which results in candidates getting selected against Open/General category with less merit than the other available candidates will certainly be opposed to principles of equality. There can be special dispensation when it comes to candidates being considered against seats or quota meant for reserved categories and in theory it is possible that a more meritorious candidate coming from Open/General category may not get selected. But the converse can never be true and will be opposed to the very basic principles which have all the while been accepted by this Court. Any view or process of interpretation which will lead to incongruity as highlighted earlier, must be rejected.

32. The second view will thus not only lead to irrational results where more meritorious candidates may possibly get sidelined as indicated above but will, of necessity, result in acceptance of a postulate that Open / General seats are reserved for candidates other than those coming from vertical reservation categories. Such view will be completely opposed to the long line of decisions of this Court.

\end{indquote}

The following quote from \textit{Saurav Yadav (2020)}, is also critical, because it implies that
the \textit{compliance with VR protections} axiom  is enforced in its stronger form with Condition 3:

\begin{indquote}
36. Finally, we must say that the steps indicated by the High Court of Gujarat in para 56 of its judgment in Tamannaben Ashokbhai Desai contemplate the correct and appropriate procedure for considering and giving effect to both vertical and horizontal reservations. The illustration given by us deals with only one possible dimension. There could be multiple such possibilities. Even going by the present illustration, the first female candidate allocated in the vertical column for Scheduled Tribes may have secured higher position than the candidate at Serial No.64. In that event said candidate must be shifted from the category of Scheduled Tribes to Open / General category causing a resultant vacancy in the vertical column of Scheduled Tribes. Such vacancy must then enure to the benefit of the candidate in the Waiting List for Scheduled Tribes -- Female.
\end{indquote}

\section{Examples Omitted from the Main Text} \label{appendix:Chileanexamples}

This section includes examples omitted from the main text.  
Example \ref{ex:inefficient} shows that \textit{Saurav Yadav (2020)} axioms are incompatible with Pareto efficiency even when there is a single underlying merit ranking at each institution. 
Example \ref{ex1} shows that the Chilean school choice mechanism discussed in Section \ref{sec:chile} fails \textit{no justified envy}, whereas Example \ref{ex2}  shows that it fails
\textit{maximal accommodation of HR protections}.

\begin{example} \label{ex:inefficient}
There are two jobs $x$, $y$, with one position each at open category.
The only position at job $x$ is HR-protected for trait $t_1$, and the only position at job $y$ is HR-protected for trait $t_2$.
There are three individuals $a$, $b$, $c$  with traits $\tau(a) = \{t_1\}$ and  $\tau(b) = \tau(c) = \{t_2\}$.
Preferences of the individuals,  and their merit rankings at jobs are given as follows.
\[
	\begin{array}{ccc}
	\succ_{a} & \succ_{b}& \succ_{c} \\ \hline
	y       & x    & x \\
	x        & y    & y\\
	\emptyset & \emptyset & \emptyset
	\end{array}
\qquad \qquad \qquad
	\begin{array}{cc}
	\sigma_{x} \; & \sigma_{y}\\ \hline
	a       &  a    \\
	b        &  b    \\
	c         & c\\
	\end{array}
	\]
Consider the following two assignments:
\[ \alpha = \left(\begin{array}{ccc}
	a & b & c\\
	(x,o)  &  (y,o) & \emptyset
	\end{array}\right)
\qquad \mbox{and} \qquad	
 \beta = \left(\begin{array}{ccc}
	a & b & c\\
	(y,o)  &  (x,o) & \emptyset
	\end{array}\right).
\]
Since (i) all individuals prefer either of the two jobs to remaining unmatched, (ii)
individuals $b$ and $c$ each have trait $t_2$, and (iii) the only position at job $y$ is HR-protected for trait $t_2$,
individual $a$ cannot receive his first choice position at job $y$ under any assignment that satisfies the axiom of maximal accommodation of
HR protections. Therefore, assignment $\alpha$ is the only assignment that satisfies all five axioms, even though
it is Pareto dominated by assignment $\beta$.
\qed
\end{example}

\begin{example} \label{ex1}
There is only one school $s$ with three seats. There are four students $i_1, i_2, i_3, i_4$ who are merit-ranked as follows:
\[ \sigma(i_1) > \sigma(i_2) > \sigma(i_3) > \sigma(i_4) \]

One of the seats is HR-protected for students with the financially disadvantaged trait $t_d$, and one of the seats is HR-protected for students with the high-performing trait $t_h$. 
Student $i_1$ is a regular student with neither of the traits, whereas student $i_2$ has both traits, student $i_3$ has the high-performing trait $t_h$ only, and student $i_4$ has the financially disadvantaged trait $t_d$ only.

Let $s_o$ denote the open seat, $s_d$ denote the reserve seat for financially disadvantaged students,
and $s_h$ denote the reserve seat for high-achieving students.
Hence student $i_2$ receives preferential treatment for both seat $s_d$ and seat $s_h$,
whereas student $i_3$ receives preferential treatment for seat $s_h$ only, and
student $i_4$ receives preferential treatment for seat $s_d$ only.
Hence, student priorities for each seat are given as follows:
\begin{align*}
&  i_1 \; \pi_{s_o} \; i_2 \; \pi_{s_o} \; i_3 \; \pi_{s_o} \; i_4\\
&  i_2 \; \pi_{s_h} \; i_3 \;  \pi_{s_h} \; i_1 \;  \pi_{s_h} \; i_4\\
&  i_2 \; \pi_{s_d} \; i_4 \;  \pi_{s_d} \; i_1 \;  \pi_{s_d} \; i_3
\end{align*}

As for the tie-breaking under the Chilean mechanism,
let us assume HR-protected seat $s_h$ is preferred to
HR-protected seat $s_d$ for any student who either receives preferential treatment  for both HR-protected seats or
for neither of the HR-protected seats.
Since by the Chilean design each student is also assumed to prefer seats at one of her traits to open seats,
this results in the following preferences over seats for the students:
\begin{align*}
& s_o \; \succ_{i_1} \;  s_h \; \succ_{i_1} \; s_d\\
& s_h \; \succ_{i_2} \; s_d \; \succ_{i_2} \; s_o\\
& s_h \; \succ_{i_3} \; s_o \; \succ_{i_3} \; s_d\\
& s_d \; \succ_{i_4} \; s_o \; \succ_{i_4} \; s_h
\end{align*}

Therefore, under the individual-proposing DA algorithm, at Step 1 student $i_1$ applies
to the open seat $s_o$, whereas students $i_2$ and $i_3$ both apply for the HR-protected seat $s_h$,
and student $i_4$ applies to the HR-protected seat $s_d$.
HR-protected seat $s_h$ holds student $i_2$ and rejects student $i_3$, whereas the open seat $s_o$
and the HR-protected seat $s_d$ each hold their only applicant.
At Step 2, student $i_3$ applies to the open seat $s_o$, only to be rejected again since
student $i_1$ who is on hold for the open seat has higher priority for the open seat.
Finally at Step 3, student $i_3$ applies to the reserve seat $s_d$, and gets rejected for a third time since
student $i_4$ who is on hold for the HR-protected seat $s_d$ has
higher priority for the seat $s_d$ having the financially disadvantaged trait.
This results in the following matching
\[ \left( \begin{array}{ccc}
s_o & s_h & s_d \\
i_1 & i_2 & i_4
\end{array} \right)\]
of students to seats, and hence the set of students admitted to school $s$ are
$\{i_1,i_2,i_4\}$.

This outcome fails to satisfy \textit{no justified envy}: Observe that student $i_2$ has the highest priority not only for seat $s_h$ but also for $s_d$. 
Therefore, if they were not artificially assumed to prefer seat $s_h$ over $s_d$, they could instead be assigned seat $s_d$, allowing student $i_3$ to receive seat $s_h$, resulting in the matching:

\[ \left( \begin{array}{ccc}
s_o & s_h & s_d \\
i_1 & i_3 & i_2
\end{array} \right).\]

Thus, compared to the outcome of the Chilean mechanism, it is possible to admit
the third priority student $i_3$ instead of the fourth priority student $i_4$,
while still honoring both HR-protected positions.
This outcome is indeed the result of the meritorious horizontal choice rule,
which is agnostic about which type of HR-protected seat an individual receives when they have multiple traits.
\qed
\end{example}

\begin{example} \label{ex2}

There is only one school $s$ with three seats. There are four students $i_1, i_2, i_3, i_4$ who are merit-ranked as follows:
\[ \sigma(i_1) > \sigma(i_2) > \sigma(i_3) > \sigma(i_4) \]

One of the seats is HR-protected for students with the financially disadvantaged trait $t_d$, and one of the seats is HR-protected for students with the high-performing trait $t_h$.
Students $i_1, i_4$ are both regular students with neither of the traits,
whereas student $i_2$ has both traits, and
student $i_3$ has the financially disadvantaged trait $t_d$ only.

Let $s_o$ denote the open seat, $s_d$ denote the reserve seat for financially disadvantaged students, and $s_h$ denote the reserve seat for high-achieving students. 
Hence, student $i_2$ receives preferential treatment for both seats $s_d$ and $s_h$, whereas student $i_3$ receives preferential treatment for seat $s_d$ only. 
Therefore, student priorities for each seat are given as follows:

\begin{align*}
&  i_1 \; \pi_{s_o} \; i_4 \; \pi_{s_o} \; i_2 \; \pi_{s_o} \; i_3\\
&  i_2 \; \pi_{s_h} \; i_1 \;  \pi_{s_h} \; i_4 \;  \pi_{s_h} \; i_3\\
&  i_2 \; \pi_{s_d} \; i_3 \;  \pi_{s_d} \; i_1 \;  \pi_{s_d} \; i_4.
\end{align*}

As for the tie-breaking under the Chilean school choice mechanism,
let us assume that the HR-protected seat $s_d$ is preferred to
HR-protected seat $s_h$ for any student who either receives preferential treatment  for both HR-protected seats or
for neither of the HR-protected seats.
Since by the Chilean design each student is also assumed to prefer seats at one of her traits to open seats,
this results in the following preferences over seats for the students:

\begin{align*}
& s_o  \; \succ_{i_1} \;  s_d \; \succ_{i_1} \; s_h \\
& s_d \; \succ_{i_2} \; s_h \; \succ_{i_2} \; s_o\\
& s_d \; \succ_{i_3} \; s_o \; \succ_{i_3} \; s_h\\
& s_o \; \succ_{i_4} \; s_d \; \succ_{i_4} \; s_h.
\end{align*}

So under the individual-proposing DA algorithm, at Step 1 students $i_1$ and
$i_4$ both apply to the open seat $s_o$, whereas students $i_2$ and $i_3$ both apply for the
HR-protected seat $s_d$.
Open seat $s_o$ holds student $i_1$ and rejects student $i_4$, whereas the
HR-protected seat $s_d$ holds student $i_2$ and rejects student $i_3$.
At Step 2, student $i_4$ applies to the HR-protected seat $s_d$ and student $i_3$ applies to the open seat $s_o$,
and both students are rejected since these seats are each holding higher-priority students.
Finally at Step 3, both students $i_3$ and $i_4$ apply to the HR-protected seat $s_h$, which holds student $i_4$ and
rejects student $i_3$. Since student $i_3$ is rejected from all seats, the algorithm terminates at the end
of Step 3 finalizing all assignments.
This results in the following matching
\[\mu =  \left( \begin{array}{ccc}
s_o & s_h & s_d \\
i_1 & i_4 & i_2
\end{array} \right)\]
of students to seats, and hence the set of students admitted to school $s$ are
$\{i_1,i_4, i_2\}$.

This outcome fails to satisfy \textit{maximal accommodation of HR protections}: While it is possible to honor  both HR protections by assigning them to their target students  through the matching 
\[\nu = \left( \begin{array}{ccc}
s_o & s_h & s_d \\
i_1 & i_2 & i_3
\end{array} \right),\]
the matching $\mu$ only honors the HR protections for the financially disadvantaged. As a result, it effectively ``converts'' the HR-protected seat for high-achieving students to an open position. 
Hence, the tie-breaking rule forces student $i_2$, who has the flexibility to receive either of the two HR-protected seats, to rigidly accept the HR-protected seat $s_d$. 
This means no one else can benefit from the HR protections for high-achieving. Consequently, only one of the HR-protected seats is honored for affirmative action under the Chilean school choice mechanism, 
even though both could have been honored. In contrast, both HR-protected seats are honored under the meritorious horizontal choice rule, 
which is agnostic about which type of HR-protected seat a student receives when they have multiple traits.
\qed
\end{example}

\section{Additional Terminology and Auxiliary Results}
In this appendix, we provide additional terminology and auxiliary lemmas that we use in the proofs.

\subsection{Additional Terminology}
In this section, we define some terminology to use in the rest of the appendix.

\begin{definition}
A \emph{job matching} $\mu: \cali \to \calj \medcup \{\emptyset\}$ is a function such that, for every $j\in \calj$, $|\mu^{-1}(j)|\leq q_j$.
\end{definition}
If $\mu(i)=\emptyset$ for some individual $i\in \cali$, then the individual is unmatched.

\begin{definition}
A \emph{choice rule} is a function $C: 2^{\cali} \rightarrow 2^{\cali}$ such that,  for any $I\subseteq \cali$,
\[ C(I) \subseteq I.\]
\end{definition}
Note that, any single-category choice rule (introduced in Definition \ref{def:single-category}) is a choice rule.
Similarly, any aggregate choice rule  (introduced in Definition \ref{def:aggregate}) is also a choice rule.

\begin{definition}
A job matching $\mu$ is \emph{stable} with respect to a
profile of choice rules $(C_j)_{j\in \calj}$ if the following three conditions hold:
    \begin{enumerate}
        \item \textit{Individual rationality\/}: For each $i\in \cali$, \;  $\mu(i) \succeq_i \emptyset$,
        \item \textit{Job rationality\/}: For each  $j\in \calj$,  \; $C_j(\mu^{-1}(j))=\mu^{-1}(j)$.
        \item \textit{No blocking pairs\/}: There exist no $i\in \cali$ and $j\in \calj$
        such that $j \succ_i \mu(i)$ and $i \in C_j(\mu^{-1}(j) \medcup \{i\})$.
    \end{enumerate}
\end{definition}

Given an assignment $\alpha\in\cala$,  the \textit{job matching $\mu$ induced by assignment $\alpha$\/} is constructed as follows:
For each $i\in \cali$,
  \[
    \mu(i) = \begin{cases}
        j, & \text{if } \alpha(i)=(j,v) \text{ for some } (j,v) \in \calj\times\calv,\\
        \emptyset, & \text{if } \alpha(i)=\emptyset.\\
        \end{cases}
  \]

It is easy to check that $\mu$ is a job matching given that $\alpha$ is an  assignment.

Likewise, for an assignment mechanism, there is an induced job matching mechanism where, for every preference profile of
individuals, the outcome of the matching mechanism is the matching induced by the outcome of the assignment mechanism for the preference profile.

Given the profile of aggregate choice rules $(\widehat{C}_{\circled{M},j}^{2s})_{j\in\calj}$,
consider the individual-proposing deferred acceptance mechanism:
At each step of the mechanism, if $I$ is the set of individuals considered for job $j$,  job $j$ tentatively accepts $\widehat{C}_{\circled{M},j}^{2s}(I)$ without specifying any category and permanently rejects $I\setminus \widehat{C}_{\circled{M},j}^{2s}(I)$. Call this  job matching mechanism the \emph{aggregate
meritorious deferred-acceptance mechanism (AM-DA)}, and denote it by $\widehat{\varphi}^{2s}_{\circled{M}}$.
For any preference profile $\succ_{\cali}\in\calp$,
the outcome $\widehat{\varphi}^{2s}_{\circled{M}}(\succ_{\cali})$
of AM-DA is a job matching.

Strategy-proofness of a job-matching mechanism is defined analogously as
the strategy-proofness of an assignment mechanism.

\subsection{Choice Rule Properties}
In this section, we define choice rule properties and establish some
lemmas that we use in our proofs.

\begin{definition}\citep{kelso82}
A choice rule $C$ satisfies the \emph{substitutes} condition, if, for each $I\subseteq \cali$ and $i,i'\in I$,
\[i\in C(I) \text{ and } i'\neq i \; \implies \;  i\in C(I\setminus \{i'\}).\]
\end{definition}

\begin{definition}\citep{aygson12a}
A choice rule $C$ satisfies the \emph{irrelevance of rejected individuals} condition, if, for each $I\subseteq \cali$,
\[i\in I \text{ and } i\notin C(I) \; \implies \;  C(I\setminus \{i\})=C(I).\]
\end{definition}

For any job $j\in\calj$ and category $v\in\calv$,
we next show that the meritorious horizontal choice rule $C_{\circled{M},j}^{v}$
satisfies the substitutes condition and the irrelevance of rejected individuals condition.

\begin{lemma}\label{lem:MH-substitutes}
For each $j\in \calj$ and $v\in \calv$, the single-category choice rule $C_{\circled{M},j}^{v}$
satisfies the substitutes condition.
\end{lemma}

\begin{proof}
Fix a job $j\in\calj$ and a category $v\in\calv$. Let
 $C_{\circled{M},j}^{v,1}(I)$ be the set of individuals who are selected in Step 1, and  $C_{\circled{M},j}^{v,2}(I)$ be the set of individuals who are selected in Step 2
of the  choice rule  $C_{\circled{M},j}^{v}$. Hence, for any  $I\subseteq\cali$,
\[ C_{\circled{M},j}^{v}(I) = C_{\circled{M},j}^{v,1}(I) \medcup C_{\circled{M},j}^{v,2}(I).
\]

Choice rule $C_{\circled{M},j}^{v,1}$ satisfies the substitutes condition: \cite{fleiner2001}
shows that the greedy rule defined on a matroid satisfies the substitutes condition. In \cite{sonmez/yenmez:22},
we make the observation that $C_{\circled{M},j}^{v,1}(I)$ is equivalent to the greedy rule for the
transversal matroid on the HR graph of job $j$ and category $v$ with rank function $n_j^v$.

Let $I\subseteq \cali$, $i\in C_{\circled{M},j}^{v}(I)$, and $i'\in I\setminus\{i\}$. For substitutability
of $C_{\circled{M},j}^{v}$, we need to show that $i\in C_{\circled{M},j}^{v}(I\setminus\{i'\})$. We consider 
three cases.

\noindent
\textbf{Case 1:} If $i\in C_{\circled{M},j}^{v,1}(I)$, then $i\in C_{\circled{M},j}^{v,1}(I\setminus\{i'\})$ by substitutability
of $C_{\circled{M},j}^{v,1}$, which implies $i\in C_{\circled{M},j}^{v}(I\setminus\{i'\})$ as desired.

\noindent
\textbf{Case 2:} If  $i\in C_{\circled{M},j}^{v,2}(I)$ and $i\in C_{\circled{M},j}^{v,1}(I\setminus\{i'\})$, then we also have $i\in C_{\circled{M},j}^{v}(I\setminus\{i'\})$ as desired.

\noindent
\textbf{Case 3:} Finally, let  $i\in C_{\circled{M},j}^{v,2}(I)$ and $i\not\in C_{\circled{M},j}^{v,1}(I\setminus\{i'\})$.
By substitutability of $C_{\circled{M},j}^{v,1}$,
\[I \setminus C_{\circled{M},j}^{v,1}(I) \supseteq  \left(I\setminus\{i'\}\right) \setminus C_{\circled{M},j}^{v,1}(I\setminus\{i'\}).
\]
Furthermore, since $n^v_j$ is monotone, 
\[n_j^v(I)\geq n_j^v(I\setminus \{i'\}),\]
which is equivalent to 
\[\left|C_{\circled{M},j}^{v,1}(I)\right| \geq \left|C_{\circled{M},j}^{v,1}(I\setminus\{i'\})\right|.\]
Therefore, because $i\in C_{\circled{M},j}^{v,2}(I)$, individual $i$ is one of the $\left(q^v_j - \left|C_{\circled{M},j}^{v,1}(I) \right|\right)$ highest merit ranking individuals in $I\setminus C_{\circled{M},j}^{v,1}(I)$
under $\sigma_j$. Then individual $i$ also has to be one of the 
$\left(q^v_j -  \left|C_{\circled{M},j}^{v,1}(I\setminus\{i'\})\right|\right)$ highest merit ranking individuals in
$\left(I\setminus\{i'\}\right) \setminus \left(C_{\circled{M},j}^{v,1}(I\setminus\{i'\})\right)$ under $\sigma_j$ 
because $I \setminus C_{\circled{M},j}^{v,1}(I) \supseteq  \left(I\setminus\{i'\}\right) \setminus C_{\circled{M},j}^{v,1}(I\setminus\{i'\})$ and 
$\left(q^v_j -  \left|C_{\circled{M},j}^{v,1}(I\setminus\{i'\})\right|\right) \geq 
\left(q^v_j - \left|C_{\circled{M},j}^{v,1}(I) \right|\right)$. We conclude
$i\in C_{\circled{M},j}^{v,2}(I\setminus\{i'\})$, which in turn  implies
$i\in C_{\circled{M},j}^{v}(I\setminus\{i'\})$ as desired.

This establishes the desired relation for all three cases, and completes the proof.
\end{proof}

\begin{lemma}\label{lem:MH-iri}
For each $j\in \calj$ and $v\in \calv$, the single-category choice rule $C_{\circled{M},j}^{v}$
satisfies the irrelevance of rejected individuals condition.
\end{lemma}

\begin{proof}
Fix a job $j\in \calj$, category $v\in \calv$, and $I \subseteq \cali$. Let $i\in I$ be such that $i\not\in  C_{\circled{M},j}^{v}(I)$.
Since  $i\notin C_{\circled{M},j}^{v,1}(I)$ implies $n^v_j(I)=n^v_j(I\setminus \{i\})$, the same individual will be selected in each sub-step of
Step 1 of the  choice rule $C_{\circled{M},j}^{v}$ for both sets of individuals $I$ and $I\setminus\{i\}$, and therefore we have
$C_{\circled{M},j}^{v,1}(I\setminus\{i\}) = C_{\circled{M},j}^{v,1}(I)$.
Moreover, an individual $i'$ is one of the $\left(q^v_j - \left|C_{\circled{M},j}^{v,1}(I) \right|\right)$ highest merit ranking individuals in 
$I\setminus C_{\circled{M},j}^{v,1}(I)$ if and only if he is one of the $\left(q^v_j - \left|C_{\circled{M},j}^{v,1}(I\setminus\{i\}) \right|\right) =
\left(q^v_j - \left|C_{\circled{M},j}^{v,1}(I) \right|\right)$ highest
merit ranking individuals in $\left(I\setminus\{i\}\right)\setminus C_{\circled{M},j}^{v,1}(I\setminus\{i\})$. Therefore, we also have
 $C_{\circled{M},j}^{v,2}(I\setminus\{i\}) = C_{\circled{M},j}^{v,2}(I)$. Hence, we have $C_{\circled{M},j}^{v}(I\setminus\{i\}) = C_{\circled{M},j}^{v}(I)$,
 establishing that the choice rule $C_{\circled{M},j}^{v}$ satisfies the irrelevance of rejected individuals condition.
\end{proof}

\begin{definition}
A choice rule $C$ is \emph{path independent} if, for each $I,I'\subseteq \cali$,
\[C(I\medcup I')=C\Big(C(I)\medcup C(I')\Big).\]
\end{definition}

\begin{lemma}[\citet{aizmal81}]\label{lem:pi}
A choice rule satisfies path independence if, and only, if it satisfies both the substitutes condition and the irrelevance of rejected individuals condition.
\end{lemma}

\begin{definition}
A choice rule $C$ satisfies the law of aggregate demand if, for every $I,I'\subseteq \cali$
\[I' \supseteq I \; \implies \; |C(I')|\geq |C(I)|.\]
\end{definition}

\begin{lemma}\label{lem:lad}
For each $j\in \calj$ and $v\in \calv$,  the single-category choice rule $C_{\circled{M},j}^{v}$ satisfies the law of aggregate demand.
\end{lemma}

\begin{proof}
By construction, for any $I \subseteq \cali$, we have
$|C_{\circled{M},j}^{v}(I)|=\min\{r_j^v,|I\medcap \cali^v|\}$.
Fix a set of individuals $I\subseteq \cali$ and let $I'\subseteq I$.
Then, we have $\min\{r_j^v,|I'\medcap \cali^v|\} \leq \min\{r_j^v,|I\medcap \cali^v|\}$, or equivalently
$|C_{\circled{M},j}^{v}(I')|\leq |C_{\circled{M},j}^{v}(I)|$. Therefore, $C_{\circled{M},j}^{v}$ satisfies
the law of aggregate demand.
\end{proof}

\subsection{Properties of 2-Step Meritorious Horizontal Choice Rule}
Fix a job $j\in \calj$.
In this section, we establish some properties of the 2-step meritorious horizontal  choice rule
$\vec{C}^{2s}_{\circled{M},j} = (C^{2s,v}_{\circled{M},j})_{v \in \calv}$ that will be instrumental to prove our main results in Appendix \ref{App:MainProofs}.

\begin{lemma}\label{lem:ircmulti}
Let $I\subseteq \cali$ and $i\in I$. If $i\notin \widehat{C}^{2s}_{\circled{M},j}(I)$, then, for every $v\in \calv$,
\[ C^{2s,v}_{\circled{M},j}(I\setminus \{i\})=C^{2s,v}_{\circled{M},j}(I). \]
\end{lemma}

\begin{proof}
First, we establish the desired relation for the open category. Since $C^{2s,o}_{\circled{M},j}=C^{o}_{\circled{M},j}$ and $C^{o}_{\circled{M},j}$
satisfies the irrelevance of rejected individuals condition by Lemma \ref{lem:MH-iri},
we have $C^{2s,o}_{\circled{M},j}(I\setminus \{i\})=C^{2s,o}_{\circled{M},j}(I)$.

Next we establish the desired relation for  any VR-protected category in $\calr$.
Let $c\in \calr$. Then,
\begin{flalign*}
C^{2s,c}_{\circled{M},j}(I) & = C^{c}_{\circled{M},j}\bigg(\Big(I\setminus C^{o}_{\circled{M},j}(I)\Big) \medcap \cali^c\bigg) \\
     &= C^{c}_{\circled{M},j}\bigg(\Big((I\setminus\{i\}) \setminus C^{o}_{\circled{M},j}(I)\Big) \medcap \cali^c\bigg) \\
     &= C^{c}_{\circled{M},j}\bigg(\Big((I\setminus\{i\}) \setminus C^{o}_{\circled{M},j}(I\setminus \{i\})\Big) \medcap \cali^c\bigg) \\
     &= C^{2s,c}_{\circled{M},j}(I\setminus \{i\}),
\end{flalign*}
where the first equation holds by definition of $C^{2s,c}_{\circled{M},j}$, the second equation
holds because $i\notin C^{2s,c}_{\circled{M},j}(I)$ and $C^{c}_{\circled{M},j}$
satisfies the irrelevance of rejected individuals condition by Lemma \ref{lem:MH-iri},
the third equation holds because $i\notin C^{2s,o}_{\circled{M},j}(I)=C^{o}_{\circled{M},j}(I)$ and $C^{o}_{\circled{M},j}$
satisfies the irrelevance of rejected individuals condition by Lemma \ref{lem:MH-iri}, and the last equation holds by definition of
$C^{2s,c}_{\circled{M},j}$.
\end{proof}

The following result is a direct implication of Lemma \ref{lem:ircmulti}.

\begin{corollary}\label{cor:iri}
The aggregate choice rule $\widehat{C}^{2s}_{\circled{M},j}$ satisfies the irrelevance of rejected individuals condition.
\end{corollary}

\begin{lemma}\label{lem:cenvlad}
The aggregate choice rule $\widehat{C}^{2s}_{\circled{M},j}$ satisfies the law of aggregate demand.
\end{lemma}

\begin{proof}
Let $I',I\subseteq \cali$ be such that $I' \supseteq I$. Since $C^{o}_{\circled{M},j}$ satisfies the law of aggregate demand by Lemma \ref{lem:lad},
\[\left|C^{2s,o}_{\circled{M},j}(I')\right|  =  \left|C^{o}_{\circled{M},j}(I') \right|  \geq \left| C^{o}_{\circled{M},j}(I) \right| = \left|C^{2s,o}_{\circled{M},j}(I)\right|.
\]
Furthermore, because $C^{o}_{\circled{M},j}$ satisfies the substitutes condition by Lemma \ref{lem:MH-substitutes},
we have
\[\left(I\setminus C^{o}_{\circled{M},j}(I)\right)  \subseteq \left(I'\setminus C^{o}_{\circled{M},j}(I')\right).\]
Consequently, for each $c \in \calr$,
\[ \left|C^{2s,c}_{\circled{M},j}(I')\right| =  \left|C^{c}_{\circled{M},j}\Big(\big(I'\setminus C^{o}_{\circled{M}}(I')\big)\medcap{\cali}^c\Big)\right|
\geq  \left|C^{c}_{\circled{M},j}\Big(\big(I\setminus C^{o}_{\circled{M}}(I)\big)\medcap{\cali}^c\Big)\right|
= \left|C^{2s,c}_{\circled{M},j}(I)\right|,\]
because $C^{c}_{\circled{M},j}$ satisfies the law of aggregate demand (Lemma \ref{lem:lad}).
We conclude that
\[ \left|\widehat{C}^{2s}_{\circled{M},j}(I')\right| = \sum_{v\in \calv}\left|C^{2s,v}_{\circled{M},j}(I')\right| \geq
\sum_{v\in \calv}\left|C^{2s,v}_{\circled{M},j}(I)\right| = \left|\widehat{C}^{2s}_{\circled{M},j}(I)\right|.\]
Therefore, $\widehat{C}^{2s}_{\circled{M},j}$ satisfies the law of aggregate demand.
\end{proof}

\begin{lemma}\label{lem:cenvsub}
The aggregate choice rule $\widehat{C}^{2s}_{\circled{M},j}$ satisfies the substitutes condition.
\end{lemma}

\begin{proof}
Let $I\subseteq \cali$, $i,i'\in I$, $i\neq i'$, and $i\in \widehat{C}^{2s}_{\circled{M},j}(I)$.
Since $i\in \widehat{C}^{2s}_{\circled{M},j}(I)$, then either $i\in C^{2s,o}_{\circled{M},j}(I)$ or $i\in C^{2s,c}_{\circled{M},j}(I)$
for some $c\in \calr$. If $i\in C^{2s,o}_{\circled{M},j}(I)$, then we have $i\in C^{2s,o}_{\circled{M},j}(I\setminus \{i'\})$,
because $C^{2s,o}_{\circled{M},j}=C^{o}_{\circled{M},j}$ satisfies the substitutes condition by
Lemma \ref{lem:MH-substitutes}. If $i\in C^{2s,c}_{\circled{M},j}(I)$ for some $c\in \calr$,
then either (1) $i\in C^{2s,o}_{\circled{M},j}(I\setminus \{i'\})$ or
(2) $i\in (I\setminus \{i'\}) \setminus C^{2s,o}_{\circled{M},j}(I\setminus \{i'\})$ which implies that
$i\in C^{2s,c}_{\circled{M},j}(I\setminus \{i'\})$ because
\begin{enumerate}
\item $C^{c}_{\circled{M},j}$ satisfies the substitutes condition,
\item $I \setminus C^{2s,o}_{\circled{M},j}(I) \supseteq (I\setminus \{i'\}) \setminus C^{2s,o}_{\circled{M},j}(I\setminus \{i'\})$, and
\item $i\in C^{2s,c}_{\circled{M},j}(I)$.
\end{enumerate}
Therefore, $i\in \widehat{C}^{2s}_{\circled{M},j}(I\setminus \{i'\})$, and hence $\widehat{C}^{2s}_{\circled{M},j}$ satisfies the substitutes condition.
\end{proof}

\begin{lemma}\label{lem:aggregatepi}
The aggregate choice rule $\widehat{C}_{\circled{M},j}^{2s}$ is path independent.
\end{lemma}

\begin{proof}
The proof directly follows from  Corollary \ref{cor:iri}, Lemma \ref{lem:pi}, and Lemma \ref{lem:cenvsub}.
\end{proof}

\begin{definition}\label{def:resnonwaste}
Let $I\subseteq \cali$. 
A multi-category choice rule
$\vec{C}_j=(C_j^{v})_{v \in \calv}$ satisfies \emph{non-wastefulness for I\/} if,
for  each $v \in \calv$ and $i \in I$,
\[i \not\in  \widehat{C}_j(I) \; \mbox{ and } \;  |C^v_j(I)| < r^v_j   \quad \implies \quad    i \not\in \cali^v.\]
\end{definition}

\begin{definition}\label{def:reshor}
Let $I\subseteq \cali$. 
A multi-category choice rule $\vec{C}_j=(C_j^{v})_{v \in \calv}$ satisfies
\emph{maximal accommodation of HR protections for I}, if for each
$v \in \calv$,\; and
$i \in (I\medcap \cali^v) \setminus \widehat{C}_j(I)$,
\[ n^v_j\Big(C_j^v(I)\Big) =  n_j^v\Big(C_j^v(I) \medcup \{i\}\Big).
\]
\end{definition}

\begin{definition}\label{def:reseli}
Let $I\subseteq \cali$. 
A multi-category choice rule $\vec{C}_j=(C^v_j)_{v \in \calv}$ satisfies \emph{no justified envy for I\/} if, for each
$v \in \calv$,\; $i\in C^v_j(I)$, and
$i' \in \big(I\medcap \cali^v\big) \setminus \widehat{C}_j(I)$, 
\[ \sigma_j(i') \, > \, \sigma_j(i)    \; \implies \;  n_j^v\bigg(\Big(C_j^v(I)\setminus \{i\}\Big) \medcup \{i'\}\bigg) < n_j^v\Big(C_j^v(I)\Big).
\]
\end{definition}

\begin{definition}\label{def:resvert}
Let $I\subseteq \cali$. 
A multi-category choice rule $\vec{C}_j=(C^v_j)_{v \in \calv}$ satisfies
\emph{compliance with VR protections for I} if,
for  every
$c \in \calr$ and $i\in C_j^c(I)$,
\begin{enumerate}
\item $|C_j^o(I)| = r^o_j$,
\item for every $i' \in C^o_j(I)$,
\[ \sigma_j(i') < \sigma_j(i) \quad \implies \quad n^o_j\Big(C^o(I)\Big) > n^o_j\bigg(\Big(C^o(I)\setminus\{i'\}\Big) \medcup \{i\}\bigg), \mbox{ and} \]
\item $n^o_j\Big(C_j^o(I) \medcup \{i\}\Big) = n_j^o\Big(C^o(I)\Big)$.
\end{enumerate}
\end{definition}

\begin{lemma}[\citealp{sonmez/yenmez:22}]\label{lem:reschar}
Let $I\subseteq \cali$. 
A multi-category choice rule $\vec{C}_j$ satisfies
(i) non-wastefulness for $I$, (ii) maximal accommodation of HR protections for $I$,
(iii) no justified envy for $I$,  and
(iv) compliance with VR protections for $I$ 
if and only if $\vec{C}_j(I)=\vec{C}_{\circled{M},j}^{2s}(I)$.
\end{lemma}

Lemma \ref{lem:reschar} is originally given as Theorem 3 in \cite{sonmez/yenmez:22}.

\section{Proofs of Theorems \ref{thm:resrulechar}-\ref{thm:jointchar}} \label{App:MainProofs}

We first prove Theorem \ref{thm:resrulechar}, followed by 
Theorem \ref{thm:jointchar} and Theorem \ref{thm:dominate}.

\subsection{Proof of Theorem \ref{thm:resrulechar}} 

\mbox{}\\
\noindent \textbf{Only if direction:} Let $\succ_{\cali}\in\calp$ be the profile of individual preferences and
$\alpha\in\cala$ be an assignment that is stable with respect to $(\vec{C}_{\circled{M},j}^{2s})_{j\in \calj}$. 

\textit{(i) Individual rationality}: By stability, $\alpha$ is individually rational. 

\textit{(ii) Non-wastefulness}: Consider any $j\in \calj$, $v\in \calv$, and $i\in \cali$ such that $j \succ_i \alpha(i)$ and $\big|\alpha^{-1}(j,v)\big|<r_j^v$. Since $\alpha$ is stable there are no blocking pairs for it and, hence, 
$i \notin \widehat{C}^{2s}_{\circled{M},j}(\alpha^{-1}(j) \cup \{i\})$.   
By Lemma \ref{lem:ircmulti}, $C^{2s,v}_{\circled{M},j}(\alpha^{-1}(j) \cup \{i\})=C^{2s,v}_{\circled{M},j}(\alpha^{-1}(j))$. 
Furthermore, by stability of $\alpha$, it also satisfies job-category rationality:  $C^{2s,v}_{\circled{M},j}(\alpha^{-1}(j))=\alpha^{-1}(j,v)$. 
Therefore, $\abs{C^{2s,v}_{\circled{M},j}(\alpha^{-1}(j) \cup \{i\})}=
\abs{C^{2s,v}_{\circled{M},j}(\alpha^{-1}(j))}=\abs{\alpha^{-1}(j,v)}<r_j^v$. Since $\vec{C}_{\circled{M},j}^{2s}$ satisfies  non-wastefulness for $\alpha^{-1}(j) \cup \{i\}$ (Lemma \ref{lem:reschar}), 
$i \notin \widehat{C}^{2s}_{\circled{M},j}(\alpha^{-1}(j) \cup \{i\})$  and 
$\abs{\widehat{C}^{2s}_{\circled{M},j}(\alpha^{-1}(j) \cup \{i\})} <r_j^v$, we get that $i\notin \cali^v$. Therefore, 
$\alpha$ satisfies non-wastefulness.

\textit{(iii) Maximal accommodations of HR protections}: Consider any $j\in \calj$, $v\in \calv$, and $i\in \cali^v$ such that $j\succ_i \alpha(i)$. Since $\alpha$ is stable there are no blocking pairs for it and, hence, 
$i \notin \widehat{C}^{2s}_{\circled{M},j}(\alpha^{-1}(j) \cup \{i\})$. 
By Lemma \ref{lem:ircmulti}, $C^{2s,v}_{\circled{M},j}(\alpha^{-1}(j) \cup \{i\})=C^{2s,v}_{\circled{M},j}(\alpha^{-1}(j))$. 
Since $\alpha$ is stable it satisfies job-category-rationality and, therefore,  
$C^{2s,v}_{\circled{M},j}(\alpha^{-1}(j))=\alpha^{-1}(j,v)$ and, so we get 
$C^{2s,v}_{\circled{M},j}(\alpha^{-1}(j) \cup \{i\})=\alpha^{-1}(j,v)$.
Since $(\vec{C}_{\circled{M},j}^{2s})_{j\in \calj}$ satisfies 
maximal accommodations of HR protections for $\alpha^{-1}(j)\cup \{i\}$ and 
$i\in \left( \left( \alpha^{-1}(j)\cup \{i\} \right) \cap \cali^v \right) \setminus \widehat{C}^{2s}_{\circled{M},j}(\alpha^{-1}(j) \cup \{i\})$, we get 
\[n^v_j\left(C^{2s}_{\circled{M},j}(\alpha^{-1}(j)\cup \{i\})\right) = n^v_j\left(C^{2s}_{\circled{M},j}(\alpha^{-1}(j)\cup \{i\}) \cup \{i\}\right).\] 
Since $C^{2s,v}_{\circled{M},j}(\alpha^{-1}(j) \cup \{i\})=\alpha^{-1}(j,v)$, the displayed equation can be written as 
\[n^v_j(\alpha^{-1}(j,v))= n^v_j(\alpha^{-1}(j,v) \cup \{i\}).\]
Therefore, $n_j^v\Big(\alpha^{-1}(j,v) \medcup \{i\}\Big) \not > n_j^v\Big(\alpha^{-1}(j,v)\Big)$, which implies that 
$\alpha$ satisfies maximal accommodations of HR protections.

\textit{(iv) No justified envy}: Suppose, for contradiction, that there exist $i\in \cali$, $j\in \calj$, $v\in\calv$, and $i'\in\cali^v$ such that $\alpha(i)=(j,v)$, $j \succ_{i'} \alpha(i')$, and $\sigma_j(i')>\sigma_j(i)$. 
Since $\alpha$ is stable there are no blocking pairs for it and, hence, 
$i' \notin \widehat{C}^{2s}_{\circled{M},j}(\alpha^{-1}(j) \cup \{i'\})$. By Lemma \ref{lem:ircmulti}, 
$C^{2s,v}_{\circled{M},j}(\alpha^{-1}(j) \cup \{i'\})=C^{2s,v}_{\circled{M},j}(\alpha^{-1}(j))$. Furthermore, since 
$\alpha$ is stable it satisfies job-and-category rationality: 
$C^{2s,v}_{\circled{M},j}(\alpha^{-1}(j))=\alpha^{-1}(j,v)$. Therefore, 
$C^{2s,v}_{\circled{M},j}(\alpha^{-1}(j) \cup \{i'\})=\alpha^{-1}(j,v)$. 
Since $\vec{C}_{\circled{M},j}^{2s}$ satisfies no justified envy for $\alpha^{-1}(j) \cup \{i'\}$, 
$i\in C^{2s,v}_{\circled{M},j}(\alpha^{-1}(j) \cup \{i'\})$, $i'\notin C^{2s,v}_{\circled{M},j}(\alpha^{-1}(j) \cup \{i'\})$, 
and $\sigma_j(i')>\sigma_j(i)$, we get 
\[n_j^v\bigg(\Big(C^{2s,v}_{\circled{M},j}(\alpha^{-1}(j) \cup \{i'\})\setminus \{i\}\Big) \medcup \{i'\}\bigg) < n_j^v\Big(C^{2s,v}_{\circled{M},j}(\alpha^{-1}(j) \cup \{i'\})\Big).
\]
The equation $C^{2s,v}_{\circled{M},j}(\alpha^{-1}(j) \cup \{i'\})=\alpha^{-1}(j,v)$ simplifies the above as:
\[n_j^v\bigg(\Big(\alpha^{-1}(j,v) \setminus \{i\}\Big) \medcup \{i'\}\bigg) < 
n_j^v\Big(\alpha^{-1}(j,v) \Big).
\]
Therefore, $\alpha$ satisfies no justified envy.

\textit{(v) Compliance with VR protections}: Let $j\in \calj$, $c\in\calr$, and $i\in\cali^c$ be such that $\alpha(i)=(j,c)$. By job-category rationality, $\vec{C}^{2s}_{\circled{M},j}(\alpha^{-1}(j))=(\alpha^{-1}(j,v))_{v\in \calv}$. Since $\vec{C}_{\circled{M},j}^{2s}$ satisfies compliance with VR protections for $\alpha^{-1}(j)$, we get the following three properties.

The first property is $\abs{C^{2s,o}_{\circled{M},j}(\alpha^{-1}(j))}=r^o_j$, which implies $\abs{\alpha^{-1}(j,o)}=r^o_j$ because 
$C^{2s,o}_{\circled{M},j}(\alpha^{-1}(j))=\alpha^{-1}(j,o)$. This is item (1) in Definition \ref{def:VR}.

The second property is that, for every $i'\in C^{2s,o}_{\circled{M},j}(\alpha^{-1}(j)) = \alpha^{-1}(j,o)$, we have 
\[ \sigma_j(i') < \sigma_j(i) \quad \implies \quad n^o_j\Big(C^{2s,o}_{\circled{M},j}(\alpha^{-1}(j))\Big) > n^o_j\bigg(\Big(C^{2s,o}_{\circled{M},j}(\alpha^{-1}(j)) \setminus\{i'\}\Big) \medcup \{i\}\bigg),\]
which is equivalent to
\[ \sigma_j(i') < \sigma_j(i) \quad \implies \quad n^o_j\left(\alpha^{-1}(j,o)\right) > n^o_j\bigg(\Big(\alpha^{-1}(j,o) \setminus\{i'\}\Big) \medcup \{i\}\bigg).\]
Therefore, we get that, for every $i'\in C^{2s,o}_{\circled{M},j}(\alpha^{-1}(j)) = \alpha^{-1}(j,o)$, 
either $\sigma_j(i') > \sigma_j(i)$ or 
$n^o_j\left(\alpha^{-1}(j,o)\right) > n^o_j\bigg(\Big(\alpha^{-1}(j,o) \setminus\{i'\}\Big) \medcup \{i\}\bigg)$. 
This is item (2) in Definition \ref{def:VR}.

The third property is that 
$n^o_j\Big(C^{2s,o}_{\circled{M},j}(\alpha^{-1}(j)) \medcup \{i\}\Big) = n_j^o\Big(C^{2s,o}_{\circled{M},j}(\alpha^{-1}(j))\Big)$. 
Since $C^{2s,o}_{\circled{M},j}(\alpha^{-1}(j)) = \alpha^{-1}(j,o)$, this equation can be written as 
$n^o_j\Big(\alpha^{-1}(j,o) \medcup \{i\}\Big) = n_j^o\Big(\alpha^{-1}(j,o)\Big)$. Therefore, 
$n_j^o\Big(\alpha^{-1}(j,o)\medcup \{i\}\Big) \ngtr n_j^o\Big(\alpha^{-1}(j,o)\Big)$.
This is item (3) in Definition \ref{def:VR}.

We conclude that $\alpha$ satisfies compliance with VR protections. 

\medskip
\noindent
\textbf{If direction:}
Let $\succ_{\cali}\in\calp$ and $\alpha\in\cala$ be an assignment that satisfies the axioms in the statement of the lemma. 
Then $\alpha$ satisfies individual rationality by assumption. To establish stability with respect to $(\vec{C}_{\circled{M},j}^{2s})_{j\in \calj}$, 
we need to show job-and-category rationality and the absence of blocking pairs when each job $j\in \calj$ is endowed with the multi-category choice rule $\vec{C}_{\circled{M},j}^{2s}$.

For each $j\in\calj$,  define
\[\tilde I_j=\{\tilde i\in\cali  \;:\;  j \succeq_{\tilde i} \alpha(\tilde i)\}.\]
Since $\alpha$ is individually rational, for every $\tilde i \in \tilde I_j$, job $j$
is acceptable to individual $\tilde i$.

\begin{claim}\label{claim:I-tilde}
For each $j\in\calj$ and $v\in\calv$,
\[ C_{\circled{M},j}^{2s,v}(\tilde I_j) = \alpha^{-1}(j,v),\]
and, for each $j\in\calj$,
\[ \widehat{C}_{\circled{M},j}^{2s}(\tilde I_j) = \alpha^{-1}(j).\]
\end{claim}

\begin{proof}\renewcommand{\qedsymbol}{$\blacksquare$}
Given a job $j\in\calj$ and category $v\in \calv$, construct a single-category choice rule $C_j^v$ as follows.
For each $I \subseteq \cali$,
\[
     C_j^v(I) = \begin{cases}
        C_{\circled{M},j}^{2s,v}(I), & \text{if } I \neq \tilde I_j\\
        \alpha^{-1}(j,v), & \text{if } I = \tilde I_j.\\
        \end{cases}
  \]
Define $\vec{C}_j=(C^v_j)_{v\in \calv}$.
Since $\alpha$ is an assignment and $\vec{C}_{\circled{M},j}^{2s}$ is a
multi-category choice rule, $\vec{C}_j=(C^v_j)_{v\in \calv}$ is also a multi-category choice rule.

We next show that, for each job $j \in \calj$, $\vec{C}_j$ satisfies
non-wastefulness for $\tilde I_j$, maximal accommodation of HR protections for $\tilde I_j$,
no justified envy for $\tilde I_j$, and compliance with VR protections for $\tilde I_j$.

\textit{Non-wastefulness for $\tilde I_j$\/}:
Let $v\in \calv$, $i\in \tilde I_j \setminus \widehat{C}_j(\tilde I_j)$, and $|C_j^v(\tilde I_j)|<r_j^v$.
By construction, we have  $j \succ_{i} \alpha(i)$ and $|\alpha^{-1}(j,v)|<r_j^v$.
Therefore, since assignment $\alpha$ satisfies non-wastefulness, we must have $i\notin \cali^v$.
Hence, $\vec{C}_j$ satisfies non-wastefulness for $\tilde I_j$.

\textit{Maximal accommodation of HR protections for $\tilde I_j$\/}:
Let $v\in \calv$ and $i\in (\tilde I_j \medcap \cali^v)\setminus \widehat{C}_j(\tilde I_j)$. By construction, we have $j\succ_i \alpha(i)$.
Since $\alpha$ satisfies maximal accommodation of HR protections and function $n_j^v$ is monotone, we have
\[n_j^v\Big(\alpha^{-1}(j,v)\Big)=n_j^v\Big(\alpha^{-1}(j,v) \medcup \{i\}\Big),\]
or equivalently
\[n_j^v\Big(C_j^v(\tilde I_j)\Big)=n_j^v\Big(C_j^v(\tilde I_j) \medcup \{i\}\Big).\]
Therefore, $\vec{C}_j$ satisfies maximal accommodation of HR protections for $\tilde I_j$.

\textit{No justified envy for $\tilde I_j$\/}:
Let $v\in \calv$, $i\in C^v_j(\tilde I_j)=\alpha^{-1}(j,v)$, and
$i'\in (\tilde I_j\medcap \cali^v) \setminus \widehat{C}_j(\tilde I_j)$.
By construction, we have $j\succ_{i'} \alpha(i')$.
Since $\alpha$ satisfies  no justified envy, we have
\[\sigma_j(i')>\sigma_j(i) \; \implies \; n_j^v\left(\alpha^{-1}(j,v)\right) > n_j^v\left((\alpha^{-1}(j,v)\setminus \{i'\})\medcup \{i\}\right),\]
or equivalently
\[\sigma_j(i')>\sigma_j(i) \; \implies \; n_j^v\left(C_j^v(\tilde I_j)\right) > n_j^v\left((C_j^v(\tilde I_j)\setminus \{i'\})\medcup \{i\}\right).\]
Therefore, $\vec{C}_j$ satisfies no justified envy for $\tilde I_j$.

\textit{Compliance with VR protections for $\tilde I_j$\/}:
Let $c\in \calr$ and $i\in C_j^c(\tilde I_j)$. By construction, $i\in \alpha^{-1}(j,c)$.
Since $\alpha$ satisfies condition (1) of the axiom compliance with VR protections, we have
 $|\alpha^{-1}(j,o)| =  r^o_j$, or equivalently
 \[ |\underbrace{C^o_j(\tilde I_j)}_{=  \alpha^{-1}(j,o)}| =  r^o_j. \]
Furthermore, for each $i'\in C_j^o(\tilde I_j)$, we have $\alpha(i')=(j,o)$, and since $\alpha$
satisfies condition (2) of the axiom compliance with VR protections, we have
\[\sigma_j(i') > \sigma_j(i) \; \mbox{ or }\; n_j^o\left(\alpha^{-1}(j,o)\right)>n_j^o\left((\alpha^{-1}(j,o)\setminus \{i'\})\medcup \{i\}\right),\]
or equivalently
\[\sigma_j(i) > \sigma_j(i') \; \implies \;  n_j^o\Big(\underbrace{C^o_j(\tilde I_j)}_{=  \alpha^{-1}(j,o)}\Big) >
n_j^o\Big((\underbrace{C^o_j(\tilde I_j)}_{=  \alpha^{-1}(j,o)}\setminus \{i'\})\medcup \{i\}\Big).\]
Finally, since $\alpha$ satisfies condition (3) of the axiom compliance with VR protections, we have
$n_j^o\Big(\alpha^{-1}(j,o)\medcup \{i\}\Big) \ngtr n_j^o\Big(\alpha^{-1}(j,o)\Big)$,
which in turn  implies
that $n_j^o\Big(\alpha^{-1}(j,o)\medcup \{i\}\Big) = n_j^o\Big(\alpha^{-1}(j,o)\Big)$ since function $n_j^o$ is monotone. Therefore,
\[ n_j^o\Big(\underbrace{C_j^o(\tilde I_j)}_{= \alpha^{-1}(j,o)}\Big) = n_j^o\Big(\underbrace{C_j^o(\tilde I_j)}_{= \alpha^{-1}(j,o)} \medcup \{i\}\Big). \]
Hence, $\vec{C}_j$ complies with VR protections for $\tilde I_j$.\\

We have established that, for any job $j\in\calj$,
the multi-category choice rule $\vec{C}_j$ satisfies
non-wastefulness for $\tilde I_j$, maximal accommodation of HR protections for $\tilde I_j$,
no justified envy for $\tilde I_j$, and compliance with VR protections for $\tilde I_j$.
By Lemma \ref{lem:reschar}, $C_j^v(\tilde I_j) = C_{\circled{M},j}^{2s,v}(\tilde I_j)$ for each $v\in \calv$.
Therefore, for each $j\in\calj$ and $v\in \calv$,
\[\alpha^{-1}(j,v) = C_j^v(\tilde I_j) = C_{\circled{M},j}^{2s,v}(\tilde I_j),\]
and, so, for each $j\in\calj$,
\[ \alpha^{-1}(j) = \bigcup_{v\in\calv} \alpha^{-1}(j,v) =  \bigcup_{v\in\calv} C_{\circled{M},j}^{2s,v}(\tilde I_j) = \widehat{C}_{\circled{M},j}^{2s}(\tilde I_j),\]
completing the proof of Claim \ref{claim:I-tilde}.
\end{proof}

Fix a job $j\in \calj$ and category $v\in \calv$. By construction, we have $\alpha^{-1}(j) \subseteq \tilde I_j$.
Therefore, since
removing a rejected individual does not change the outcome of $\vec{C}_{\circled{M},j}^{2s}$ by Lemma \ref{lem:ircmulti}
and $\alpha^{-1}(j,v)=C_{\circled{M},j}^{2s,v}(\tilde I_j)$ by Claim \ref{claim:I-tilde},  we  have
\[ C_{\circled{M},j}^{2s,v}(\alpha^{-1}(j)) =  \alpha^{-1}(j,v).\]
Hence, $\alpha$ satisfies job-and-category rationality.

To show that there are no blocking pairs, consider an individual-job pair $(i,j) \in \cali\times\calj$ such that $j \succ_i \alpha(i)$.
By the choice of the pair $(i,j)$, we have $j \succ_i \alpha(i)$, and, therefore, by construction we have $i \in {\tilde I_j}=\{\tilde i\in\cali  \;:\;  j \succeq_{\tilde i} \alpha(\tilde i)\}$.
By the choice of the pair $(i,j)$, we also have  $i\notin  \alpha^{-1}(j)$.
Since $\alpha^{-1}(j) = \widehat{C}_{\circled{M},j}^{2s}(\tilde I_j)$  by Claim \ref{claim:I-tilde} and
the aggregate choice rule $\widehat{C}_{\circled{M},j}^{2s}$ satisfies
the irrelevance of rejected individuals condition by Corollary \ref{cor:iri},
we have
\[\widehat{C}_{\circled{M},j}^{2s}(\alpha^{-1}(j)) =
\widehat{C}_{\circled{M},j}^{2s}(\alpha^{-1}(j) \medcup \{i\}) =  \widehat{C}_{\circled{M},j}^{2s}(\tilde I_j) = \alpha^{-1}(j) .\]
Since $i\notin  \alpha^{-1}(j)$, we get $i\notin \widehat{C}_{\circled{M},j}^{2s}(\alpha^{-1}(j) \medcup \{i\})$.
Therefore, there are no blocking pairs.

Hence, we conclude that assignment $\alpha$ is stable with respect to $(\vec{C}_{\circled{M},j}^{2s})_{j\in \calj}$, completing the proof.

\subsection{Proof of Theorem \ref{thm:jointchar}}
We provide the proof in several lemmas. Lemmata \ref{lem:indrash}-\ref{lem:charstrategy}
establish that 2SMH-DA satisfies the five axioms, whereas Theorem \ref{thm:resrulechar} and
Lemmata \ref{lem:stable}-\ref{lem:reschoice} 
establish that it is the only assignment mechanism to do so.

\begin{lemma}\label{lem:indrash}
2SMH-DA satisfies individual rationality.
\end{lemma}

\begin{proof} Fix a preference profile $\succ_{\cali} \, = (\succ_i)_{i\in\cali} \in \calp$.
Let assignment $\alpha = \varphi^{2s}_{\circled{M}}(\succ_{\cali})$ be the outcome of mechanism 2SMH-DA for $\succ_{\cali}$.
Let $i\in\cali$ be any individual.
Since no individual proposes to an unacceptable job under 2SMH-DA,  either $\alpha(i) = \emptyset$ or
 $\alpha(i)=(j,v)$ for a job $j\in \calj$ with $j \succ_i \emptyset$ and category
$v\in \calv$. Therefore, $\alpha(i) \succeq_i \emptyset$, and hence the assignment $\alpha$ satisfies individual rationality.
\end{proof}

\begin{lemma}\label{lem:danonwaste}
2SMH-DA satisfies non-wastefulness.
\end{lemma}

\begin{proof}
Fix a preference profile $\succ_{\cali} \, = (\succ_i)_{i\in\cali} \in \calp$.
Let assignment $\alpha = \varphi^{2s}_{\circled{M}}(\succ_{\cali})$ be the outcome of the mechanism 2SMH-DA for $\succ_{\cali}$.
Suppose that $j\in \calj$, $v\in \calv$, and $i\in \cali$ are such that $j \succ_i \alpha(i)$ and $|\alpha^{-1}(j,v)|<r^v_j$. To show non-wastefulness,
we need to establish that $i\notin \cali^v$.

Let $I$ be the set of individuals who are considered for job $j$ at the last
step of 2SMH-DA.
Since  $j \succ_i \alpha(i)$ (by assumption) and
$\alpha(i) \succeq_i \emptyset$ (by Lemma \ref{lem:indrash}),  we have $j \succ_i \emptyset$.
Therefore, individual $i$ must have applied
to job $j$ at some step of 2SMH-DA, and he must have been rejected by job  $j$ prior to the termination of the algorithm.
Since $\widehat{C}_{\circled{M},j}^{2s}$ satisfies path independence by Lemma \ref{lem:aggregatepi}, we have
$\widehat{C}_{\circled{M},j}^{2s}(I \medcup \{i\})= \widehat{C}_{\circled{M},j}^{2s}(I)$,
which in turn implies $i\notin \widehat{C}_{\circled{M},j}^{2s}(I\medcup \{i\})$. Finally, since
$\vec{C}_{\circled{M},j}^{2s}$ satisfies non-wastefulness by Lemma \ref{lem:reschar}, the relations
$i\notin \widehat{C}_{\circled{M},j}^{2s}(I\medcup \{i\})$ and
$|C_{\circled{M},j}^{2s,v}(I \medcup \{i\})|=|C_{\circled{M},j}^{2s,v}(I)|= |\alpha^{-1}(j,v)|<r^v_j$
imply that $i\notin \cali^v$.
Hence, the assignment $\alpha$ satisfies non-wastefulness.
\end{proof}

\begin{lemma}\label{lem:hrprotect}
2SMH-DA satisfies maximal accommodation of HR protections.
\end{lemma}

\begin{proof}
Fix a preference profile $\succ_{\cali} \, = (\succ_i)_{i\in\cali} \in \calp$.
Let assignment $\alpha = \varphi^{2s}_{\circled{M}}(\succ_{\cali})$ be the outcome of mechanism 2SMH-DA for $\succ_{\cali}$.
Consider a job $j\in \calj$,
category $v\in \calv$, and $i\in \cali^v$ such that $j\succ_i \alpha(i)$. To prove
that 2SMH-DA satisfies maximal accommodation of HR protections, we need to establish that
\[ n_j^v\Big(\alpha^{-1}(j,v) \medcup \{i\}\Big) \ngtr n_j^v\Big(\alpha^{-1}(j,v)\Big).\]

Let $I$ be the set of individuals who are considered for job $j$ at the last step of 2SMH-DA.
Since  $j \succ_i \alpha(i)$ (by assumption) and
$\alpha(i) \succeq_i \emptyset$ (by Lemma \ref{lem:indrash}),  we have $j \succ_i \emptyset$.
Therefore, individual $i$ must have applied
to job $j$ at some step of 2SMH-DA, and he must have been rejected by job $j$ prior to the termination of the algorithm.
Since $\widehat{C}_{\circled{M},j}^{2s}$ satisfies path independence
by Lemma \ref{lem:aggregatepi}, we have
$\widehat{C}_{\circled{M},j}^{2s}(I \medcup \{i\})= \widehat{C}_{\circled{M},j}^{2s}(I)$,
which in turn implies $i\notin \widehat{C}_{\circled{M},j}^{2s}(I\medcup \{i\})$.
Since $\vec{C}_{\circled{M},j}^{2s}$ satisfies maximal accommodation of HR protections
by Lemma \ref{lem:reschar}, the relation $i\in \Big((I\medcup \{i\}) \medcap \cali^v\Big)\setminus \widehat{C}_{\circled{M},j}^{2s}\Big(I\medcup \{i\}\Big)$ implies
\[n_j^v(\alpha^{-1}(j,v))
=n_j^v\Big(C_{\circled{M},j}^{2s,v}(I \medcup \{i\})\Big)
=n_j^v\Big(C_{\circled{M},j}^{2s,v}(I \medcup \{i\}) \medcup \{i\}\Big)
=n_j^v\Big(\alpha^{-1}(j,v) \medcup \{i\}\Big).\]
Therefore,
\[n_j^v\Big(\alpha^{-1}(j,v) \medcup \{i\}\Big) \ngtr n_j^v\Big(\alpha^{-1}(j,v)\Big),\]
which establishes that assignment $\alpha$ satisfies maximal accommodation of HR protections.
\end{proof}

\begin{lemma}
2SMH-DA satisfies no justified envy.
\end{lemma}

\begin{proof}
Fix a preference profile $\succ_{\cali} \, = (\succ_i)_{i\in\cali} \in \calp$.
Let assignment $\alpha = \varphi^{2s}_{\circled{M}}(\succ_{\cali})$ be the outcome of mechanism 2SMH-DA for $\succ_{\cali}$.
Consider $i\in \cali$, $j\in \calj$, $v\in \calv$,
and $i'\in \cali^v$ such that $\alpha(i)=(j,v)$ and
$j \succ_{i'} \alpha(i')$. To prove
that 2SMH-DA satisfies no justified envy,
we need to establish that,
\[ \sigma_j(i) > \sigma_j(i') \;\; \mbox{ or }  \;\;
n_j^v\Big(\alpha^{-1}(j,v)\Big) > n_j^v\Big(\alpha^{-1}(j,v) \setminus \{i\}\big) \medcup \{i'\}\Big).\]
If $\sigma_j(i) > \sigma_j(i')$, then
we are done. For the rest of the proof assume that $\sigma_j(i)<\sigma_j(i')$.

Let $I$ be the set of individuals who are considered for job $j$ at the last step of 2SMH-DA.
Since  $j \succ_{i'} \alpha(i')$ (by assumption) and
$\alpha(i') \succeq_{i'} \emptyset$ (by Lemma \ref{lem:indrash}),  we have $j \succ_{i'} \emptyset$.
Therefore, individual $i'$ must have applied
to job $j$ at some step of 2SMH-DA, and he must have been rejected by job  $j$ prior to the termination of the algorithm.
Since $\widehat{C}_{\circled{M},j}^{2s}$ satisfies path independence
by Lemma \ref{lem:aggregatepi}, we have
$\widehat{C}_{\circled{M},j}^{2s}(I \medcup \{i'\})= \widehat{C}_{\circled{M},j}^{2s}(I)$,
which in turn implies $i'\notin \widehat{C}_{\circled{M},j}^{2s}(I\medcup \{i'\})$.
Therefore,  by Lemma \ref{lem:ircmulti}, we have
\[C_{\circled{M},j}^{2s,v}(I\medcup \{i'\})=C_{\circled{M},j}^{2s,v}(I)=\alpha^{-1}(j,v),\]
which in turn implies $i\in C_{\circled{M},j}^{2s,v}(I\medcup \{i'\})$.

Since $\vec{C}_{\circled{M},j}^{2s}$ satisfies no justified envy by Lemma \ref{lem:reschar}, the relations
$i\in C_{\circled{M},j}^{2s,v}(I\medcup \{i'\})$, \;
$i'\in \Big((I \medcup\{i'\}) \medcap \cali^v\Big) \setminus \widehat{C}_{\circled{M},j}^{2s}(I\medcup \{i'\})$, and
$\sigma_j(i') > \sigma_j(i)$ imply
\[n_j^v\bigg(\Big(\underbrace{C_{\circled{M},j}^{2s,v}(I\medcup \{i'\})}_{=\alpha^{-1}(j,v)}\setminus\{i\}\Big) \medcup \{i'\}\bigg)
< n_j^v\Big(\underbrace{C_{\circled{M},j}^{2s,v}(I\medcup \{i'\})}_{=\alpha^{-1}(j,v)}\Big),\]
which establishes that assignment $\alpha$ satisfies no justified envy.
\end{proof}

\begin{lemma}\label{lem:charver}
2SMH-DA satisfies compliance with VR protections.
\end{lemma}

\begin{proof}
Fix a preference profile $\succ_{\cali} \, = (\succ_i)_{i\in\cali} \in \calp$.
Let assignment $\alpha = \varphi^{2s}_{\circled{M}}(\succ_{\cali})$ be the outcome of mechanism 2SMH-DA for $\succ_{\cali}$.
Suppose that $i\in \cali$ is such that $\alpha(i)=(j,c)$ for some $j\in \calj$ and  $c\in \calr$.
Let $I$ be the set of individuals who are considered for job $j$ at the last step of 2SMH-DA. Then $i\in I$ and $\cenvcat(I)=\alpha^{-1}(j,v)$ for each
$v\in \calv$. Since $\vec{C}^{2s}_{\circled{M},j}$ complies with VR protections by Lemma \ref{lem:reschar}, we have
\begin{enumerate}
\item $|\alpha^{-1}(j,o)|=|C_{\circled{M},j}^{2s,o}(I)|=r^o_j$,
\item for each $i'\in \cali$ with $\alpha(i')=(j,o)$, we have
\[\sigma_j(i) > \sigma_j(i') \; \implies \; n_j^o\Big(C_{\circled{M},j}^{2s,o}(I)\Big)>n_j^o\Big((C_{\circled{M},j}^{2s,o}(I)\setminus \{i'\})\medcup \{i\}\Big),\]
or equivalently
\[\sigma_j(i') > \sigma_j(i) \; \mbox{ or } \; n_j^o\Big(\underbrace{\alpha^{-1}(j,o)}_{=\; C_{\circled{M},j}^{2s,o}(I)}\Big) >
n_j^o\Big((\underbrace{\alpha^{-1}(j,o)}_{= \; C_{\circled{M},j}^{2s,o}(I)} \setminus \{i'\})\medcup \{i\}\Big), \mbox{ and }\]
\item $n^o_j\Big(\alpha^{-1}(j,o)\Big)=n^o_j\Big(C_{\circled{M},j}^{2s,o}(I)\Big)=n^o_j\Big(C_{\circled{M},j}^{2s,o}(I) \medcup \{i\}\Big)=n^o_j\Big(\alpha^{-1}(j,o) \medcup \{i\}\Big)$,
which in turn  implies
\[ n^o_j\Big(\alpha^{-1}(j,o) \medcup \{i\}\Big) \ngtr n^o_j\Big(\alpha^{-1}(j,o)\Big).\]
\end{enumerate}
Therefore, assignment $\alpha$ satisfies compliance with VR protections.
\end{proof}

\begin{lemma}\label{lem:charstrategy}
2SMH-DA satisfies strategy-proofness.
\end{lemma}

\begin{proof}
For any preference profile $\succ_{\cali} \in \calp$ and
individual $i\in \cali$, the job matching mechanism AM-DA assigns individual $i$ to a job
$j\in \calj$ if and only if the (assignment) mechanism 2SMH-DA assigns individual $i$ to a
pair $(j,v)$ where $v\in \calv$ and $i\in \cali^v$. Likewise, the job matching mechanism
AM-DA keeps individual $i\in \cali$ unassigned if and only if the mechanism 2SMH-DA keeps individual $i$ unassigned.
Hence,  for any preference profile $\succ_{\cali} \in \calp$,
the job matching $\widehat{\varphi}_{\circled{M}}^{2s}(\succ_{\cali})$ (that is
generated by  AM-DA) is equal to the job matching that is induced by the
assignment $\varphi_{\circled{M}}^{2s}(\succ_{\cali})$
(which is generated by 2SMH-DA).  Therefore,  for any preference profile $\succ_{\cali} \in \calp$ and individual $i\in \cali$,
\[   \widehat{\varphi}_{\circled{M}}^{2s}(\succ_{\cali})(i) \; \sim_i \;  \varphi_{\circled{M}}^{2s}(\succ_{\cali})(i).
\]

Since the aggregate choice rule $\widehat{C}_{\circled{M},j}^{2s}$ satisfies the substitutes
condition (Lemma \ref{lem:cenvsub}) and the law of aggregate demand
(Lemma \ref{lem:cenvlad}) for each job $j\in\calj$, strategy-proofness of the
job matching mechanism AM-DA follows from Theorem 11 in \cite{hatfield/milgrom:05}.
Finally, since each individual $i\in \cali$ is indifferent between the outcomes of AM-DA and 2SMH-DA for any given preference profile,
strategy-proofness of the job matching mechanism AM-DA implies the strategy-proofness of the assignment mechanism 2SMH-DA as well.
\end{proof}

\begin{lemma}\label{lem:stable}
Let $\alpha$ be an assignment that is stable with respect
to $(\vec{C}_{\circled{M},j}^{2s})_{j\in \calj}$.
Then, the job matching $\mu$ induced by the assignment $\alpha$ is stable with respect to $(\widehat{C}_{\circled{M},j}^{2s})_{j\in \calj}$.
\end{lemma}

\begin{proof}
Let assignment $\alpha$ be stable with respect
to $(\vec{C}_{\circled{M},j}^{2s})_{j\in \calj}$, and $\mu$  be the job matching that is  induced by $\alpha$.
Individual rationality of $\alpha$ implies individual
rationality of $\mu$. Job-and-category rationality of $\alpha$ implies that, for each job $j\in \calj$
and category $v\in \calc$, we have $C_{\circled{M},j}^{2s,v}(\alpha^{-1}(j))=\alpha^{-1}(j,v)$, which implies
$\widehat{C}_{\circled{M},j}^{2s}(\alpha^{-1}(j))=\alpha^{-1}(j)$.
Since $\alpha^{-1}(j)=\mu^{-1}(j)$ by definition,
the last equation is equivalent to $\widehat{C}_{\circled{M},j}^{2s}(\mu^{-1}(j))=\mu^{-1}(j)$.
Hence, $\mu$ satisfies job rationality.  Finally, consider an individual-job pair $(i,j)\in \cali\times\calj$ such that $j\succ_i \mu(i)$. Since $\alpha$ has no
blocking pairs and $\alpha^{-1}(j)=\mu^{-1}(j)$, we have
$i\notin \widehat{C}_{\circled{M},j}^{2s}(\alpha^{-1}(j)\medcup \{i\})=\widehat{C}_{\circled{M},j}^{2s}(\mu^{-1}(j)\medcup \{i\})$. Therefore, there
are no blocking pairs for $\mu$. Hence, $\mu$ is stable with respect to
$\widehat{C}_{\circled{M},j}^{2s}$.
\end{proof}

\begin{lemma}\label{lem:strategy-proof}
Let  $\phi$ be a strategy-proof assignment mechanism and $\widehat{\phi}$ be the job matching mechanism
induced by $\phi$.
Then the job matching mechanism $\widehat{\phi}$ is also strategy-proof.
\end{lemma}

\begin{proof}
Strategy-proofness of  $\widehat{\phi}$  follows from the simple observation that if an individual has a
profitable deviation at the induced job matching mechanism for a given preference profile,
then she has the same profitable deviation at the assignment
mechanism for the same preference profile, because, an individual is
indifferent between the categories of any given job but otherwise have strict preferences
over the set of  jobs and remaining unmatched.
\end{proof}

The following result is a direct implication of Lemma \ref{lem:stable} along with
Lemma \ref{lem:strategy-proof}.

\begin{corollary}\label{lem:stablestrategy}
Let $\phi$ be an assignment mechanism that is strategy-proof and stable with respect
to $(\vec{C}_{\circled{M},j}^{2s})_{j\in \calj}$
and $\widehat{\phi}$ be the job matching mechanism induced by $\phi$.
Then the job matching mechanism $\widehat{\phi}$ is strategy-proof
and stable with respect to $(\widehat{C}_{\circled{M},j}^{2s})_{j\in \calj}$.
\end{corollary}

The following lemma is a generalization of Theorem 3 in \cite{alcalde/barbera:94}.\footnote{There are also similar results in \cite{Hirata/Kasuya:17} and \cite{Hatfield/Kominers/Westkamp:21}.}
\begin{lemma}\label{lemma:dachoice}
Let $\phi$ be a job matching mechanism that is strategy-proof and stable with respect to $(\widehat{C}_{\circled{M},j}^{2s})_{j\in \calj}$.
Then, $\phi=\widehat{\varphi}_{\circled{M}}^{2s}$.
\end{lemma}

\begin{proof}
Towards a contradiction, suppose that mechanism $\phi$ is strategy-proof
and stable with respect to $(\widehat{C}_{\circled{M},j}^{2s})_{j\in \calj}$,
but it differs than the mechanism AM-DA. Then there exists a preference
profile $\succ_{\cali}=(\succ_i)_{i\in \cali}$ such that
$\phi(\succ_{\cali})$ is different than the outcome $\widehat{\varphi}_{\circled{M}}^{2s}(\succ_{\cali})$ of AM-DA.
Therefore, there exists an individual $i\in \cali$ such that $\phi(\succ_{\cali})(i) \neq \widehat{\varphi}_{\circled{M}}^{2s}(\succ_{\cali})(i)$.
Since, for every job $j$, the choice rule $\widehat{C}_{\circled{M},j}^{2s}$ satisfies the substitutes condition (Lemma \ref{lem:cenvsub}) and the law of aggregate
demand (Lemma \ref{lem:cenvlad}), AM-DA produces the individual-optimal stable matching \citep{hatfield/milgrom:05,aygson12a}. Therefore,
\[ \widehat{\varphi}_{\circled{M}}^{2s}(\succ_{\cali})(i) \; \succ_i \; \phi(\succ_{\cali})(i).\]

Since $\phi$ is individually rational, we have
$\phi(\succ_{\cali})(i) \succeq_i \emptyset$. Therefore,
$\widehat{\varphi}_{\circled{M}}^{2s}(\succ_{\cali})(i) \succ_i \emptyset$, which in turn implies
$\widehat{\varphi}_{\circled{M}}^{2s}(\succ_{\cali})(i) \in \calj$. Let $\succ'_i$ be a preference relation
where only job $\widehat{\varphi}_{\circled{M}}^{2s}(\succ_{\cali})(i)$ is acceptable. Since $\widehat{\varphi}_{\circled{M}}^{2s}(\succ_{\cali})$ is
stable under $\succ_{\cali}$, it is also
stable under $(\succ'_i,\succ_{\cali \setminus \{i\}})$. For every job $j$, the choice rule $\widehat{C}_{\circled{M},j}^{2s}$ satisfies
the substitutes condition (Lemma \ref{lem:cenvsub}) and the law of aggregate
demand (Lemma \ref{lem:cenvlad}).
Therefore, by Theorem 8 in \cite{hatfield/milgrom:05} (which is also known as the rural hospitals theorem),  the job matching $\phi(\succ'_i,\succ_{\cali \setminus \{i\}})$
assigns individual $i$ the same number of partners as in job matching $\widehat{\varphi}_{\circled{M}}^{2s}(\succ_{\cali})$.
Since  $\widehat{\varphi}_{\circled{M}}^{2s}(\succ_{\cali})(i) \in \calj$ and the only acceptable job for $i$ under $\succ'_i$ is
$\widehat{\varphi}_{\circled{M}}^{2s}(\succ_{\cali})(i)$, we have
$\phi(\succ'_i,\succ_{\cali \setminus \{i\}})(i)=\widehat{\varphi}_{\circled{M}}^{2s}(\succ_{\cali})(i)$.
Hence,
\[\underbrace{\phi(\succ'_i, \succ_{\cali \setminus \{i\}})(i)}_{ = \widehat{\varphi}_{\circled{M}}^{2s}(\succ_{\cali})(i)} \; \succ_i \; \phi(\succ_{\cali})(i),\]
contradicting strategy-proofness of mechanism $\phi$, and completing the proof of the lemma.
\end{proof}

\begin{lemma}\label{lem:reschoice}
Fix a preference profile $(\succ_i)_{i\in \cali}$. Let  the job matching
$\mu=\widehat{\varphi}_{\circled{M}}^{2s}(\succ_{\cali})$ be the outcome of the job matching mechanism AM-DA under $(\succ_i)_{i\in \cali}$.
Let assignment $\alpha$ be such that,
\begin{enumerate}
\item $\alpha^{-1}(j)=\mu^{-1}(j)$ for each job $j\in \calj$, and
\item $\alpha$ satisfies non-wastefulness, maximal accommodation of HR protections, no justified envy, and
compliance with VR protections.
\end{enumerate}
Then $\alpha=\varphi_{\circled{M}}^{2s}(\succ_{\cali})$.
\end{lemma}

\begin{proof}
Fix a preference profile $(\succ_i)_{i\in \cali}$. Let the job matching $\mu=\widehat{\varphi}_{\circled{M}}^{2s}(\succ_{\cali})$ and the assignment $\alpha$
be given as in the statement of the lemma.
Observe that, the mechanics of the
job matching mechanism AM-DA  is identical to the mechanics of the assignment mechanism 2SMH-DA,
and the two procedures only differ in the structure of their outcomes.
AM-DA only specifies individuals' job assignments.
In addition to specifying individuals' job assignments,
2SMH-DA also specifies their category assignments.
Since the set of individuals under consideration by any given job $j\in\calj$ at the last step of both procedures
is $\alpha^{-1}(j)=\mu^{-1}(j)$, all we have to show is,  for any job $j\in\calj$ and category $v\in\calv$,
\[  C_{\circled{M},j}^{2s,v}(\alpha^{-1}(j)) = \alpha^{-1}(j,v).
\]

Since the job matching $\mu$ satisfies individual rationality, so does the assignment $\alpha$.
Fix a job  $j\in \calj$.
Let $\tilde I_j=\{\tilde i\in\cali  :   j \succeq_{\tilde i} \alpha(\tilde i)\}$.
Since assignment $\alpha$ satisfies individual rationality,  non-wastefulness, maximal accommodation of HR protections, no justified envy, and
compliance with VR protections, by Claim \ref{claim:I-tilde} (in the proof of Theorem \ref{thm:resrulechar}), we have
\[ C_{\circled{M},j}^{2s,v}(\tilde I_j) = \alpha^{-1}(j,v) \qquad \mbox{ for each } v\in\calv. \]
Furthermore, since (i) $\alpha^{-1}(j)\subseteq \tilde I_j$ by construction and
(ii) $\vec{C}_{\circled{M},j}^{2s}$ does not depend on the rejected individuals by repeated application of
Lemma \ref{lem:ircmulti},  for any category  $v\in\calv$,
we have  $C_{\circled{M},j}^{2s,v}(\alpha^{-1}(j))=\alpha^{-1}(j,v)$ as desired.
Hence, $\alpha=\varphi_{\circled{M}}^{2s}(\succ_{\cali})$.
\end{proof}

We are ready to establish that 2SMH-DA is the unique assignment mechanism that satisfies the five axioms.
Theorem \ref{thm:resrulechar} shows that any assignment mechanism
that satisfies the axioms has to be  stable with respect to $(\vec{C}_{\circled{M},j}^{2s})_{j\in \calj}$. Corollary \ref{lem:stablestrategy} shows
that for any strategy-proof assignment mechanism which is stable with respect to  $(\vec{C}_{\circled{M},j}^{2s})_{j\in \calj}$,
the induced matching mechanism is strategy-proof and
stable with respect to $(\widehat{C}_{\circled{M},j}^{2s})_{j\in \calj}$.  Therefore, for any
assignment mechanism  that satisfies the axioms, the induced matching
mechanism is strategy-proof and  stable with respect to
$(\widehat{C}_{\circled{M},j}^{2s})_{j\in \calj}$. Lemma \ref{lemma:dachoice} shows that AM-DA is
the unique job matching mechanism that is strategy-proof and  stable with respect to
$(\widehat{C}_{\circled{M},j}^{2s})_{j\in \calj}$. Finally, Lemma \ref{lem:reschoice} establishes that
for any assignment that satisfies the axioms, the induced job matching is
the outcome of AM-DA only if the assignment is the outcome of 2SMH-DA.
This concludes the proof of Theorem  \ref{thm:jointchar}.

\subsection{Proof of Theorem \ref{thm:dominate}}
Let $\varphi$ be a mechanism that satisfies
individual rationality,  non-wastefulness,  maximal accommodation of HR protections,
no justified envy,  and  compliance with VR protections. Fix a preference profile  $\succ_{\cali} \in \calp$.
Let  $\mu$ be the job matching that is induced by the assignment $\varphi(\succ_{\cali})$.

Since job matching $\mu$ is induced by the assignment $\varphi(\succ_{\cali})$,
for each $i\in\cali$, we have
\[ \mu(i) \; \sim_i \; \varphi(\succ_{\cali})(i).
\]
Similarly, since job matching $\widehat{\varphi}_{\circled{M}}^{2s}(\succ_{\cali})$ is induced by the
assignment $\varphi_{\circled{M}}^{2s}(\succ_{\cali})$,
for each $i\in\cali$, we have
\[   \widehat{\varphi}_{\circled{M}}^{2s}(\succ_{\cali})(i) \; \sim_i \;  \varphi_{\circled{M}}^{2s}(\succ_{\cali})(i).
\]
By Theorem \ref{thm:resrulechar}, the assignment  $\varphi(\succ_{\cali})$ is stable with respect to
$(\vec{C}_{\circled{M},j}^{2s})_{j\in \calj}$.
Therefore, the job matching $\mu$ that is induced by the assignment $\varphi(\succ_{\cali})$ is stable with respect to
$(\widehat{C}_{\circled{M},j}^{2s})_{j\in \calj}$ by Lemma \ref{lem:stable}.
For any job $j\in\calj $ the choice rule $\widehat{C}_{\circled{M},j}^{2s}$ satisfies the
substitutes condition by Lemma \ref{lem:cenvsub} and the law of aggregate demand by Lemma \ref{lem:cenvlad}.
Therefore, stability of the job matching $\mu$ and  Theorem 4 in \cite{hatfield/milgrom:05} together imply,\footnote{Strictly speaking
\cite{hatfield/milgrom:05} states this result under an implicit assumption of irrelevance of rejected individuals condition,
which is implied by the substitutes condition together with the law of aggregate demand. See \cite{aygson12a} for
further details.}  for each $i\in\cali$,
\[   \widehat{\varphi}_{\circled{M}}^{2s}(\succ_{\cali})(i) \; \succeq_i \; \mu(i) .
\]
Hence, for each $i\in\cali$,
\[ \varphi_{\circled{M}}^{2s}(\succ_{\cali})(i) \; \sim_i \;  \widehat{\varphi}_{\circled{M}}^{2s}(\succ_{\cali})(i) \; \succeq_i \; \mu(i)  \; \sim_i \; \varphi(\succ_{\cali})(i),
\]
establishing that, for any preference profile  $\succ_{\cali} \in \calp$, either
\begin{enumerate}
\item the job matching  $\widehat{\varphi}_{\circled{M}}^{2s}(\succ_{\cali})$ is equal to the job matching $\mu$ that is induced
by the assignment $\varphi(\succ_{\cali})$,  or
\item the assignment $\varphi_{\circled{M}}^{2s}(\succ_{\cali})$ Pareto dominates the assignment $\varphi(\succ_{\cali})$.
\end{enumerate}
However, since the assignment $\varphi(\succ_{\cali})$ satisfies all five axioms, under the first possibility
it must be equal to the assignment $\varphi_{\circled{M}}^{2s}(\succ_{\cali})$ by Lemma  \ref{lem:reschoice}.
This establishes that the assignment mechanism  $\varphi_{\circled{M}}^{2s}$ Pareto dominates
any other assignment mechanism that satisfies the five axioms, concluding the proof of Theorem   \ref{thm:dominate}.

\end{document}